\documentclass[prx,amssymb,showkeys,floatfix,superscriptaddress,
reprint%,
]{revtex4-2}
\usepackage{bibentry}
\usepackage{xcolor}
\usepackage{graphicx}% Include figure files
\usepackage{dcolumn}% Align table columns on decimal point
\usepackage{bm}% bold math
\usepackage{xfrac}
\usepackage{amsmath}
\usepackage{amssymb}
\usepackage{amsthm} % for variations on Theorem environment
\usepackage{physics}
\usepackage{cancel}
\usepackage{mathtools}
\usepackage[normalem]{ulem}
\newtheorem{theorem}{Theorem}
\theoremstyle{definition}
\newtheorem*{definition}{Definition}

\newcommand{\Operator}[2]{\ensuremath{#1_{\textnormal{#2}}}}
\newcommand{\OperatorSuper}[3]{\ensuremath{#1_{\textnormal{#2}}^{\textnormal{#3}}}}
\newcommand{\Hmixer}{\Operator{H}{mix}}
\newcommand{\Hcost}{\Operator{H}{cost}}
\newcommand{\Hinst}{\Operator{H}{inst}}
\newcommand{\Uinst}{\Operator{U}{inst}}
\newcommand{\HQAOA}{\Operator{H}{\textsc{qaoa}}}
\newcommand{\Ucost}{\Operator{U}{cost}}
\newcommand{\UQAOA}{\Operator{U}{\textsc{qaoa}}}
\newcommand{\RQAOA}{\Operator{R}{\textsc{qaoa}}} % residual matrix, used instead of $R_{\text{QAOA}}$ for consistency with \UQAOA.
\newcommand{\DeltaCrit}{\Operator{\Delta}{crit}}
\newcommand{\cEff}{\OperatorSuper{c}{eff}{x,y}}
\newcommand{\cEffspecific}{\Operator{c}{eff}}
\newcommand{\region}[1]{\textsc{#1}}

\newcommand{\Fig}[1]{Fig.~\ref{f:#1}}

\newcommand{\Eq}[1]{Eq.~\eqref{e:#1}}
\newcommand{\Eqs}[2]{Eqs.~\eqref{e:#1} and~\eqref{e:#2}}

\newcommand{\Section}[1]{Section~\ref{s:#1}}

\newcommand{\Appendix}[1]{Appendix~\ref{s:#1}}

\renewcommand{\thesection}{\arabic{section}}
\renewcommand{\thesubsection}{\thesection.\arabic{subsection}}
\renewcommand{\thesubsubsection}{\thesubsection.\arabic{subsubsection}}
\makeatletter
\renewcommand{\p@subsection}{}
\renewcommand{\p@subsubsection}{}
\makeatother

\usepackage[english]{babel}
\usepackage{hyphenat}
\usepackage{graphicx}% Include figure files
\usepackage{dcolumn}% Align table columns on decimal point
\usepackage{bm}% bold math
\usepackage{xfrac}
\usepackage{colortbl}

\begin{document}
%\title{Exotic holonomies in the Quantum Alternating Operator Ansatz (QAOA) and their applications}
\title{Quantum Alternating Operator Ansatz (QAOA) beyond \\ low depth with gradually changing unitaries }

\author{Vladimir Kremenetski }
\affiliation{Applied and Engineering Physics, Cornell University, Ithaca, NY 14853}
%\affiliation{Quantum Artificial Intelligence Laboratory (QuAIL), Exploration Technology Directorate, NASA Ames Research Center, Moffett Field, CA 94035, USA}
\affiliation{NASA Ames Research Center, Moffett Field, CA 94035, USA}
\affiliation{USRA Research Institute for Advanced Computer Science, Mountain View, CA 94043, USA}

\author{Anuj Apte}
%\affiliation{Quantum Artificial Intelligence Laboratory (QuAIL), Exploration Technology Directorate, NASA Ames Research Center, Moffett Field, CA 94035, USA}
\affiliation{NASA Ames Research Center, Moffett Field, CA 94035, USA}
\affiliation{Department of Physics, University of Chicago, Chicago, IL 60637, USA}

\author{Tad Hogg}
\affiliation{NASA Ames Research Center, Moffett Field, CA 94035, USA}
%\affiliation{Quantum Artificial Intelligence Laboratory (QuAIL), Exploration Technology Directorate, NASA Ames Research Center, Moffett Field, CA 94035, USA}

\author{Stuart Hadfield}
\affiliation{NASA Ames Research Center, Moffett Field, CA 94035, USA}
%\affiliation{Quantum Artificial Intelligence Laboratory (QuAIL), Exploration Technology Directorate, NASA Ames Research Center, Moffett Field, CA 94035, USA}
\affiliation{USRA Research Institute for Advanced Computer Science, Mountain View, CA 94043, USA}

\author{Norm M. Tubman}
\email{norman.m.tubman@nasa.gov}
\affiliation{NASA Ames Research Center, Moffett Field, CA 94035, USA}
%\affiliation{Quantum Artificial Intelligence Laboratory (QuAIL), Exploration Technology Directorate, NASA Ames Research Center, Moffett Field, CA 94035, USA}
\date{\today}

\begin{abstract}
% fixed intros

The Quantum Approximate Optimization Algorithm and its generalization to Quantum Alternating Operator Ans\"atze (QAOA) is a promising approach for applying quantum computers to  challenging problems such as combinatorial optimization and computational chemistry. In this paper, we study the underlying mechanisms governing the behavior of QAOA circuits beyond shallow depth in the practically relevant setting of gradually varying unitaries. We use the discrete adiabatic theorem, which complements and generalizes the insights obtained from the continuous-time adiabatic theorem primarily considered in prior work. Our analysis explains some general properties that are conspicuously depicted in the recently introduced \textit{QAOA performance diagrams}. For parameter sequences derived from continuous schedules (e.g. linear ramps), these diagrams capture the algorithm's performance over different parameter sizes and circuit depths. Surprisingly, they have been observed to be qualitatively similar across different performance metrics and application domains. Our analysis explains this behavior as well as entails some unexpected results, such as connections between the eigenstates of the cost and mixer QAOA Hamiltonians changing based on parameter size and the possibility of reducing circuit depth without sacrificing performance.

%\tad{Careful with claims about performance in the abstract: late in the paper (Section 7) we note we don't have theory for absolute performance on the ridge. Instead the ``comparable performance'' comes from numerical studies such as previously done in our chemistry paper.}

\end{abstract}
\maketitle

%temp!
%\tableofcontents

\section{Introduction}

Quantum computing presents numerous promising approaches to challenging computational problems across a wide variety of applications. Among these approaches are the Quantum Approximate Optimization Algorithm~\cite{farhi_qaoa} and Quantum Alternating Operator Ansatz (QAOA)~\cite{hadfield2019quantum}, developed to solve classical combinatorial optimization problems, and later extended to more general tasks such as state preparation~\cite{ho2019efficient,kremenetski21diagram}. 
In its simplest realization, given a cost Hamiltonian to optimize, the protocol starts with the ground state of a mixer Hamiltonian and then proceeds by alternately evolving the state under the two Hamiltonians $p$ times each. While there are results on QAOA behavior for small $p$~\cite{farhi_qaoa_p1_supremacy, wang2018quantum, wurtz_and_love_lowp, vikst2020, garciaripoli2022,akshay2022, streif2019} and in some cases $p \gg 1$~\cite{farhi_qaoa,jiang2017, rosa2021, deepcircuits2022}, there remains the challenge of understanding the behavior of QAOA more broadly. Generally, QAOA is characterized by parameters that represent how long the Hamiltonians are simulated at each step in the protocol, and therefore parameter setting is expected to become challenging as $p$ grows~\cite{zhou2020,leng2022,wurtz_and_lhykov,brady_et_al,deepcircuits2022}.
%Some work has been done on this such as numerical benchmarking of polynomial-time protocols for quasi-optimal settings of parameters~\cite{zhou2020,leng2022}, numerical evidence for the fixed angle conjecture~\cite{wurtz_and_lhykov}, or broad theoretical descriptions of optimal schedules~\cite{brady_et_al} and problem settings~\cite{deepcircuits2022}.

To better understand the behavior of QAOA, we introduce an approach to analyzing deep circuits with gradually varying unitaries. This extends beyond the small-parameter regime of QAOA considered as Trotterized annealing in prior work~\cite{farhi_qaoa, kocia2022, sack2021, pelofske2023}. Through our analysis, we show such deep circuits exhibit surprising phenomena. One is the ground state of the mixer Hamiltonian directly connecting to a highly excited state of the cost Hamiltonian, resulting in poor QAOA performance even in an adiabatic limit. Another is shallower circuits outperforming deeper ones at large parameters due to \emph{inherently} non-adiabatic effects (performing comparably to deeper circuits at smaller parameters). We also show how these phenomena, along with small-parameter approximations, explain a general qualitative feature~\cite{kremenetski21diagram} in performance for deep QAOA circuits with slowly varying parameters. 

Specifically, we apply the discrete adiabatic theorem~\cite{dranov98}. This theorem is a generalization of the continuous adiabatic theorem, where a state evolving in time tracks an eigenstate across a product of gradually varying unitaries rather than the eigenstate of a slowly changing Hamiltonian. Since unitary eigenvalues can wrap around the complex unit circle and produce avoided crossings, when QAOA parameters are large, this wrap-around leads to the mixer ground state becoming connected to a high-lying cost excited state. This change in connectivity is described by the mathematical theory of holonomy~\cite{cheon09} that generalizes the well-known phenomenon of Berry phase \cite{Berry1984, Wilczek1984}. Consequently, the state resulting from QAOA has high support on the high-lying excited state in the discrete adiabatic limit (arbitrarily large $p$). Outside of this limit, we characterize how well the state tracks the changing eigenstate through small gaps using a discretized Landau-Zener formula.  

\begin{figure}
    \centering
    \includegraphics[width=0.45\textwidth]{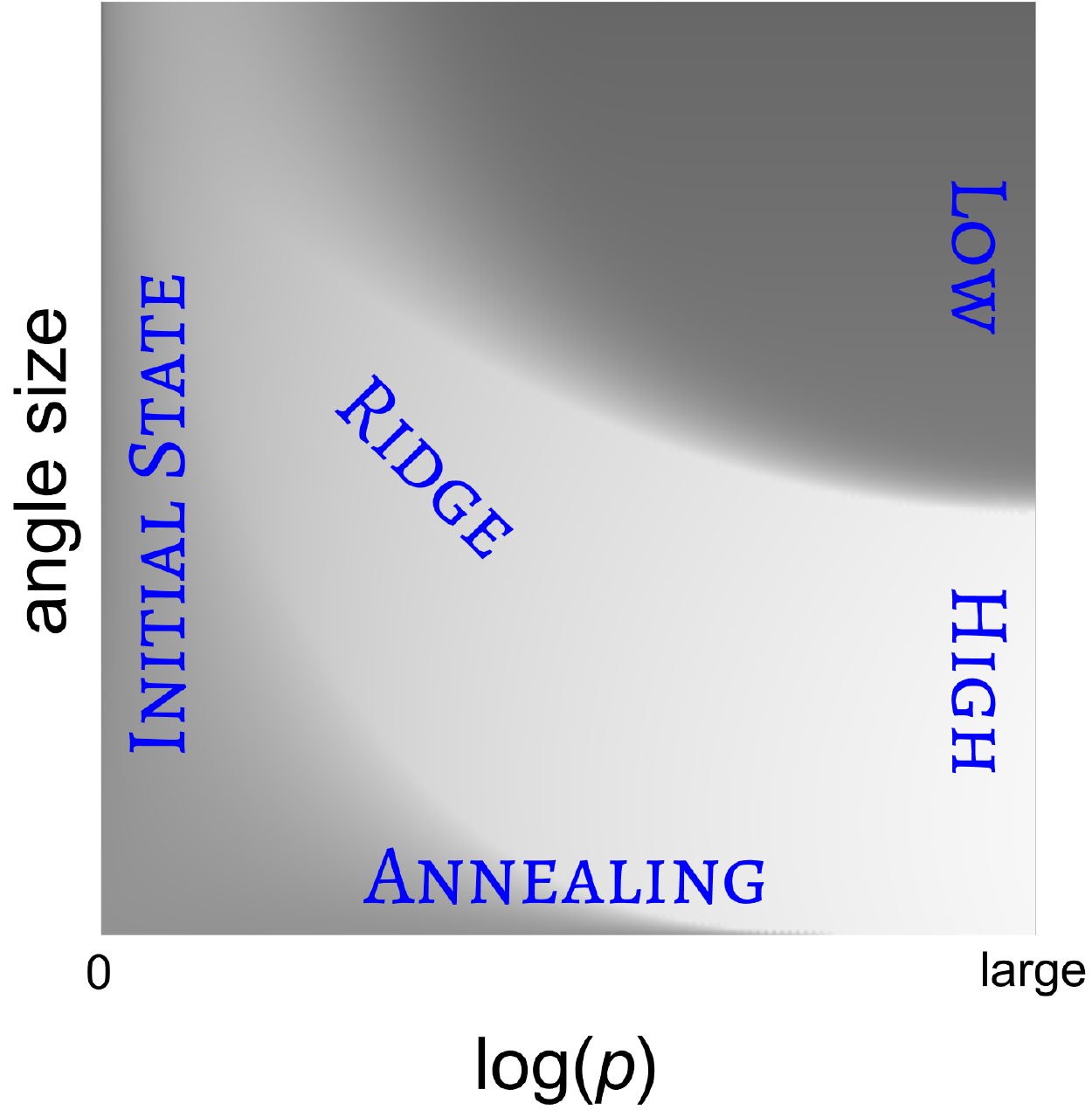}
    \caption{Schematic QAOA performance diagram. The vertical axis characterizes the sum of the pairs of angles' magnitudes in the QAOA schedule. Shading indicates performance, with lighter shading corresponding to higher performance, e.g., higher overlap with the target state or lower expected cost. The blue labels indicate the qualitative behavior regions described in \Section{performance_diagrams}.
    %\sh{[SH: We should make different regions we consider (in Sec 5) more explicit here. Can we make this figure color too? Current shading seems off as mid right looks better than bottom right?]}
    }
    \label{f:general_behavior}
\end{figure}

This new analysis explains the performance features of QAOA seen in a recent study~\cite{kremenetski21diagram} that applied QAOA with parameters derived from a linear ramp to electronic structure problems. Qualitatively similar behavior in performance was observed for intermediate $p$ and in the $p \gg 1$ limit for large angles across a wide variety of problems, a generality that persisted with nonlinear, smooth schedules. The general nature of this behavior is sketched in Fig.~\ref{f:general_behavior}. The labels in \Fig{general_behavior} indicate regions with qualitatively different performance, though the regions are not entirely disjoint and the boundaries between them are not sharp.
To the left and bottom of the diagram (labeled \region{Initial State}), the performance is low and close to that of the starting state; while in the lower-right region (\region{High}), the performance is high and approaches continuous adiabatic behavior.
The upper-right region (\region{Low}) has very poor performance.
At the bottom, QAOA in the \region{Annealing} region closely approximates continuous annealing.
Toward the upper left, the \region{Ridge} region has relatively high performance. At large values of $p$, there is a sudden drop in performance when the angle size exceeds a threshold value: an abrupt change between \region{High} and \region{Low} regions. In this paper, we argue that this performance drop is due to a change in connectivity of eigenstates associated with QAOA operators, which causes the QAOA state vector to end in an excited state of the cost Hamiltonian. 

The paper is structured as follows. We introduce QAOA in Section~\ref{s:qaoa_overview}. Section~\ref{s:dat} introduces the discrete adiabatic theorem. Section~\ref{s:QAOA operator behavior}\ studies the connections between eigenstates of the product pairs of unitaries that make up a QAOA protocol. Section~\ref{s:performance_diagrams}  introduces and explains the qualitative generality of QAOA performance diagrams. Section~\ref{s:examples} gives examples with more complex performance diagram behavior, and Section~\ref{s:generality} illustrates how our results apply to schedules with gradually changing angles beyond those easily captured by performance diagrams and how these diagrams scale with problem size. Additional technical derivations are provided as appendices.

%\tad{The above paragraph uses the term ``topological  connections''.  Is that a (well-known?) distinct type of connection different from some other sorts? Earlier in the intro we just use ``connection'', e.g., between mixer ground state and high-lying excited cost state. Or do we mean ``topological'' in the sense of study how connections change is part of the approach discussed in this paper; rather than a type of connection? Or delete  ``topological''?}

\section{QAOA with gradually changing parameters}\label{s:qaoa_overview}
%RESOLVED#
%\tad{Notation: later in the paper with use $n$ for number of spins and variables. Any reason to keep $N$ here instead of change to $n$?}

For QAOA, a given optimization problem on $n$ variables is encoded as a cost Hamiltonian $\Hcost$ on $n$ qubits.
Level-$p$ QAOA uses $\Hcost$ and a %``mixer'' 
suitably chosen mixer Hamiltonian $\Hmixer$ to apply $p$ parameterized unitary operators
\begin{equation}
\UQAOA^{(j)} = e^{-i\beta_j \Hmixer} e^{-i\gamma_j  \Hcost}~,
\label{e:general unitary}
\end{equation}
for $j=1,\ldots,p$, with parameters $\vec{\beta}=(\beta_1,\ldots,\beta_p)$ and $\vec{\gamma}=(\gamma_1,\ldots,\gamma_p)$, also referred to as ``angles". We refer to the operators $\exp(-i\beta_j\Hmixer)$ and $\exp(-i\gamma_j\Hcost)$ as the mixer and cost unitaries, respectively. 
% moved the followiong sentence to after we first mention mixer
We consider the general setting for QAOA, where the mixer and the cost Hamiltonians do not commute but are otherwise arbitrary~\cite{hadfield2019quantum}, which includes a wide range of problems from combinatorial optimization as well as quantum chemistry~\cite{kremenetski21diagram}. 
%SH: revised and moved to end of subsection
%\sh{%We primarily study 
%Our primary focus is the %setting of varying depth $p$ for fixed problems of %given size $n$.}
%In this work our primraa

QAOA starts with an easily prepared initial state $\ket{\psi_0}$, which is typically the ground state of  $\Hmixer$. For example, for the  originally proposed transverse-field mixer (also known as the $X$-mixer)  $\Hmixer=-(\sigma^x_1+\dots+\sigma^x_n)$, $\ket{\psi_0} = \ket{+}^{\otimes n}$, an equally weighted superposition over the computational basis.
QAOA produces the output state 
\begin{equation} \label{e:QAOAstate}
    \ket{\psi_p(\vec{\gamma},\vec{\beta})} = \UQAOA^{(p)} \ldots \UQAOA^{(1)} \ket{\psi_0}~.
    %e^{-i\beta_pH_B}e^{-i\gamma_pH_C}...e^{-i\beta_1H_B}e^{-i\gamma_1H_C}\ket{\psi_0}.
\end{equation}
Ideally, this QAOA output state has high overlap with the ground state of \Hcost. For combinatorial optimization, it suffices to have non-negligible support on the ground state in order to solve the problem with repeated state preparation and measurement, though for challenging problems we often must settle for an approximate solution i.e., obtaining sufficient support on a low-lying eigenstate. On the other hand, chemistry applications such as electronic structure problems often require a much higher overlap with the true ground state~ \cite{kremenetski21diagram}.

%When evaluating the effectiveness of a protocol, it is common in QAOA research to use the expected cost energy of a prepared state instead of squared overlap with the target ground state~\cite{exact_vs_approximate}, since it is not expected that the ground states of hard optimization problems can be efficiently prepared. This performance metric will be  additionally useful for our purposes since expectation value reflects which states the state vector gains support on, as opposed to the fidelity of ground state preparation on its own. 
A common measure of the effectiveness of QAOA is the expected cost (or energy) $\langle \Hcost \rangle$ of the output state $\ket{\psi_p(\vec{\gamma},\vec{\beta})}$. Typically, it may be efficiently estimated via repeated preparations and measurements. %of the QAOA state.  
In some cases this quantity is used as a proxy for the true overlap with the ground state, which may be as difficult to compute as the underlying problem itself. On the other hand, for small or moderately sized benchmarking problems we may often compute the ground state overlap directly \cite{kremenetski21diagram,exact_vs_approximate}. 
We provide empirical evidence that both metrics produce similar phenomena, though we primarily use ground state overlap as the QAOA performance metric.

We consider QAOA parameter schedules with gradually changing parameters, i.e., schedules for which $|\gamma_j-\gamma_{j+1}|,|\beta_j-\beta_{j+1}|$ are small (relative to the inverse of the norms of $\Hmixer$ and $\Hcost$). Specifically, we consider
a \textit{continuous schedule} to be one generated through samples from continuous functions $\beta(f), \gamma(f)$ on $[0,1]$, so that $\beta_j=\beta(f_j)$, $\gamma_j=\gamma(f_j)$ where
\begin{equation}
    f_j = \frac{j}{p+1},
    \label{e:f}
\end{equation}
 %\sh{[SH: need to make sure 'continuous QAOA parameter schedules' is precisely defined. Having clear definitions here makes later section 5+6 easier to read/write]}
%Our results apply to general continuous QAOA parameter schedules. 
for $j=1,\ldots,p$, and $p\gg 1$. For simplicity, we focus on linear ramps
\begin{equation}
    \gamma(f) = \Delta f  \quad\mbox{and}\quad  \beta(f) = \Delta(1-f)~,
    \label{e:gamma and beta}
\end{equation}
where $\Delta$ is a constant.
Section~\ref{s:generality} discusses more general schedules.

%The version of QAOA considered here uses, at each layer, 
For the linear schedule of \Eqs{f}{gamma and beta}, 
the \emph{QAOA unitary} of \Eq{general unitary} becomes
\begin{equation}
\UQAOA(\Delta, f) = e^{-i\Delta (1-f) \Hmixer} e^{-i\Delta f \Hcost}~,
\label{e:unitary}
\end{equation}
for fixed $\Delta$. 
% Tad: no need to say U varies slowly for large p here; that repeats what we say  at start of Discreate Adiabatic section
%For large $p$, this sequence of QAOA unitaries is slow-varying. %, even as $\Delta$ is increased away from the usual adiabatic limit.  
% 
%, \sh{with $f_0=0$ and $f_{p+1}=1$.}
Similar linear schedules are frequently considered for QAOA in optimization settings~\cite{hogg2000,crooks2018performance,mbeng2019quantum,zhou2020,shaydulin2020classical,sack2021quantumannealing, streif2020}. In particular, this choice %of schedule avoids 
alleviates the parameter setting problem, which generally suffers from the \textit{curse of dimensionality} as $p$ becomes large~\cite{mcclean2018}, and 
%as in~\cite{kremenetski21diagram} 
so enables the study of much deeper QAOA circuits~\cite{kremenetski21diagram} 
than is possible for general parameter choices.  
%REVISIT
Hence, our primary focus in this work will be the effect of increasing the depth $p$ for fixed problems of given size $n$.

\begin{figure}
    \centering
    \includegraphics[width=0.4\textwidth]{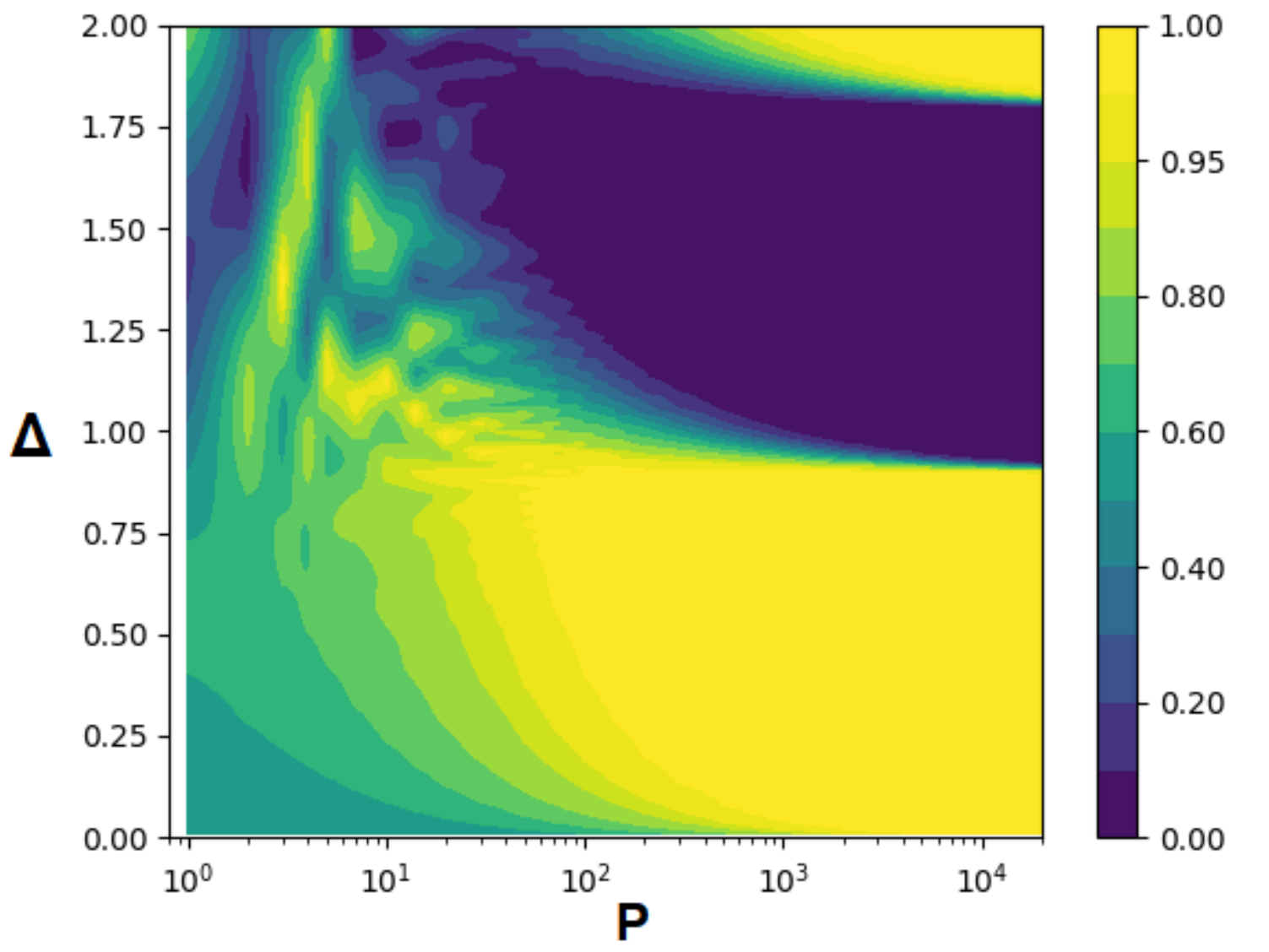}
    \caption{Performance diagram displaying squared overlap of the QAOA output state with the cost ground state, for the pair of Hamiltonians defined in Eq.~\ref{e:2-level example}.}
    \label{f:qaoa_diagram_1qubit_example}
\end{figure}

With this choice of schedule %conveniently reduces 
the number of free QAOA parameters is conveniently reduced to just two: $p$ and $\Delta$. Plotting performance vs.~these parameters gives a 2D plot we call a ``performance diagram"~\cite{kremenetski21diagram}, such as shown in \Fig{qaoa_diagram_1qubit_example} and illustrated above in \Fig{general_behavior}.\footnote{In~\cite{kremenetski21diagram} the performance diagrams are also referred to as ``QAOA phase diagrams''.}

%\sh{[SH: I think we need at least a short subsection here spelling out for reader definition and motivation of QAOA performance diagrams. Description in Intro is quite sparse...]}

\section{The Discrete Adiabatic Regime}
\label{s:dat}

For parameters derived from continuous schedules discussed in \Section{qaoa_overview}, the QAOA unitary changes very little between each step when $p$ is large. This section describes QAOA in this regime from the viewpoint of the discrete adiabatic theorem. 
% this sentence repeats the statement made in Section 3.2
%In particular, discrete sequences of slowly changing unitaries can yield qualitatively different behavior from that of a slowly varying Hamiltonian.

\begin{figure*}[t]
\centering 
\begin{tabular}{cc}
\includegraphics[width=0.45\textwidth]{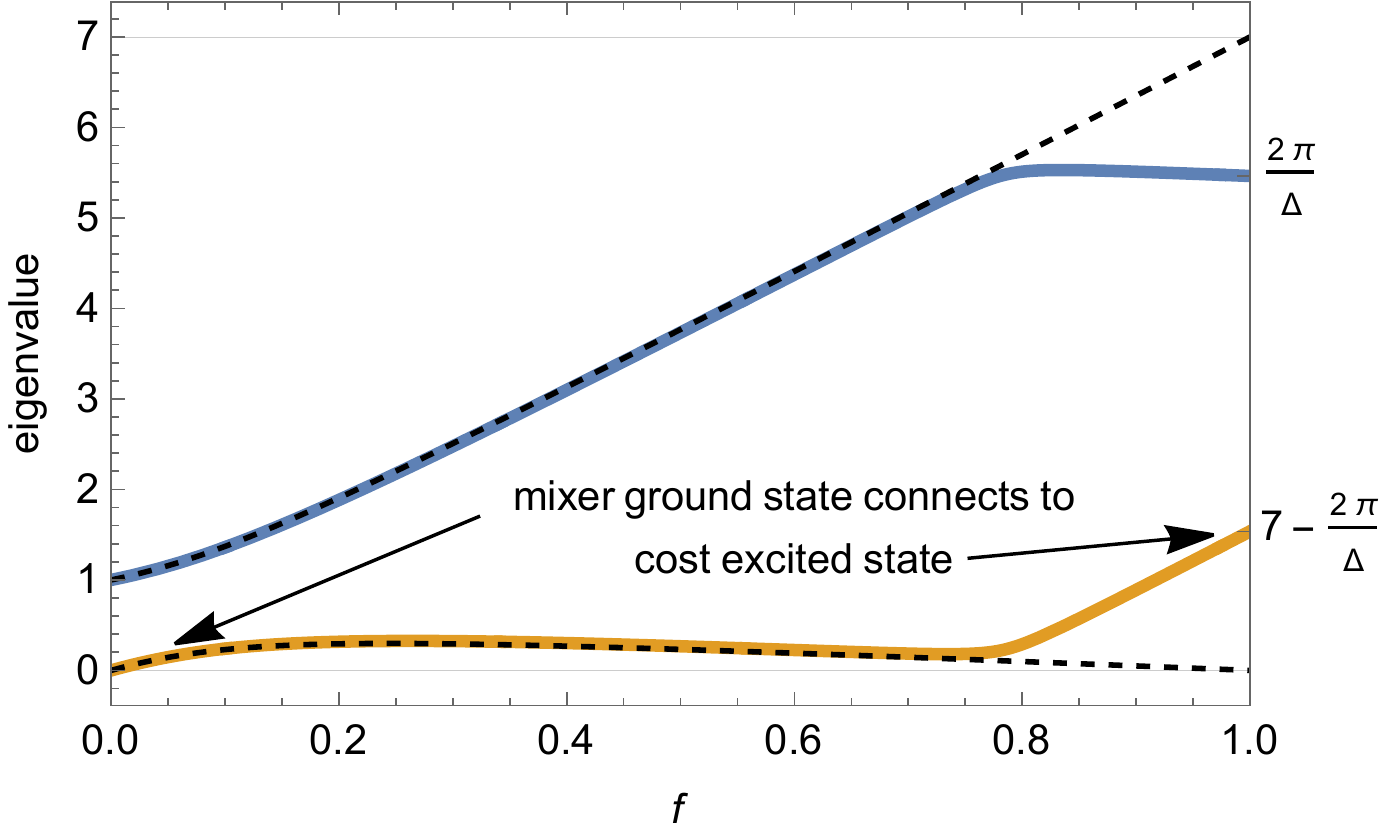}	&	\includegraphics[width=0.45\textwidth]{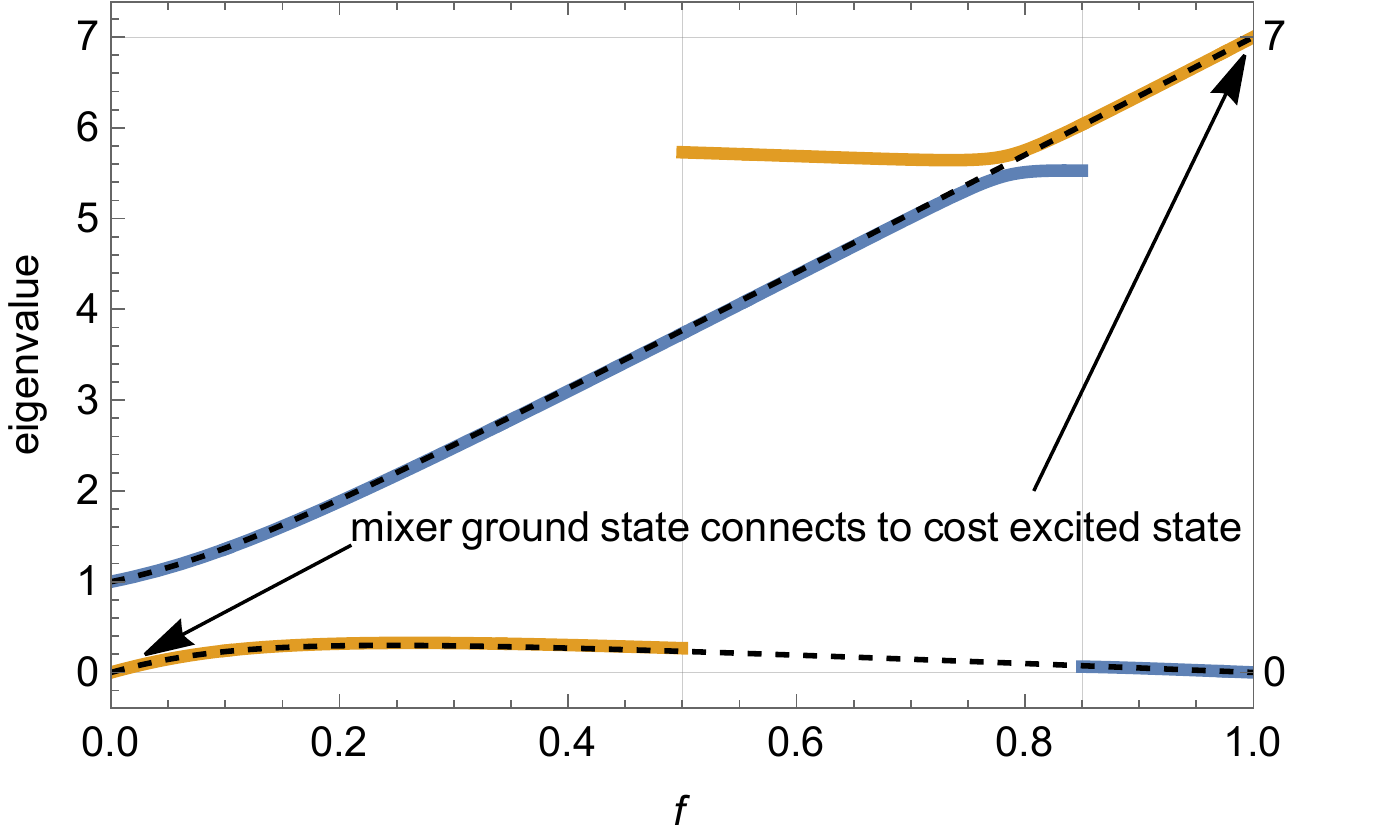} \\
(a) & (b)
\end{tabular}
\caption{Eigenvalues of \HQAOA\ (solid) and \Hinst\ (dashed) as a function of $f$ for the 2-level example of \Eq{2-level example} with $\Delta = 1.15$. \Fig{2-level eigenvalues}b shows the corresponding eigenvalues of the unitary \UQAOA. (a) \HQAOA\ chosen to be continuous and match \Hinst\ at $f=0$. (b) \HQAOA\ chosen to match \Hinst\ at both $f=0$ and $f=1$.}
\label{f:Hinst vs HQAOA}
\end{figure*}

\subsection{Discrete Adiabatic Evolution}

Consider a unitary operator $U(s)$ that is a continuous function of a parameter
$s\in [0,1]$.
Starting with an eigenvector $\ket{v_0}$ of $U(0)$, define
\begin{equation}
    \ket{v_j} = U(s_j) \ket{v_{j-1}}~,
\end{equation}
for $j=1,..,L$ where $s_j$ uniformly samples the interval, and the unitary varies only slightly from one iteration to the next when $L$ is large: $\left||U(s_{j+1})-U(s_j)|\right| = \mathcal{O}(1/L)$, where $||\cdot||$ denotes the spectral norm.

%\tad{To set the stage for generalizing to nonuniform, smooth schedules, can we generalize from ``$s_j$ uniformly samples'' to something like ``maximum distance between successive $s_j$ goes to zero as $L$ increases''? Or is this already implied as a consequence of rescaling the definition of $s$ by any monotonic continuous mapping?}

For sufficiently large $L$, $\ket{v_j}$ remains close to an eigenvector of $U(s_j)$ for each $j$, assuming these eigenvectors are not degenerate for any $s_j$. In particular, the final vector $\ket{v_{L}}$ will be close to an eigenvector of $U(s=1)$, within an error of size $\mathcal{O}(1/L)$~\cite{dranov98, costa2021}. 

Thus, products of gradually changing unitary matrices lead to the discrete adiabatic behavior of state vectors that track the evolving eigenvectors~\cite{dranov98,hogg03,tanaka11,costa2021}. As in the continuous adiabatic case, smaller gaps (Euclidean distance) between $U(s)$ eigenvalues neighboring that of the eigenvector being tracked require larger $L$ to maintain the same final overlap~\cite{dranov98,costa2021}. 

%The discrete adiabatic theorem generalizes to higher-dimensional parameter spaces. \tad{citation?}\vk{I think it follows from the theorem pretty directly? Parameterize a path through some high dimensional space - as long as the unitaries change less and less with each step along the path, the theorem should apply. Or maybe this already follows from earlier, and so is redundant.}

%\sh{SH: this seems to be first mention of "gaps", so we should make clear with respect to what - ie instantaneous Hamiltonian?}

\subsection{Application to QAOA}

The QAOA initial state is typically an eigenvector of the mixer and so also of $\UQAOA(\Delta, 0)$. 
%The desired QAOA output corresponds to a low-energy eigenvector of the cost Hamiltonian and so also of $\UQAOA(\Delta, 1)$ \sh{[SH: this was incorrect as written, please confirm this change doesn't affect anyhting below. Still not quite right, can we cut this?]}.
%RESOLVED# : \vk{yeah, this doesn't really change anything of importance - at the end of the day, we're just characterizing behavior for this regime of angle changes}
The QAOA circuit of \Eq{QAOAstate} applies a product of unitary operators defined by \Eq{unitary} to the initial state.  
At each step the QAOA unitary changes by at most $\|\UQAOA(\Delta, f_{j+1})-\UQAOA(\Delta, f_j)\|= O(\Delta (\|\Hmixer\|+\|\Hcost\|)/p)$. Consequently, for sufficiently large $p$, the discrete adiabatic theorem guarantees that, in the absence of degeneracies, the initial state evolves to \emph{some} eigenstate of the cost Hamiltonian when $p$ is sufficiently large. However, unlike the case of slowly evolving Hamiltonians, the changing unitaries can have different connections between eigenstates at different values of $\Delta$ as described in \Section{QAOA operator behavior}. This can lead to poor QAOA performance even when $p$ is large enough to be in the adiabatic regime. Thus, approaching the discrete adiabatic limit is not a sufficient condition for good QAOA performance. 

\subsection{Contrasting Discrete and Continuous Adiabatic Regimes}

The discrete adiabatic theorem is analogous to but distinct from the better-known continuous-time adiabatic theorem for evolution with slowly changing Hamiltonians, as considered in quantum annealing. For example, the ramp schedule of \Eq{gamma and beta} corresponds to a Hamiltonian
\begin{equation}\label{e:H_inst}
\Hinst(f) = (1-f) \Hmixer + f \Hcost~. 
% a more specific form using the schedule of \Eq{gamma and beta}:
%\Delta (1-f) \Hmixer + \Delta f \Hcost
\end{equation}

%\tad{If later in the paper we want to compare eigenvalues of QAOA with those from continuous annealing, we put back the $\Delta$ factor, i.e., compare $\HQAOA$ with $\Delta \Hinst$.}

The evolution of a state over a time $T$ due to a changing Hamiltonian arises from the Schr{\"o}dinger equation. With the substitution $t=f T$, this is
\begin{equation}
    \frac{d}{df} \ket{\psi(f)} = -i T \Hinst(f) \ket{\psi(f)}~.
\end{equation}
The (continuous) adiabatic theorem states that if $\ket{\psi(0)}$ is the $\text{n}^{\text{th}}$ eigenstate of $\Hinst(0)$ then,  
as $T\rightarrow \infty$, $\ket{\psi(1)}$ is arbitrarily close to the $\text{n}^{\text{th}} $ eigenstate of $\Hinst(1)$, provided the Hamiltonian changes smoothly and the evolving eigenvector of $\Hinst(f)$ is not degenerate for any $0\leq f \leq 1$~\cite{sakurai,albash2018}.

%\tad{If we define \Hinst\ without the $\Delta$ factor, then we can relate the continuous and discrete cases as follows:}
Discretizing the Schr{\"o}dinger equation over small time steps $\delta t$ produces change from the unitary operator
\begin{equation}\label{e:Uinst}
\Uinst = e^{-i \delta t \Hinst}.
\end{equation}
%This corresponds to the discrete case if $\Delta =\delta t$ from \Eq{gamma and beta}\sh{[SH: what does this sentence mean precisely?]}. %RESOLVED# Sentence unnecessary 

QAOA evolution of \Eq{QAOAstate} approximates this continuous evolution in the limit of 
\begin{equation}
\Delta \rightarrow 0,\ p\Delta=T~.
\label{e:continuous_limit}
\end{equation}
This is because when $\Delta$ is small, the QAOA unitary of \Eq{unitary} is a Trotter approximation to \Eq{Uinst} if $\Delta=\delta t$. Thus in this limit, the discrete and continuous behaviors agree.

The continuous adiabatic limit is obtained if we have 
\begin{equation} 
\Delta \rightarrow 0,\ p\Delta\rightarrow \infty~.
\end{equation}
The discrete adiabatic limit, by contrast, is more general, obtained at:
\begin{equation}
    p\rightarrow \infty~.
\end{equation}
In the context of QAOA, the continuous adiabatic limit (vanishing ``switching time'' $\Delta$ and infinitely fine discretization) is a special case of the discrete adiabatic limit (infinitely fine discretization) and in general their behaviors can differ significantly as discussed in \Section{QAOA operator behavior}.

One example of this (perhaps surprising) difference is that it is possible to connect the ground state of a mixer Hamiltonian to an excited state of the cost Hamiltonian without encountering a degeneracy. This arises because the eigenvalues of unitary operators exist on the unit circle of the complex plane, allowing for degeneracy-free connections between, for example, the mixer ground state and the cost maximal excited state (for a similar reason, excited-ground connections can emerge between the quasi-energies of Floquet operators~\cite{tanaka_quasi_energies,miyamoto_anholonomies}). In the case of non-singular, bounded Hamiltonians, where eigenvalues lie along the real line, this cannot happen.

To further elucidate the difference between the discrete and continuous cases, consider a Hamiltonian \HQAOA\ corresponding to the unitary operator of \Eq{unitary}: 
\begin{equation}
 e^{-i \Delta \HQAOA} =\UQAOA \equiv e^{-i\Delta (1-f) \Hmixer} e^{-i\Delta f \Hcost}~.
\label{e:hqaoa}
\end{equation}
The eigenvalues of $\HQAOA$ are only defined up to arbitrary integer multiples of $2\pi/\Delta$. 
By analogy with the behavior of \Hinst, one might expect to be able to choose these multiples such that $\HQAOA(f)$ is a continuous function of $f$ and matches the initial and final Hamiltonians used with annealing, i.e., $\HQAOA = \Hmixer$ at $f=0$ and $\HQAOA = \Hcost$ at $f=1$. 
In particular, for small $\Delta$, the Baker-Campbell-Hausdorff (BCH) formula~\cite{sakurai} relates \HQAOA\ to \Hinst:
\begin{equation}
\HQAOA = \Hinst -\frac{i\Delta}{2} f(1-f)[\Hmixer,\Hcost]+\dots~,
\label{e:bch}
\end{equation}
where the omitted terms to the right involve higher-order commutators of $\Hmixer$, $\Hcost$ and higher powers of
$\Delta$ and $f$. %$-i\Delta,f,1-f$. 
Each of these terms is 0 at $f=0$ and $f=1$, so at those values $\HQAOA = \Hinst$, thereby matching \Hmixer\ and \Hcost\ at those values of $f$.
If the series converges to a continuous function of $f$, then the BCH formula provides a choice for \HQAOA\ that is continuous and matches \Hmixer\ at $f=0$ and \Hcost\ at $f=1$.  
%\sh{[SH: text here confused me a bit. In any case I have some suggested improvements for this subsection..]}\vk{Any additional improvements?} \sh{SH: as I mentioned last meeting I think the unitary perspective should be emphasized more strongly, appears more scientifically interesting and novel to me. My concern with making edits in this direction is that may lead to further needed changes throughout paper...}\vk{I've tried to emphasize this point at the end of page 4/start of page 5. Is this more of the emphasis that you wanted?}

The behaviors of \HQAOA\ and \Hinst\ become categorically different for sufficiently large $\Delta$. In that case, the eigenvalues of \UQAOA\ vary (as a function of $f$ going from 0 to 1) from those corresponding to the ground state energy of the mixer to eigenvalues corresponding to a cost excited state without encountering a degeneracy. Therefore, any continuously changing \HQAOA\ no longer matches \Hcost\ at $f=1$ (if starting at \Hmixer\ at $f=0$) but rather has eigenvalues of \Hcost\ shifted by an integer multiple of $2\pi/\Delta$. For such a choice of \HQAOA, ground states still evolve to ground states for these shifted energies. Conversely, choosing to define \HQAOA\ to be the mixer Hamiltonian at $f=0$ and the cost Hamiltonian at $f=1$ results in at least one discontinuity in the spectrum of \HQAOA\ at intermediate $f$. Since the BCH formula in \Eq{bch} fixes \HQAOA\ to have such endpoints, this implies that the formula fails to converge to a continuous function of $f$ at such a value of $\Delta$. %, so for some $f$ and at most that value of $\Delta$, the formula must fail to converge.

\Fig{Hinst vs HQAOA} is an example of this behavior of \HQAOA\ for the 2-level example discussed in \Section{QAOA operator behavior}. This shows that if we choose to make \HQAOA\ continuous then $\HQAOA(1) \neq \Hinst(1)$ due to shifts in the eigenvalues by $2\pi/\Delta$, as in \Fig{Hinst vs HQAOA}a.
Alternatively, we can define \HQAOA\ to match \Hinst\ at both $f=0$ and $f=1$, but this requires a discontinuity in \HQAOA, as in \Fig{Hinst vs HQAOA}b. 

These observations indicate that focusing on the Hamiltonian corresponding to the QAOA unitary operator can lead to ambiguity in interpreting its behavior.

\section{QAOA Eigenvector Behavior}\label{s:QAOA operator behavior}

%\sh{[SH: IMO this section needs polishing. So many details hard to discern main takeways. Various technical distractions I encourage simplifying. Claims could be more precise]}

This section describes the behavior of the eigenvalues and eigenvectors of the QAOA operator defined in Eq.~\ref{e:unitary}. 
%We focus on how its eigenvalues and eigenvectors vary with these parameters.
In particular, we discuss conditions under which the ground state of $\Hmixer$ at $f=0$ does not connect to the ground state of $\Hcost$ as the QAOA operator evolves.
%starting from the ground state of $\Hmixer$ at $f=0$, the QAOA operator does not change to the ground state of $\Hcost$ at $f=1$ but rather to a different cost eigenvector. 

\begin{figure}[htb]
    \includegraphics[width=0.4\textwidth]{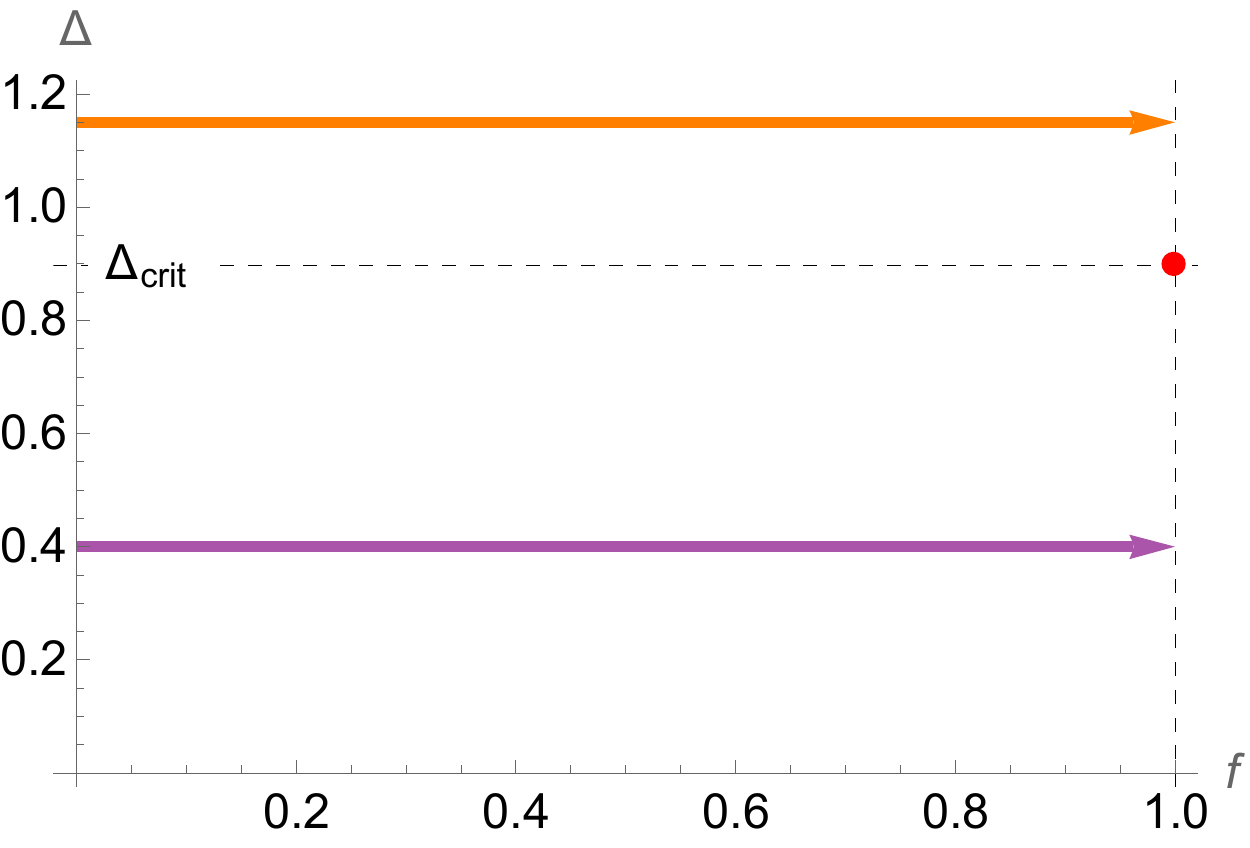}
    \caption{Two paths through the parameter space with fixed $\Delta=0.4$ (purple) and $\Delta=1.15$ (orange). The red dot indicates the parameters, $\Delta=\DeltaCrit=2\pi/7$ and $f=1$, at which the 2-level example of \Eq{2-level example} has a degeneracy due to eigenvalue wrap-around as discussed in \Section{eigenvalues: 2-level}.}
    \label{f:fixed_delta_path_diagram}
\end{figure}

Specifically, we consider paths through the space of parameters $f$ and $\Delta$, which determine the QAOA angles in \Eq{gamma and beta}. For the linear ramp schedule, $\Delta$ is constant while $f$ starts at zero and ends at one. \Fig{fixed_delta_path_diagram} illustrates two such paths.

\subsection{Eigenstate Connections and Wrap-Around}

%\sh{[SH: In several places here would be more clear to just use "eigenvalue"]}
%RESOLVED#
% We define an eigenspace of an operator as the states spanned by all the eigenvectors associated with a single eigenvalue of that operator.
% We say an eigenspace is degenerate if its associated eigenvalue is degenerate.
% %examine how sets of eigenvectors of \UQAOA\ evolve and, in particular, how the connections of the eigenspaces change as $\Delta$ increases.
\begin{definition}
For a given path in parameter space starting at $f=0$ and ending at $f=1$, we say a mixer eigenstate is \textbf{``connected"} to a cost eigenstate if it continuously changes into the cost eigenstate over the given path.%RESOLVED#
\end{definition}
In the case of degenerate eigenvalues, we can extend the notion of connection to being between eigenspaces i.e. a state in a mixer eigenspace continuously changing into a state of a cost eigenspace along the path. We consider the degenerate case briefly in Section~\ref{s:degenerate_case}.
%A degenerate mixer eigenstate may connect to more than one cost eigenstate. 

Consider the eigenstates at f=0 and f=1 as two sides of a bipartite graph. Given the set of mixer eigenstates $\{\mathcal{U}_1,...\mathcal{U}_k \}$ and the set of cost eigenstates $\{\mathcal{V}_1,...\mathcal{V}_m\}$, the pair $\{\mathcal{U}_i,\mathcal{V}_j \}$ is an edge in the graph for a particular path through parameter space if eigenstate $\mathcal{U}_i$ connects to eigenstate $\mathcal{V}_j$. As we shall show, changing $\Delta$ can change these edges.
%RESOLVED#

\begin{figure}[h]
    \centering
    \includegraphics[width=0.45\textwidth]{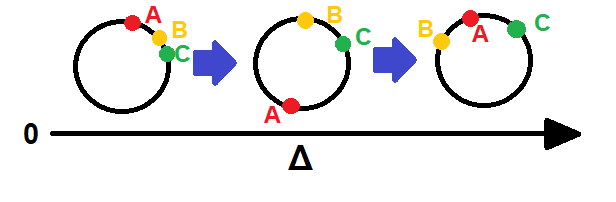}
    \caption{A sketch of the eigenvalues on the unit circle in the complex plane,
    %RESOLVED#
    for a three-level \UQAOA\ with increasing values of $\Delta$ and with fixed $0\leq f \leq 1$. For $\Delta$ close to zero, the eigenvalues are clustered near 1 on the right-most point of the circle as illustrated in the first circle. As $\Delta$ increases, eigenvalues move around the unit circle at different speeds. Eventually, the fastest-moving eigenvalue, A, overtakes and then passes the eigenvalue, C ("wraps around" with it), as shown in the third circle.}
    \label{fig:wrap_around_sketch}
\end{figure}

A key case considered in this paper are the connections for a pair of eigenstates that are the highest- and lowest-energy eigenstates of \Hmixer, though our results apply more generally. An important concept related to changing connections is ``wrap-around'', illustrated in Fig.~\ref{fig:wrap_around_sketch}

\begin{definition}
\textbf{``Wrap-around''} is the phenomenon of a pair of distinct eigenvalues of $\UQAOA(\Delta,f)$ at fixed $f$ wrapping around the unit circle in the complex plane far enough to intersect and then overtake each other as $\Delta$ increases from 0.
\end{definition}

As shown in \Section{degeneracies and eigenvalues}, wrap-around at $f=0,1$ caused by changing $\Delta$ is intimately linked to the changing connections between mixer and cost eigenstates. While wrap-around can happen at either value $f=0$ or $f=1$, we will, without loss of generality, focus on wrap-around at $f=1$ in the remaining sections. %RESOLVED#

\begin{figure*}
    \includegraphics[width=0.6\textwidth]{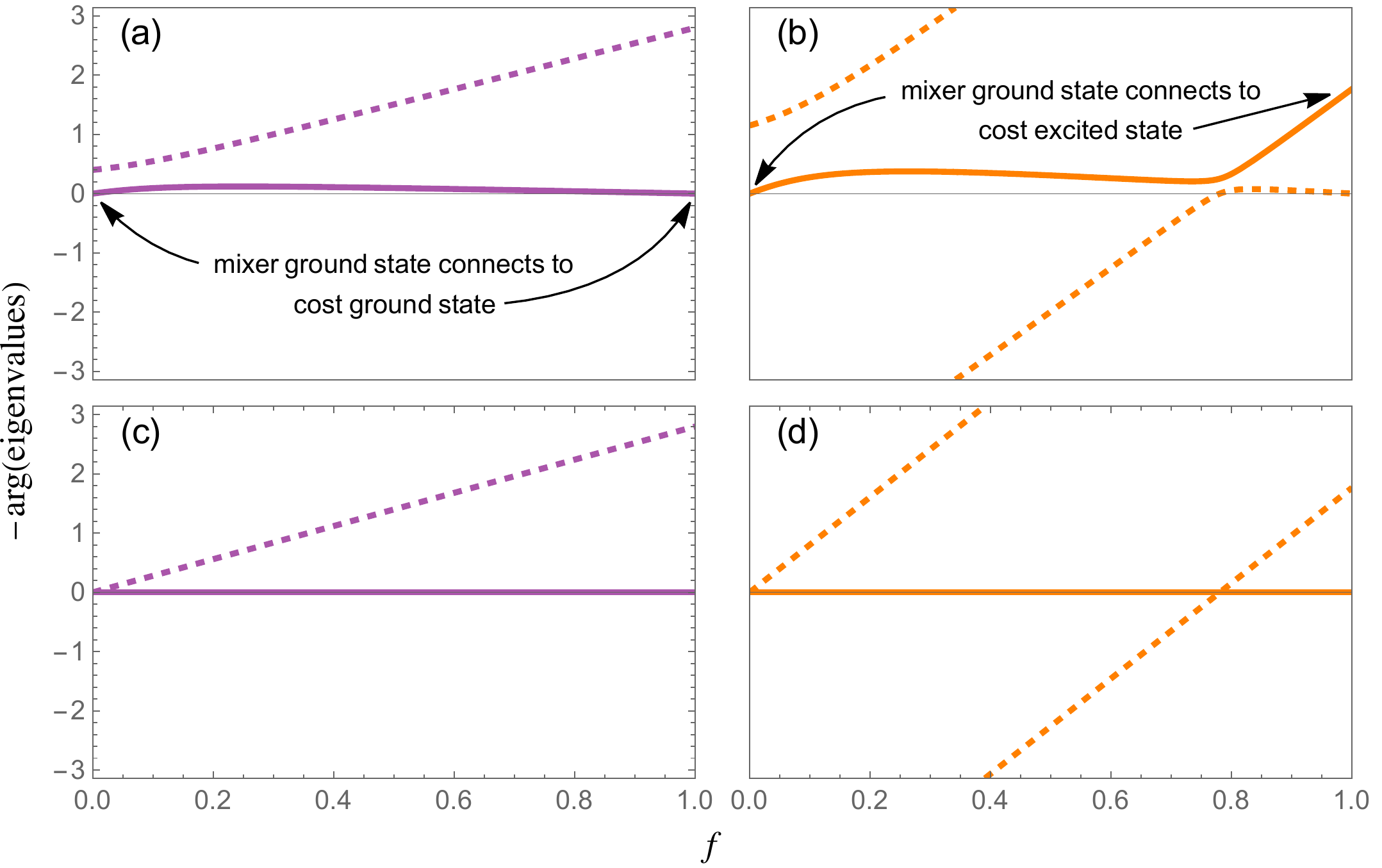}
    \caption{Negative argument of the eigenvalues of unitary operators as a function of $f$ for the 2-level example defined by \Eq{2-level example} for the paths shown in \Fig{fixed_delta_path_diagram}. (a and b) QAOA unitaries $\UQAOA$ of \Eq{unitary} for (a) $\Delta=0.4$ and (b) $\Delta=1.15$. A curve is solid (dashed) if its value at f=0 is that associated with the mixer’s ground (excited) state energy. (c and d) cost unitaries of \Eq{cost unitary} for these values of $\Delta$.}
    \label{f:2-level eigenvalues}
\end{figure*}

\begin{figure}
    \includegraphics[width=0.4\textwidth]{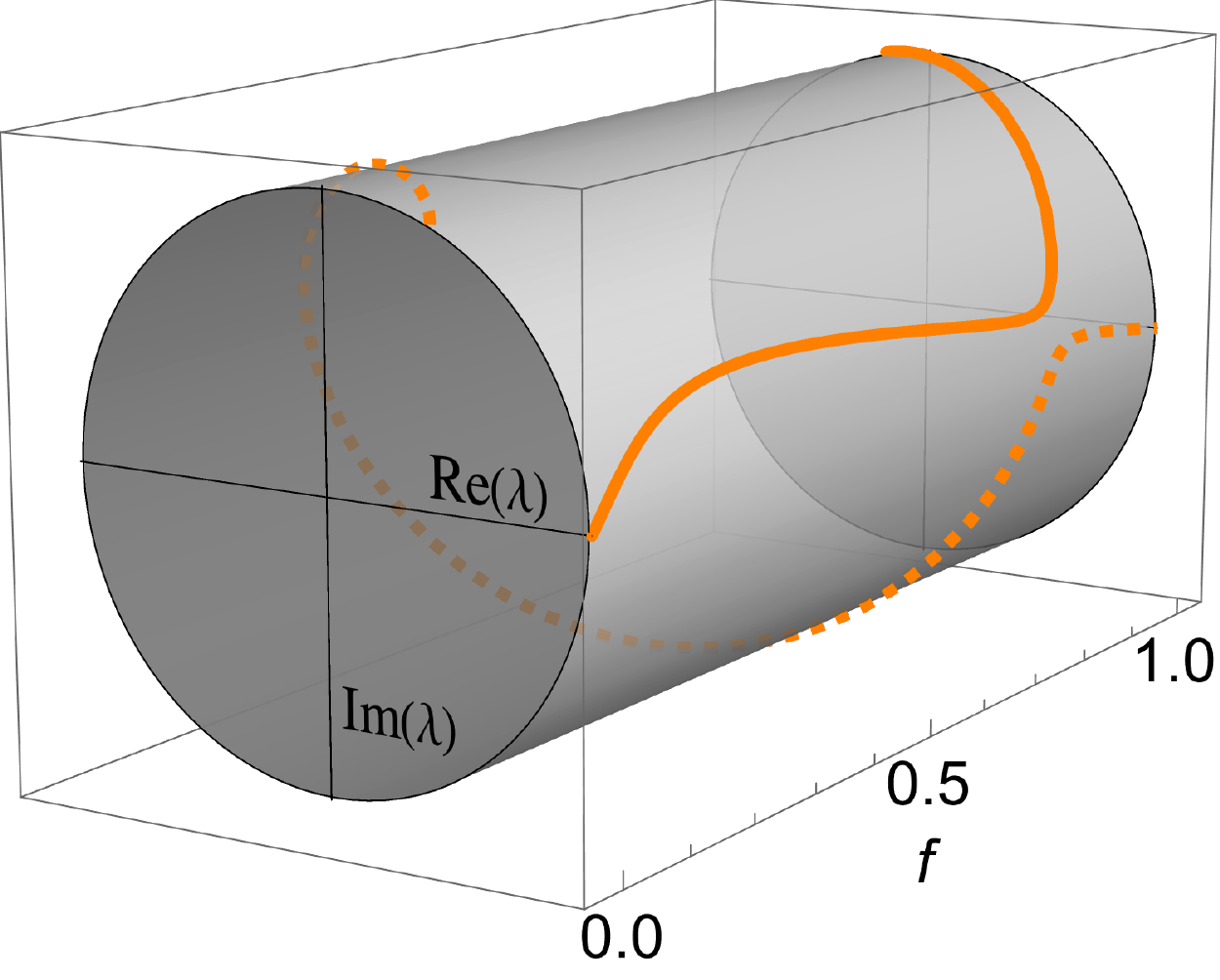}
    \caption{Evolution of the eigenvalues $\lambda$ of the 2-level example QAOA unitary operator on the unit circle in the complex plane as a function of $f$ for the case of $\Delta=1.15$. The solid and dashed curves correspond to the ground and excited states of the mixer at $f=0$, respectively. These are the same eigenvalues whose negative arguments are shown in \Fig{2-level eigenvalues}(b).}
    \label{f:2-level eigenvalues3D}
\end{figure}

\subsection{QAOA Unitary Degeneracies and Eigenvalue Wrap-around}\label{s:degeneracies and eigenvalues}
%RESOLVED#

The eigenvalues of unitary operators are on the unit circle in the complex plane. We graphically depict them by plotting the negative of the arguments of these eigenvalues in \Fig{2-level eigenvalues}. Thus for a unitary operator $U=e^{-i H}$ derived from a Hamiltonian $H$, this plot shows the eigenvalues of $H$ modulo $2\pi$ in the range $-\pi$ to $\pi$ - note that this means that $U$ can be degenerate even when $H$ isn't. 

%Since many of the concepts we discuss in this section are somewhat abstract and have not yet been directly applied to QAOA, 
We illustrate the behavior of the QAOA unitary from Eq.~\ref{e:unitary} with a two-level example defined by the paths through parameter space shown in Fig.~\ref{f:fixed_delta_path_diagram} and the Hamiltonians
\begin{equation}\label{e:2-level example}
\Hmixer = \begin{bmatrix}0.5&-0.5\\-0.5&0.5 \end{bmatrix} \quad  \text{and}  \quad
\Hcost = \begin{bmatrix}0 & 0 \\ 0 & 7 \end{bmatrix}~.
\end{equation}

\subsubsection{Eigenvalue Behavior: A Two-level Example}\label{s:eigenvalues: 2-level}

This section describes the behavior of the eigenvalues of the QAOA unitary for the 2-level example of \Eq{2-level example}.
The top half of \Fig{2-level eigenvalues} shows how eigenvalues change with $f$
along the two paths shown in \Fig{fixed_delta_path_diagram}. 

For the smaller $\Delta$ value, the QAOA unitary connects the mixer Hamiltonian ground state to the cost Hamiltonian ground state. For the larger $\Delta$ value, \Fig{2-level eigenvalues}(b) shows that the connection changes: the mixer Hamiltonian ground state connects to the cost Hamiltonian excited state.

This change in connection arises from eigenvalues of the QAOA operator at $f=0$ and $f=1$ wrapping around the complex unit circle as $\Delta$ is increased, as illustrated in \Fig{2-level eigenvalues3D}. In particular, as $\Delta$ changes along the vertical axes of \Fig{fixed_delta_path_diagram}, with $f=0$ or $f=1$, the set of eigenvectors doesn't change and the eigenvalues rotate around the unit circle linearly with $\Delta$. In \Fig{2-level eigenvalues}, this corresponds to the negative argument of one eigenvalue increasing through $\pi$, switching to $-\pi$, reaching and then passing the negative argument of the other eigenvalue.

%These eigenvalues correspond to eigenvalues for $\Hmixer$ and $\Hcost$. 
%To understand this phenomenon we must study the unitary sub-matrices of the QAOA operator: the exponentiated mixer and cost terms.\tad{unitaries from \Hmixer\ and \Hcost\ aren't ``submatrices'' of \UQAOA. Delete the sentence to avoid confusion??}
%
The first wrap-around is determined by the largest eigenvalue of either $\Hmixer$ or $\Hcost$. In this example, that maximum eigenvalue is the excited energy of $\Hcost$. Thus to better understand this wrap-around, we contrast it with that of the cost unitary on its own,
\begin{equation}\label{e:cost unitary}
 \Ucost(\Delta, f) = e^{-i\Delta f \Hcost}~.
\end{equation}
%$ \Ucost(\Delta, f) = e^{-i\Delta f \Hcost}$.
%RESOLVED#
From \Eq{2-level example}, the negative arguments of the eigenvalues of \Ucost\ are, respectively, $0$ and $7 \Delta f \bmod 2\pi$ shifted to the range $-\pi$ to $\pi$. The bottom half of \Fig{2-level eigenvalues} shows these expressions for the same $\Delta$ values used in the top half.
For the smaller $\Delta$, the values are $0$ and $7\Delta$ at $f=1$.
%  no need for \bmod since $7 \Delta < \pi$ for Delta=0.4, so can use the simpler expression to contrast with wrap-around at larger Delta
%$7\Delta\bmod{2\pi}$. %, as seen in \Fig{2-level eigenvalues}c.
On the other hand, for the larger $\Delta$, just above $2\pi / 7$, the ``excited state" eigenvalue of $\Ucost$ wraps around the unit circle and goes past the ``ground state" eigenvalue at $f=1$. Therefore, 
%purely due to geometry and the exponentiation of a diagonal unitary  
the eigenvalues cross each other at some $f$ between 0 and 1, as seen \Fig{2-level eigenvalues}(d).

%when the mixer unitary is introduced to form the QAOA unitary, the structure of the mixer unitary and its interaction with the cost results in this level crossing becoming an avoided crossing.
%To see why there is an avoided crossing, consider a value of $\Delta$ just above $2\pi / 7$, at which the crossing in \Fig{2-level eigenvalues}(d) takes place very close to $f=1$. n
Generically, the cost and mixer Hamiltonians do not commute and so the introduction of mixer unitary to form \UQAOA~lifts the degeneracy and turns the level crossing of \Ucost\ into an avoided crossing for \UQAOA.

\begin{figure}[htb]
    \centering
    \includegraphics[width=0.45\textwidth]{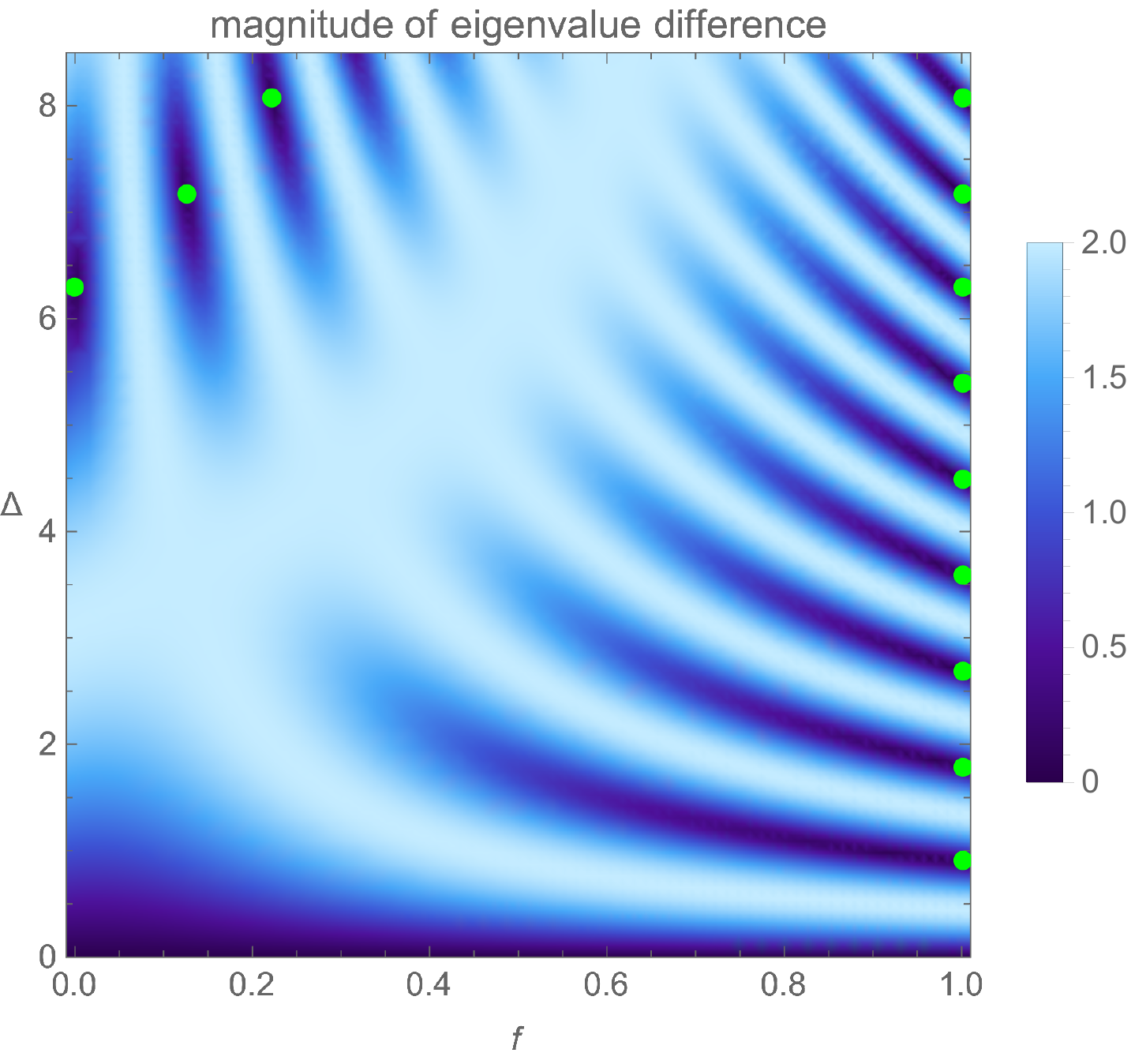}
    \caption{Magnitude of the difference between the two eigenvalues of the two-level example as a function of $f$ and $\Delta$. The green points indicate locations of degenerate eigenvalues. In addition, the entire bottom edge consists of degenerate eigenvalues, since the QAOA unitary is the identity for all $f$ when $\Delta=0$ .}
    \label{f:eigenvalue_difference}
\end{figure}

%The same phenomenon takes place for the mixer. When $\Delta$ is sufficiently large, the eigenvalues of the mixer unitary wrap around at $f=0$ and create an avoided crossing due to the degeneracy of the mixer symmetry being lifted by the presence of the cost unitary.
Additional degeneracies occur at values of $f$ other than 0 or 1 for sufficiently large $\Delta$, as illustrated for the two-level example in \Fig{eigenvalue_difference}.

\subsubsection{Eigenvalue Behavior: General Case and Definitions}

\begin{figure}[htb]
    \centering
    \includegraphics[width=0.45\textwidth]{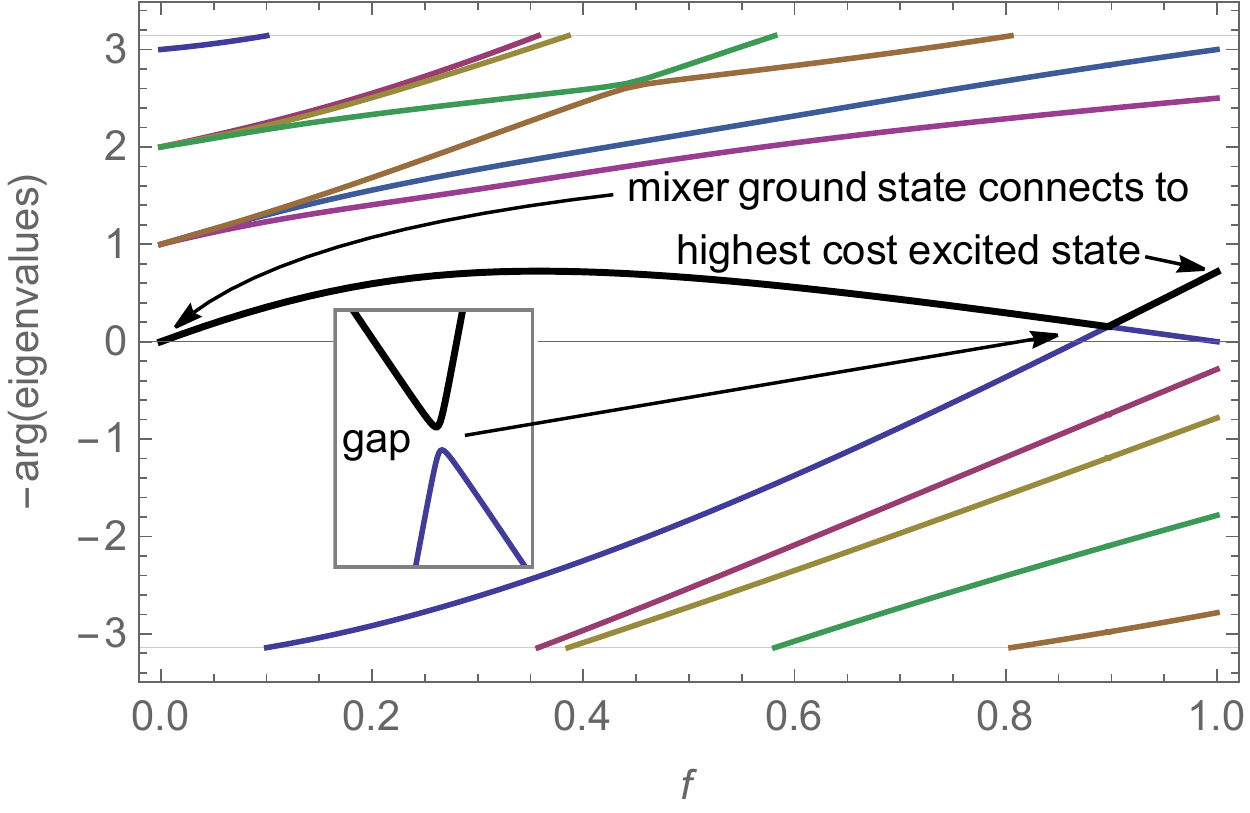}
    \caption{Negative arguments of the eigenvalues of \UQAOA\ for wrap-around leading to an avoided crossing using $\Hmixer = 0.5\sum_{i=1}^8(I-\sigma_x^i)$ and $\Hcost$ with energies $0,2.5,3,3.5,4.5,5.5,6,7$ in the computational basis, in order of bitstring value. The $0.0012\times 0.0016$ inset shows the avoided crossing between the wrap-around eigenvalues which occurs at the location indicated by the arrow. }
    \label{f:eight_level_avoided_crossing_example}
\end{figure}

This section describes the general case of two distinct, initially non-degenerate eigenvalues which become degenerate due to wrapping around the unit circle as $\Delta$ increases. The generalization to more than two (possibly initially degenerate) eigenvalues wrapping around with each other is addressed in Section~\ref{s:degenerate_case}.

Eigenvalue wrap-around at a given $(\Delta, f)$ can lead to avoided crossings at neighboring parameters. Without loss of generality, we can consider the pair of cost unitary eigenvalues corresponding to the highest and lowest energies of the cost Hamiltonian, which determine the eigenvalues of the QAOA operator at $f=1$. As $\Delta$ increases, these eigenvalues are the first pair to wrap around the unit circle and cross each other. 

Away from $f=1$, the crossing caused by the two cost eigenvalues can become avoided due to the mixer unitary- in that case, the two eigenvalues of the QAOA unitary will not cross and will continue to avoid each other even for significantly larger $\Delta$, just as in \Fig{2-level eigenvalues}(b). In this case, the mixer ground and highest eigenstates become connected to different cost eigenstates than they were at smaller values of $\Delta$. This is illustrated for an eight-level system in \Fig{eight_level_avoided_crossing_example}, where the intersection of the eigenvalues leading to the ground and highest excited states results in an avoided crossing.  
As $\Delta$ continues to increase, other pairs of eigenvalues wrap-around at $f=0$ and $f=1$ and can create avoided crossings, which would lead to more mixer eigenstates becoming connected to different cost eigenstates than they were at small $\Delta$. 

This motivates defining some common terms for wrap-around and avoided crossings. 
\begin{definition} 
We call a degeneracy between two eigenvalues \textbf{isolated} if the point in $(f, \Delta)$-space where the degeneracy takes place has no other parameter points within some neighborhood where the two eigenvalues are equal.
\end{definition} 
An example of this is the degeneracy in our two-level example at $(f=1,\Delta=2\pi/7)$ leading to an avoided crossing as $\Delta$ increases, as seen from comparing \Fig{2-level eigenvalues}(b) and \Fig{2-level eigenvalues}(d).  
\begin{definition}
We call a degeneracy between two eigenvalues \textbf{continuous} if the set of points in $(f, \Delta)$-space where the two eigenvalues are equal forms a continuous curve.
\end{definition}

As an example, \Fig{2-level eigenvalues}(d) shows a continuous degeneracy, where the eigenvalues of the cost unitary continue to cross after initially touching at $f=1$ and the resulting degeneracy moves inward to $0<f<1$ as $\Delta$ increases.

\begin{definition}
\DeltaCrit~is the smallest value of $\Delta>0$ at which the ground state eigenvalue (of the mixer or cost) wraps around with another of its eigenvalues \textbf{and} this leads to an isolated degeneracy at $f=0$ or 1, respectively.
\end{definition}

In the case of our two-level example, $\DeltaCrit = 2 \pi /7$ due to an isolated degeneracy at $f=1$ and none below that value of $\Delta$. For the same reason \DeltaCrit\ is also $2\pi/7$ for the eight level example in Fig.~\ref{f:eight_level_avoided_crossing_example}.

In summary, two conditions are jointly sufficient for two fixed-$\Delta$ paths (with $\Delta$ values $\Delta_1<\Delta_2$) to have different connections between the mixer and cost eigenstates. Firstly, that some eigenvalues wrap around at some $\Delta_1<\Delta<\Delta_2$ at $f=0$ or $f=1$, and secondly that there is an isolated degeneracy at the $\Delta$ of wrap-around and value of $f$.  

Some significant differences exist between two-level and larger systems. Multiple consecutive wrap-arounds with several different eigenvalues, both by those of \Hcost\ and \Hmixer, means that once connection between the two ground eigenstates is severed, it might not be re-established by further increases in $\Delta$. Further, since the mixer can have off-diagonal elements that couple two wrapping cost eigenstates indirectly (and vice versa for the cost), continuous degeneracies emerge in nontrivial cases where the mixer and cost don't commute. These include the cases where \DeltaCrit\ is larger than the value of $\Delta$ at which the highest excited eigenvalue first wraps around to meet the ground state eigenvalue because this first wrap-around produces a continuous degeneracy rather than an isolated one.

\subsection{Eigenvector Swap}\label{s:eigenvector swap}

Section~\ref{s:dat} states that in the discrete adiabatic limit, the QAOA state vector tracks the eigenvectors of the QAOA unitary as we go from $f=0$ to $f=1$ if starting from an eigenvector of the mixer Hamiltonian. Thus, it is important to identify how the wrap-around behavior described in the previous section affects the eigenvectors.

\subsubsection{Two-Level Example}

\begin{figure}[htb]
    \centering
    \includegraphics[width=0.35\textwidth]{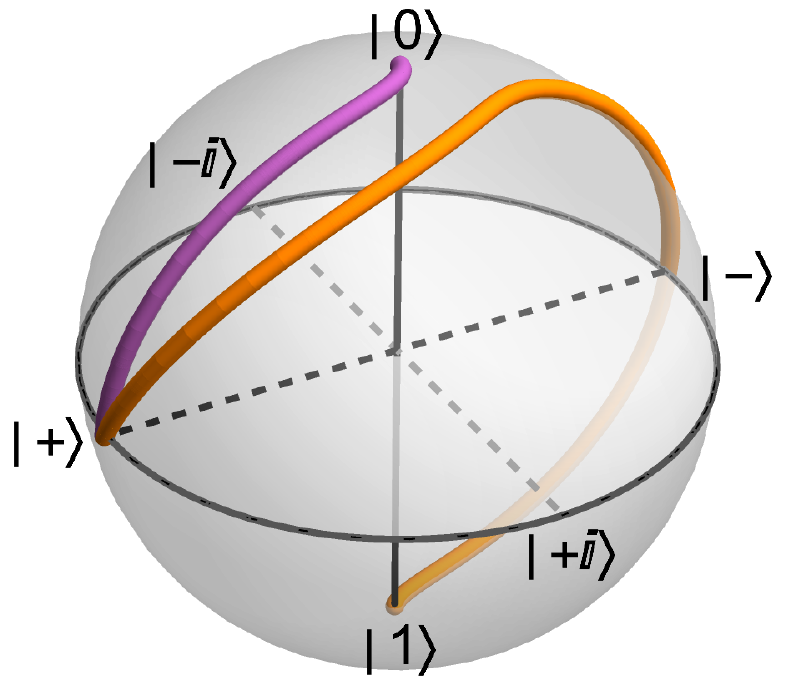}
    \caption{Paths of the QAOA eigenvector on the Bloch sphere starting from the uniform superposition as parameters change along the small and large delta paths of \Fig{fixed_delta_path_diagram}, purple and orange curves, respectively.}
    \label{f:bloch_path}
\end{figure}

We first consider our two-level example and the two paths through parameter space defined in \Fig{fixed_delta_path_diagram}.

For these paths, the evolution of the QAOA eigenvector starting at $\ket{+}$ is shown in \Fig{bloch_path}. The eigenvector trajectory for the purple, small-$\Delta$ path goes from the ground state of the mixer Hamiltonian to the ground state of the cost Hamiltonian. On the other hand, the eigenvector trajectory for the orange, large-$\Delta$ path goes from the ground state of the mixer to the excited state of the cost. Thus eigenvalue wrap-around is associated with a change in connection between initial and final eigenvectors along these paths.

\subsubsection{Eigenvector Swap from Wrap-Around at Isolated Degeneracies}

The 2-level example shows the change in eigenstate connections when $\Delta$ is large enough for the first wrap around to occur. We generalize this observation to $n$-level systems, showing that wrap around leads to the eigenvectors involved in the change in eigenvector connections directly swapping places (i.e., changing into one another along the path) without contribution from any other eigenvectors. %RESOLVED#

We do this by constructing a closed path through the $f,\Delta$ parameter space to show the swap near an isolated degeneracy from wrap-around implies the same swap occurs during QAOA when $\Delta$ is above the value at which that wrap-around takes place, but small enough to be below any other wrap arounds.

\begin{figure}[tb]
    \centering
    \includegraphics[width=0.3\textwidth]{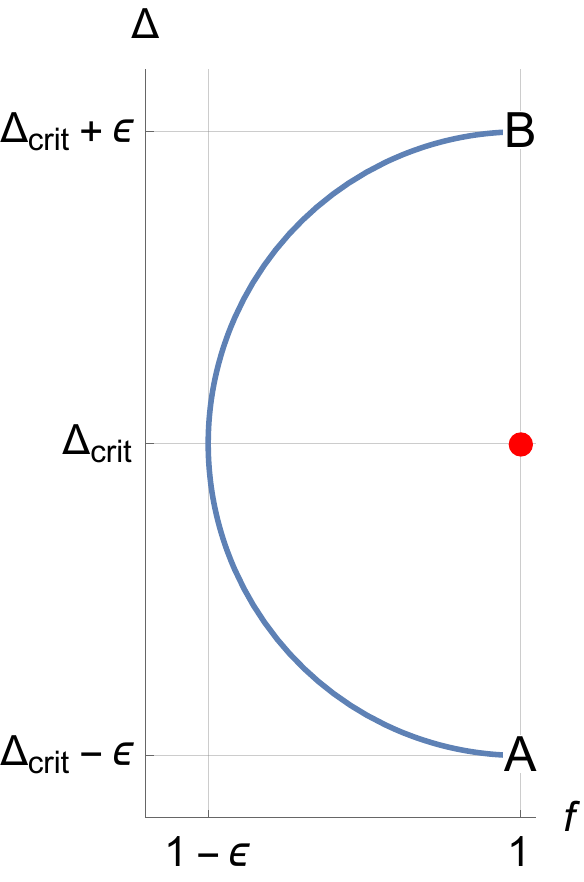}
    \caption{Example path near an isolated degeneracy (red point) at $f=1$. The path starts at point A, below the degeneracy, and ends at point B, above the degeneracy and remains within a small distance $\epsilon$ of the degeneracy.}
    \label{f:semicircle_path}
\end{figure}

To evaluate the behavior of eigenvectors near an isolated degeneracy corresponding to a wrap-around at $(f=1,\Delta=\DeltaCrit)$, consider a path within a small distance $\epsilon$ of the degeneracy that starts at $f=1$ just below the degeneracy and ends at $f=1$ above it, as illustrated by points A and B in \Fig{semicircle_path}. Such paths can pass arbitrarily close to the degeneracy without encountering other degeneracies, so the swapping \UQAOA\ eigenvectors of \UQAOA\  change continuously along the path. The wrap-around associated with this degeneracy means the eigenvalues of \UQAOA~associated with the degeneracy switch their ordering between points A and B of the path.

%As a specific case, consider starting on the path with the ground state of \Hcost. This is an eigenvector of \UQAOA\ since $f=1$. The wrap around involves an excited state of \Hcost, i.e., at the degeneracy both these states have the same eigenvalue and the ordering of those eigenvalues switches between points A and B of the path.

The entire path is near the degeneracy, so \Hmixer\ is a small perturbation to \Hcost, allowing us to describe the eigenvalues and eigenvectors of \UQAOA\ along the path with degenerate perturbation theory~\cite{shiff1968}. In the limit of $\epsilon \rightarrow 0$, this theory shows that the evolving eigenvector of \UQAOA\ is a linear combination of the eigenvectors involved in the isolated degeneracy. This linear combination starts with the ground state of \Hcost\ at point A of the path and ends with the excited state at point B. Moreover, the exchange takes place over a smaller path interval when the eigenvalue gap along the path is small. This generalizes to multiple degenerate states~\cite{shiff1968}, i.e. the evolution of eigenvectors involved in the degeneracy remains in the subspace defined by those vectors for small paths around the degeneracy.

\subsubsection{Behavior near a degeneracy: an eight-level example}

\begin{figure}[tb]
    \centering
    \includegraphics[width=0.45\textwidth]{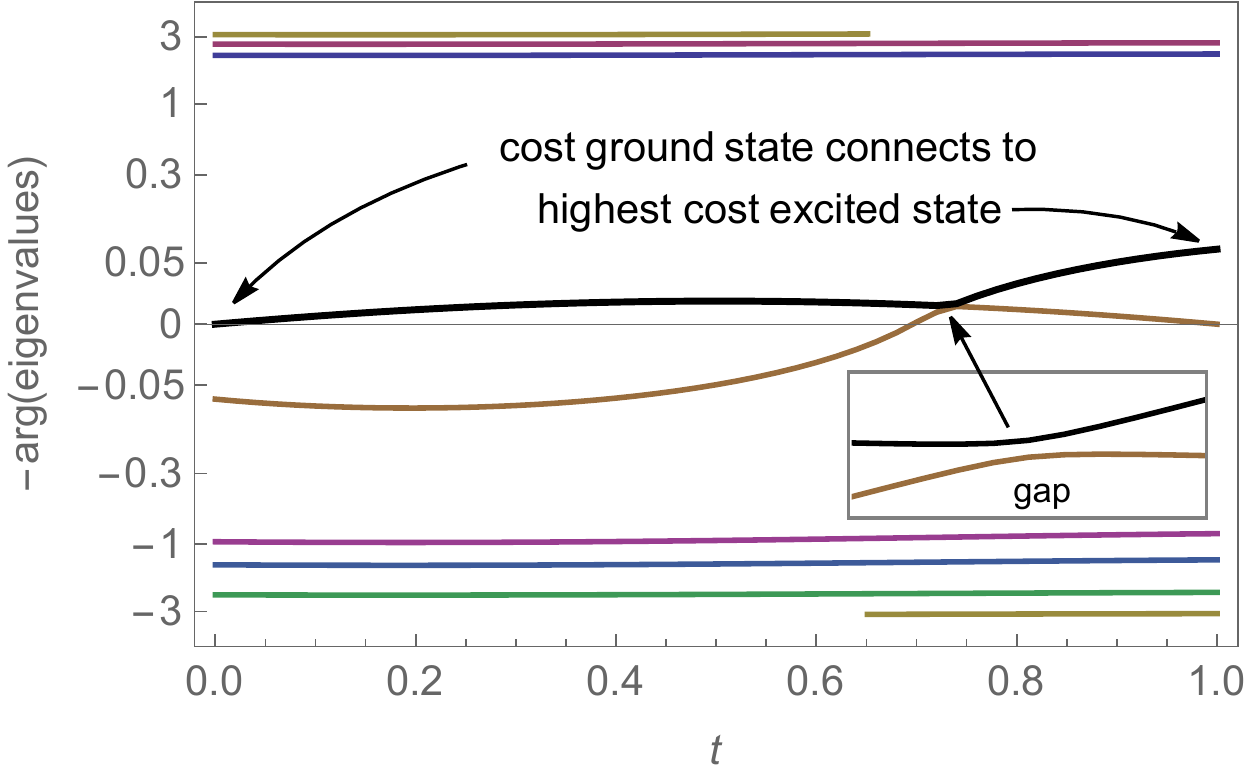}
    \caption{Negative arguments of the eigenvalues of \UQAOA\ near the wrap-around degeneracy at $\Delta = 2\pi/7$ for the 8-level example in \Fig{eight_level_avoided_crossing_example}. The values are shown as a function of $t$, ranging from 0 to 1 along the semicircle path of \Fig{semicircle_path} with $\epsilon=0.01$, starting from point A and using a symmetric logarithmic scale to emphasize the two states involved in the degeneracy. The inset shows the avoided crossing between these two eigenvalues, which occurs at the location indicated by the arrow. The inset's range is $10^{-6}$ horizontally and $0.4\times 10^{-6}$ vertically.}
    \label{f:semicircle-eigenvalues8}
\end{figure}

To illustrate the behavior near the degeneracy, we use the 8-level example shown in \Fig{eight_level_avoided_crossing_example}. Along this small path, \Fig{semicircle-eigenvalues8} shows eigenvalues change only a small amount, with wrap-around of the two eigenvalues involved in the degeneracy.

In this example, the ground state of \Hcost\ wraps around with the highest excited state. The mixer does not directly couple these two states, thereby illustrating a higher-order application of degenerate perturbation theory~\cite{shiff1968}.

\begin{figure}[tb]
    \centering
    \includegraphics[width=0.4\textwidth]{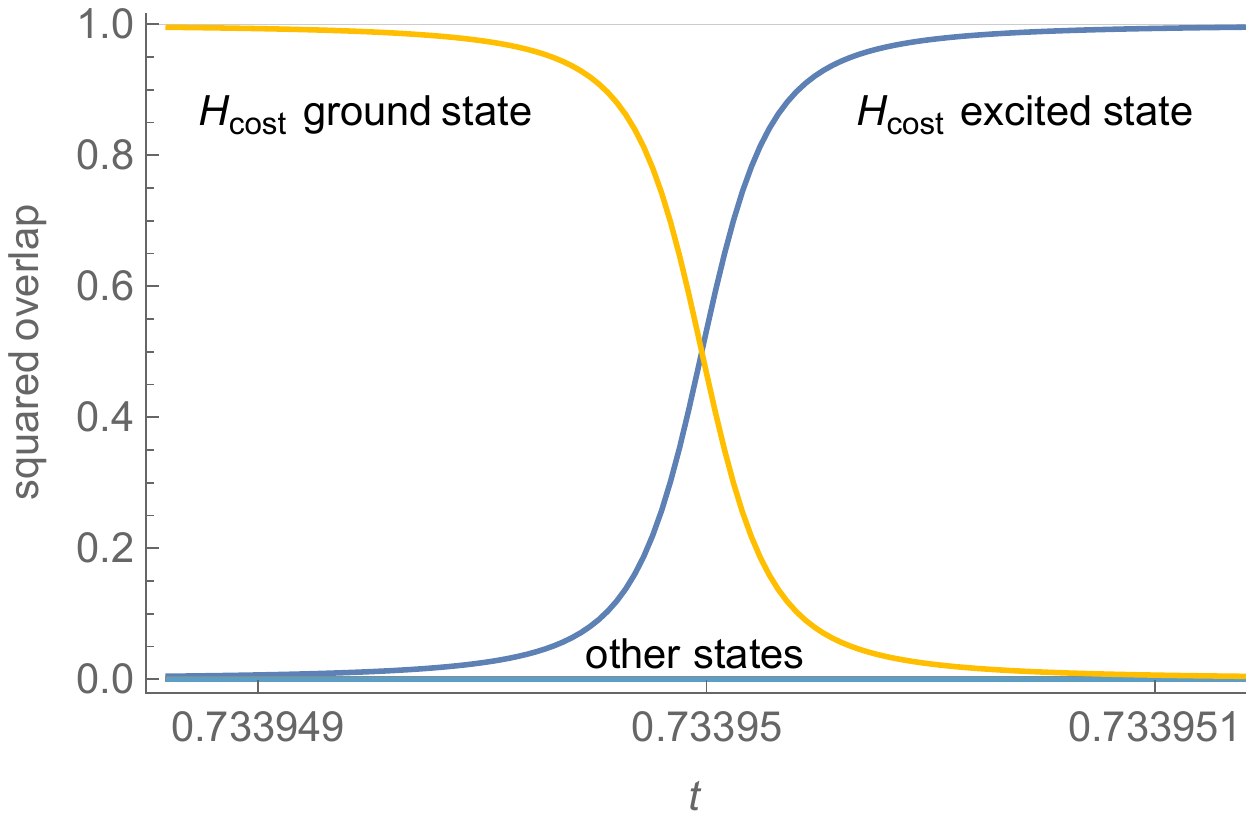}
    \caption{Eigenvector swapping in the 8-level example as a function of path parameter $t$ near the location of the avoided crossing on the semicircular path from \Fig{semicircle_path}. The curves show the squared overlap of the \UQAOA\ eigenvector with the eigenvectors of \Hcost.}
    \label{f:semicircle-eigenvectors8}
\end{figure}

\Fig{semicircle-eigenvectors8} shows how the \UQAOA\ eigenvector changes along the semicircular path, expressed as a linear combination of the eigenvectors of \Hcost. At $t=0$, i.e., point A on the path of \Fig{semicircle_path}, the eigenvector is the ground state of \Hcost. At $t=1$, i.e., point B on the path, it is the highest excited state of \Hcost. The figure shows that the change in eigenvector occurs mainly near the avoided crossing in eigenvalues along the path, and remains almost entirely in the subspace of the two eigenvectors involved in the degeneracy. The other six eigenvectors contribute only slightly to the \UQAOA\ eigenvector as it evolves from the ground to highest-excited state of \Hcost. Perturbation theory shows those slight contributions vanish in the limit $\epsilon \rightarrow 0$. Thus, the behavior of the 8-level system near the degeneracy is well-approximated by a swap within the 2-level subspace involved in the isolated degeneracy.

\subsubsection{Eigenvector Swap for QAOA with Wrap-Around}

\begin{figure}[tb]
    \centering
    \includegraphics[width=0.4\textwidth]{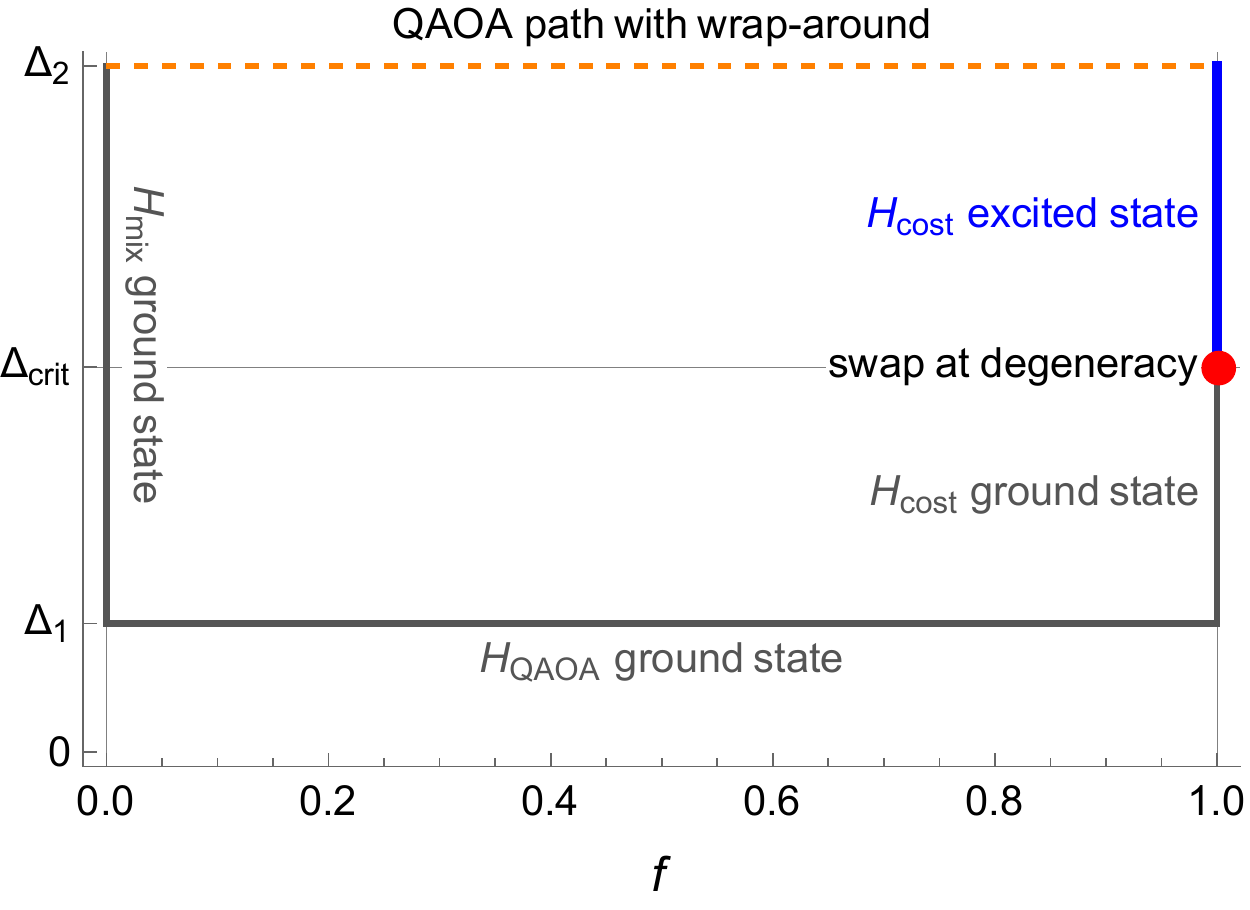}
    \caption{A closed path relating QAOA eigenvector evolution to the swap at a degeneracy at $f=1$ (indicated by the red point). The path consists of vertical segments at $f=0$ and $f=1$, and horizontal segments with values of $\Delta$ below and above the degeneracy: $\Delta_1 < \DeltaCrit < \Delta_2$. The dashed line indicates a QAOA path with wrap-around and eigenvector swap. The labels indicate the eigenvectors on various parts of the path, as described in the text.}
    \label{f:closed_path_swap}
\end{figure}

A QAOA path through parameter space, i.e., from $f=0$ to $f=1$, does not, typically, pass through a degeneracy. 
%DISCUSS#
Thus the argument establishing the eigenvector swap based on degenerate perturbation theory does not apply. Nevertheless, when $\Delta$ is increased from below $\DeltaCrit$ to above it, a swap occurs along the QAOA path.
%RESOLVED#??
To see this, we construct a closed path as illustrated in \Fig{closed_path_swap} to connect behavior along the QAOA path to the swap at the degeneracy.
Along the vertical edge at the right, with $f=1$, the eigenvector changes from \Hcost\ ground to excited state at the degeneracy. However, other than at the degeneracy, the eigenvector does not change.
Along the left vertical edge, with $f=0$, there is no change in the eigenvector, which is the ground state of \Hmixer. Along the horizontal path with $\Delta=\Delta_1<\DeltaCrit$, the eigenvector of \UQAOA\ continuously changes from the ground state of \Hcost\ to that of \Hmixer\ (there can be no interactions with high-lying excited states yet as argued in Sec.~\ref{s:nature_of_degeneracies}). 
Thus, on the horizontal portion of the path along $\Delta=\Delta_2 > \DeltaCrit$ (dashed line in \Fig{closed_path_swap}) the eigenvector continuously changes from the ground state of \Hmixer\ to the excited state of \Hcost. That is, the eigenvector swaps along this path.

\begin{figure}[tb]
    \centering
    \includegraphics[width=0.4\textwidth]{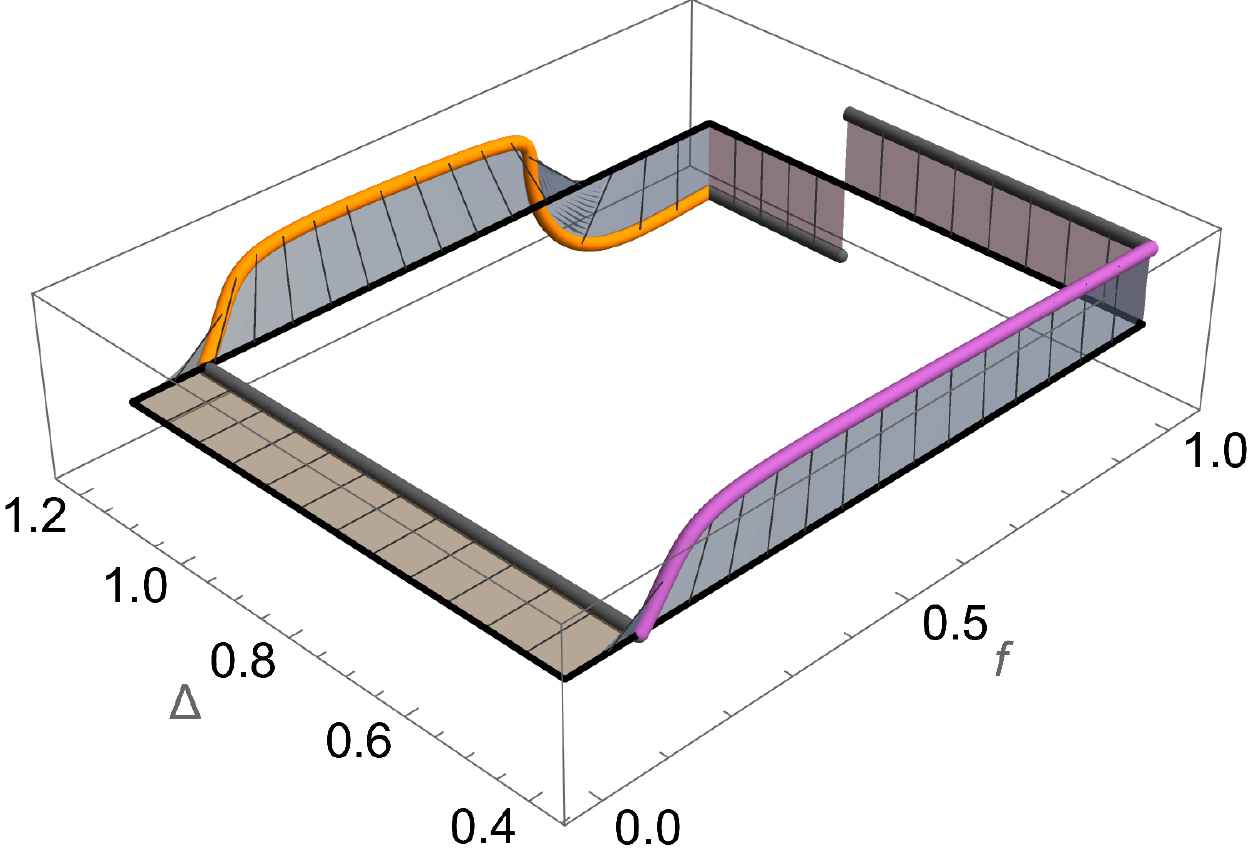}
    \caption{Evolution of the eigenvector of the QAOA operator corresponding to $\ket{+}$ at $f=0$ (i.e., the ground state of the mixer) along a closed path in $f$ and $\Delta$ for the 2-level example. At each point along the path, shown as the black rectangle, the plot indicates the location on the Bloch sphere of the eigenvector relative to that location.
    The purple and orange portions of the path correspond to the small and large $\Delta$ paths of \Fig{fixed_delta_path_diagram}, respectively. The Bloch axes are aligned so that the $\ket{+}$ state points in the direction of increasing $f$ and the $\ket{+i}$ state points in the direction of increasing $\Delta$.
  %  \sh{[SH: I'm having trouble following: how/why does the eigenvector swap on the f=1 edge? (argument operator can't change a comp basis eigenstate! i.e. U commutes with dU along that edge)]}}\tad{There can't be a change along $f=1$ \emph{except} at degeneracy, which is precisely why the degeneracy matters to account for the swap above \DeltaCrit\ on the part from $f=0$ to $f=1$.}
 %   \sh{[SH: the caption seems to describe a path starting and ending at f=1, not consistent with explanation I was given re starting and ending at degen point? (main text different..)]
    %[SH: I understand this figure to show that as the Delta-width of the square is increased we go from 0 to 2 swaps, with Delta-crit the 1-swap transisiton point, and this behaviour arising from holonomy / fact that e'state must go to same one around loop - this no longer seems to be what's communicated?]
    }
    \label{f:eigenvector_path}
\end{figure}\

We illustrate the relation between swap at the degeneracy and along the QAOA path in \Fig{eigenvector_path} for the 2-level example. \Section{degenerate_case} discusses the case of more than two states wrapping around at the degeneracy.

This argument shows the swap occurs on any path starting from the mixer ground state and ending with $\Delta$ above \DeltaCrit\ at $f=1$ provided it does not go through other degeneracies. In particular, the path need not be linear so eigenvector swaps are not specific to linear ramp of \Eq{gamma and beta}.

\subsubsection{Topological Perspective on Eigenvector Connections}
\label{s:topological_perspective}
%\tad{alternate section title since this discussion is not just about closed paths: Topological Perspective on Eigenvector Connections}

% \tad{Clarify what type of holonomy we are discussing: exotic? ``regular''? Something different because behavior is not invariant to small continuous deformations of the path (i.e., returning to a different eigenvector only happens for paths going through the degeneracy: any slight change to the path gives continuous eigenvector evolution instead.)}
% \anuj{modify this discussion slightly so that it flows better with the previous section}

% An additional perspective on the behavior seen in \Fig{closed_path_swap} and \Fig{eigenvector_path} is to note that going around the path, starting and ending at the degeneracy, results in a swap. That is, the initial eigenvector returns to a different eigenvector upon completing the circuit of the closed path. 

% As illustrated in Fig{closed_path_swap} and \Fig{eigenvector_path}, going around the closed path that starts and ends at the degeneracy takes us to an eigenvector that differs from the initial eigenvector. 

The eigenvectors of the QAOA operator~\eqref{e:unitary} change continuously along paths through parameter space. As a result, where an eigenvector connects as we move through parameter space is determined entirely by where the path begins and ends, and whether the path goes through any degeneracies. As illustrated in \Fig{closed_path_swap} and \Fig{eigenvector_path}, going around the closed path that starts and ends at the degeneracy takes us to an eigenvector that differs from the initial eigenvector. 

In the familiar case of Berry phase~\cite{Berry1984}, a closed loop in the parameter space leads to a geometrical phase but does not lead to a change in the eigenvector under consideration. The Berry (geometric) phase acquired along a closed path depends on the area of the loop in parameter space, but the eigenvector swap only depends on whether or not the path \emph{goes through} an isolated degeneracy (rather than just containing one, which also distinguishes isolated degeneracies from exceptional points in topological physics~\cite{ma2022}). 
%
% this statement repeated in below paragraph on topological perspective
%Therefore, the existence of the swap is insensitive to small deformations of the path that are away from the degeneracy. 

The analysis of eigenvector behavior can be generalized since it only relies on a closed path in the $(f,\Delta)$ space passing through a single isolated degeneracy. For the more general case, the eigenvector that becomes degenerate along the path will evolve into an orthogonal eigenvector at the end of path. When the starting point of the closed path is degenerate, then isolated degeneracy leads to an evolution into an orthogonal eigenspace. 

Given a family of parameterized Hamiltonians, cyclic evolution along a closed path can also induce interchanges of eigenvalues, with the initial and final states having different eigenvalues. This permutation of eigenstates while traversing a closed path in parameter space through a degeneracy, as described in this section, corresponds to nontrivial holonomy of the path~\cite{cheon09,tanaka_tops}. For discussion of holonomy in related applications see~\cite{tanaka11,tanaka_quasi_energies,cheon_delta_ham, ergoktas22,schumer22,patil_braiding}.

A topological perspective on QAOA may provide insight into its generic behaviors. For example, variation of the closed path does not change the eigenvector swap behavior, provided the distorted path continues to pass through the degeneracy. The portion of the path relevant for QAOA is the connection between f=0 and f=1, whereas the vertical portions are constructions that do not correspond to QAOA steps. Therefore, changes in the path due to variation in schedules or implementation noise do not alter the eigenvector swap behavior described here, provided the changes don't result the path in crossing additional degeneracies. %Thus this behavior of QAOA is unchanged by variations in schedules or noise, provided that those changes are not so large as to cross additional degeneracies. This contrasts with the focus in QAOA studies on optimizing schedules for particular problems, where even relatively small changes to the schedule can improve performance.

Another insight from topology is that as $\Delta$ increases the increasing number of wrap arounds of the eigenvalues may lead to, in effect, a connection between the mixer ground state at $f=0$ and a random cost state at $f=1$. If this occurs, the large-$p$ performance of QAOA for large $\Delta$ will be close to random guessing, in contrast to the performance for $\Delta$ just above \DeltaCrit, where QAOA connects to the highest-energy cost eigenstate, giving the worst possible expected cost. The topology of how eigenstates connect when $\Delta$ is large may be a useful guide as to how to adjust the step size (corresponding to a nonuniform schedule) to either track or avoid tracking eigenvector swaps over narrow parameter intervals, in order to increase the chances of QAOA reaching low cost eigenstates.

\subsection{Eigenvalue Gaps Associated with Wrap-Around}
\label{s:nature_of_degeneracies}

%\tad{The section title ``nature of degeneracies'' is too generic to indicate what this section is about, e.g., perhaps about wrap around of degenerate eigenvalues (the topic of Section 4.5) rather than behavior of gaps of avoided crossings associated with wrap-around of nondegenerate eigenvalues which form a two-eigenvalue degeneracy at f=0 or f=1 at the Delta value of wrap-around. Perhaps ``Eigenvalue Gaps Associated with Wrap-Around'' or ``Eigenvalue Gaps from Wrap-around of Non-degenerate Eigenstates' to distinguish from Section 4.5??}

As argued in Section~\ref{s:degeneracies and eigenvalues}, connections cost and mixer eigenstates can be reorganized if wrap-around by the eigenvalues of the cost (or mixer) unitary results in an avoided crossing. This section describes some properties of the gaps associated with these avoided crossings, with the discussion applying symmetrically to wrap-around caused by both cost and mixer eigenvalues. Derivations for these properties are in Appendix~\ref{s:terminal_degeneracies}, with their precise statements in the gap formula of Eq.~\ref{e:main_gap_equation} and Theorem 1.

Perturbation theory indicates that isolated degeneracies from wrap-arounds at $f=0,1$ occur generically for most choices of cost and mixer. Since the size of the encountered eigenvalue gaps affects how well a discrete evolution tracks an eigenvector~\cite{costa2021}, we characterize the scaling of the gap from the avoided crossing due to wrap-around. We show in Eq.~\ref{e:main_gap_equation}\ that for the specific case of wrap-around between two non-degenerate cost eigenstates $\ket{x}$ and $\ket{y}$, the gap size $|g|$ due to the avoided crossing from wrap-around at $\Delta^*,f=1$ \emph{generically} goes as 
\begin{equation}
|g| = 2\cEff \delta^k + O(\delta^{k+1})~,
\label{e:simplified gap formula}
\end{equation}
with relative error from the higher-order contribution vanishing as $\delta \rightarrow 0$. In Eq.~\ref{e:simplified gap formula}, $\delta=\Delta-\Delta^*$,  $k$ is the length of the shortest coupling path between the two wrapping cost eigenstates through the mixer (referred to in this paper as the \textbf{``coupling distance''}), $\cEff$ is an effective coupling computed through the weighted sum of products of mixer off-diagonal couplings as given in Eq.~\ref{e:effective_coupling}, and we have simplified it from Eq.~\ref{e:main_gap_equation}\ by assuming the mixer has zero diagonals. While this paper only provides a method for computing $\cEff$ for the wrap-around of non-degenerate cost eigenvalues, the degenerate case also exhibits scaling $O(\delta^k)$ in the $\delta\rightarrow 0$ limit for mixer couplings between degenerate cost eigenstates- similar to the gap scaling with low-lying excited states found for NP-Complete problem instances in the continuous adiabatic case~\cite{altshuler2009}. While this scaling holds for most Hamiltonians, the special case of the mixer and cost Hamiltonians being very sparse and local  is discussed in Appendix~\ref{s:gap size for most wrap-arounds}, where the exponent can be significantly greater than the coupling distance. 

As a clarifying example, with the $X$ mixer and a cost Hamiltonian diagonal in the computational basis, $k$ equals the Hamming distance between the states and the coupling strength between states is 1. Another example is the diffusion mixer, used in unstructured search, where all non-diagonal entries are equal (and exponentially small). In that case, $k=1$, but due to the exponentially small non-diagonal entries, $\cEff$ and the gap are exponentially small.

Our derivation for the gap size is based on the lowest-order term of a sum of contributions from different coupling paths. Thus it applies in the limit of small $\delta$. For larger $\delta$, other paths could contribute, especially those with much stronger couplings than the lowest-order term. For example, consider a slight modification to the $X$ mixer: adding a small coupling, say, $2^{-10}$, between the states $\ket{000}$ and $\ket{111}$ that are not coupled by the $X$ mixer. Near the wrap around of these two states, specifically for $\delta \ll 2^{-(10/3)}$, this coupling would dominate in its (small) contribution to the gap over that of a 3-coupling path of the $X$ mixer between the states with $\cEff=1$. However, for larger $\delta$ the $k=3$ path would dominate. 

In the case where the gap size is small compared to the gaps with neighboring cost eigenvalues of the wrapping eigenvalues, the formula is quite accurate. For example, in Fig.~\ref{f:semicircle-eigenvalues8}, Eq.~\ref{e:main_gap_equation} predicts a gap size of $4.761 \times 10^{-8}$ compared to the numerically computed gap size of $4.766 \times 10^{-8}$. Further, if $k\geq 2$, then the gap from wrap-around at $f=1$ occurs at $1-(\delta/\DeltaCrit) + O(\delta^2)$ with the $X$-mixer and a cost Hamiltonian diagonal in the computational basis. We also show in Appendix~\ref{s:gap_derivation}\ that, for $k\geq 2$, the exchange of eigenvectors occurs approximately symmetrically about the avoided crossing from  wrap-around.

Isolated degeneracies from eigenvalues wrapping around the complex unit circle can also occur for $0<f<1$, as illustrated in \Fig{eigenvalue_difference}, though these do not change connections between eigenstates at $f=0$ and $f=1$, save for paths that go directly through them. We prove through Theorem 1 that such ``intermediate'' isolated degeneracies do not occur at $\Delta$ smaller than the value for which the first pair of eigenvalues wraps around at $f=0,1$. Therefore, generically, the first isolated degeneracies from wrap-around due to increasing $\Delta$ are the ones at the endpoints $f=0,1$. Unlike most of the previous results in this section, this result holds regardless of the degeneracy of the states experiencing wrap-around. In fact, a consequence of the Theorem's proof is that the highest excited eigenvalue is closest to the ground eigenvalue at $f=0,1$ rather than in the middle, meaning no interactions between the two eigenstates can take place before wrap-around at $f=0,1$.

\subsection{Wrap-around of Degenerate Eigenvalues}
\label{s:degenerate_case}
In Section~\ref{s:nature_of_degeneracies} most of our quantitative results apply to two initially non-degenerate eigenvalues wrapping around, and this is the main focus of this paper's quantitative analysis. When the two eigenvalues are degenerate, more complicated dynamics emerge. This section describes this complexity for the wrap-around at \DeltaCrit.

The multiplicities of the cost eigenvalues are independent of $\Delta$ so there are always the same number of linearly independent eigenvectors continuously evolving into the eigenvectors of each cost eigenspace. Consider the generic case where \DeltaCrit~occurs when the highest eigenvalue wraps around with the ground eigenvalue and, without loss of generality, suppose this occurs at $f=1$. Let $\mathcal{U}_0$ be the ground eigenspace of the mixer and let $\mathcal{V}_0$ be the ground eigenspace of the cost, with $D^U_0$ and $D^V_0$ their respective dimensions. If the total multiplicity of eigenvalues wrapping around at $f=1$ ($f=0$) with $\mathcal{V}_0$ ($\mathcal{U}_0$) is greater than $D^V_0$ ($D^U_0$),  then all the states that were going to the cost ground state originally will go to some cost excited state - the cost (mixer) excited eigenspaces ``push out" all of the original states. However, if the total multiplicity of wrapping cost excited states doesn't exceed the dimension of the cost ground eigenspace (nor mixer excited states for mixer ground eigenspace), then some of the eigenvectors that continuously evolved to the cost ground eigenspace at lower $\Delta$ continue to do so. Therefore, for degenerate ground eigenvalues, the $\Delta$ at which connection between mixer and cost ground eigenspaces is completely severed is determined not necessarily by wrap-around with the maximal excited energy, but by the smallest $\Delta$ at which the multiplicity of wrapping excited eigenvalues exceeds the multiplicity of the ground eigenvalue at $f=0$ or $f=1$ (whichever comes first).

\section{Understanding QAOA performance diagrams}
\label{s:performance_diagrams}

As described in the Introduction, QAOA performance diagrams exhibit similar qualitative behaviors for a variety of schedules, mixers and problems, including those of classical optimization and quantum chemistry~\cite{kremenetski21diagram}. 
This section explains these behaviors and their generality using the \UQAOA\ eigenstate connection changes discussed in \Section{QAOA operator behavior}, conditions on when the QAOA state vector tracks the changing eigenvectors, and QAOA behavior for small angles~\cite{hadfield2021analytical}.

\begin{figure}
    \centering
    \includegraphics[width=0.45\textwidth]{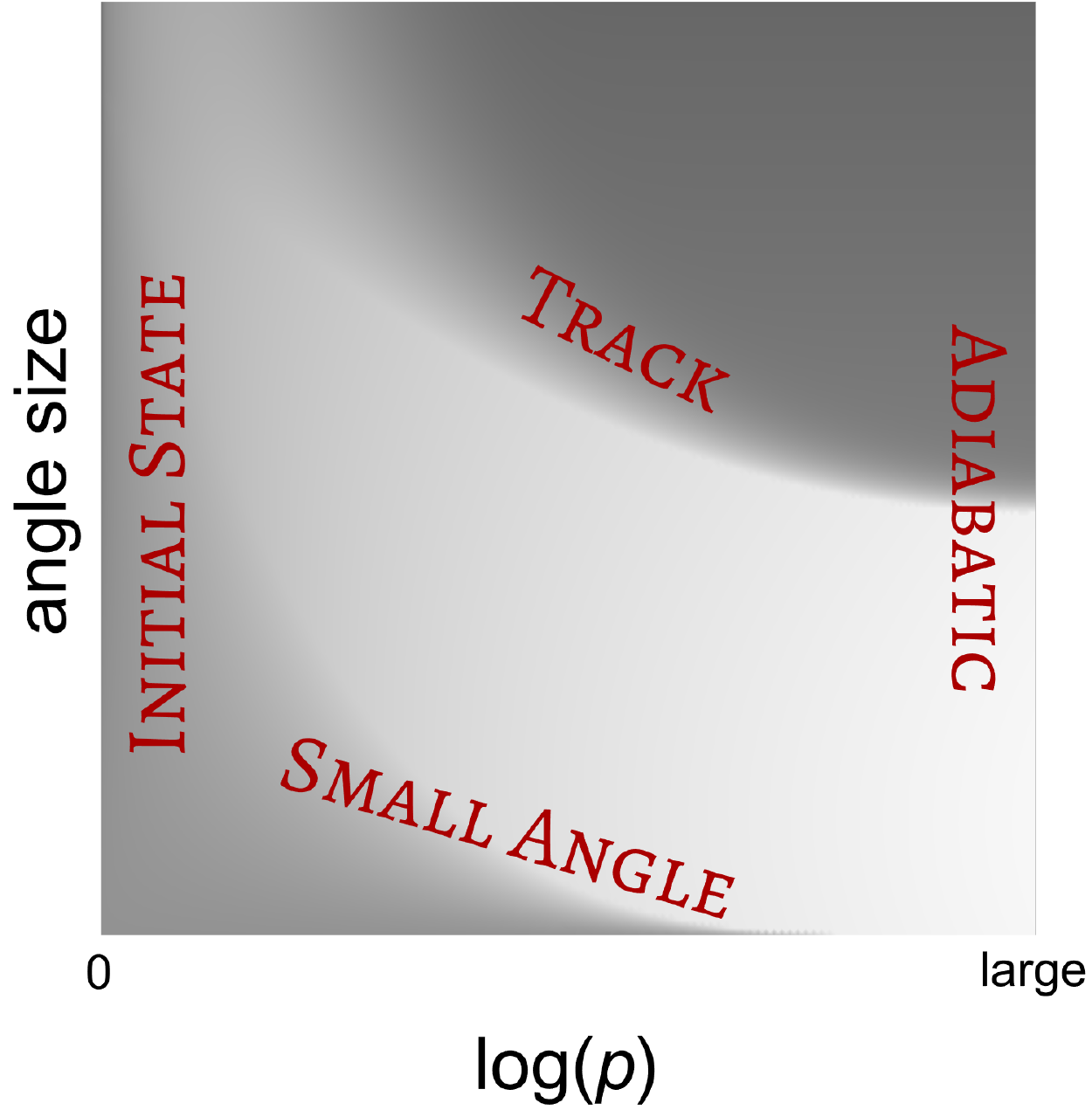}
    \caption{Schematic of our explanations of the QAOA performance diagram.
    Shading indicates performance, with lighter shading corresponding to higher performance. The labels indicate the rough locations on the diagram where each explanation applies.}
    \label{f:general_explanations}
\end{figure}

\begin{figure*}
    \centering
    \includegraphics[width=0.8\textwidth]{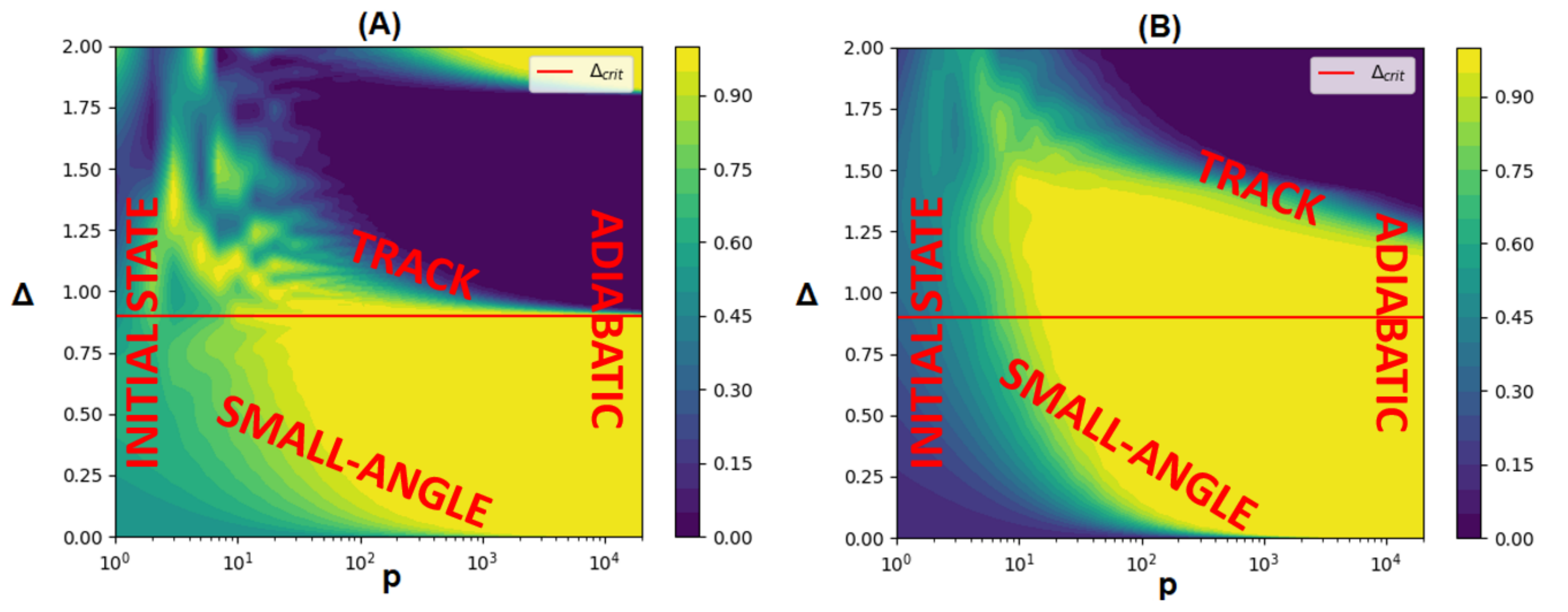}
    \caption{QAOA performance diagrams showing squared overlap with the cost ground state for (A) the two-level example given in Eq.~\ref{e:2-level example} and (B) the Hamiltonian pair from Fig.~\ref{f:eight_level_avoided_crossing_example}. For orders of magnitude of $p$ larger than the ones on the graph, the \region{Ridge} region eventually decreases to \DeltaCrit~marked by the red line in (B).}
    \label{f:qaoa_diagram_comparison}
\end{figure*}

%\Fig{general_explanations} illustrates the explanations we provide for the performance diagram in this section. For large $p$, \Section{adiabatic_region} explains the \region{Adiabatic} region. \Section{framework_application} explains the \region{Small Angle} region. \Section{performance_ridge} explains the \region{Track} region.

%\tad{We distinguish between behavior regions (e.g. \region{Low}) in Fig. 1 from labels for explanations (e.g., Adiabatic). Is it confusing to call them both regions? Alternate text for the above paragraph:}
These explanations arise from several distinct aspects of QAOA with gradually changing parameters, as summarized in \Section{behavior_sketch}.
First, \Section{adiabatic_region} explains the \region{Low} and \region{High} regions i.e. the regions at large $p$, as due to the discrete adiabatic theorem and changes in eigenstate connections.
Second, \Section{framework_application} explains the bottom portion of the diagram and the lower boundary of the \region{Ridge} region through a perturbative expansion in small angles.
Third, \Section{performance_ridge} explains that the upper boundary of the \region{Ridge} arises due to the QAOA step becoming small enough for the QAOA state vector to track the evolving \UQAOA\ eigenvector.
Finally, \Section{section5_summary} describes how these separate explanations combine to describe the full diagram.
\Fig{general_explanations} schematically illustrates the portions of the diagram to which each of these explanations applies. \Fig{qaoa_diagram_comparison} shows examples of the applicability of these explanations to two actual diagrams.

%\tad{To clarify the distinction between explanations and the behavior regions that they apply to in the text, change subsection titles to replace ``region'' by ``explanation'' or similar term. In particular, a behavior region is a (numerical) observed fact of the diagram whereas an explanation is a theory for that behavior which could be wrong or incomplete.}
%\vk{I agree that this is probably a better way to unite figures 15 + 16 with figure 1. I think to save space I will mix the above explanations with the paragraphs in 5.1}.
%\tad{I suggest using the above as the intro to this Section: so readers know they can get the short summaries in 5.1 or go directly to the detailed explanations in the later subsections (which aren't much longer than the one paragraph summaries in 5.1, except for Section 5.4). With that guidance, probably no need to repeat the guidance in each summary paragraph of Section 5.1.}

\subsection{QAOA Performance Regions}
\label{s:behavior_sketch}

%\subsubsection{General sketch of performance diagram behavior}

%\tad{I merged the first paragraph of this subsection with the 1st paragraph of Section 5, since it applies to the section as a whole, not just this short summary subsection}
%As described in the Introduction, QAOA performance diagrams exhibit similar qualitative behaviors for a variety of problems and mixers~\cite{kremenetski21diagram}. 
%\tad{I think no need to mention squared overlap since that's the default we're using in the text up to this point; expected cost comes with Ising example in Sec. 6.2}
%Here we again primarily consider the performance in terms of the squared overlap with the ground state. 

This subsection gives an overview our explanations of the features of QAOA performance diagrams. Subsequent subsections provide details of these explanations.
% the following text restates what Section 5 does as given above, no need to repeat
%We qualitatively consider different observed performance regions in turn, in particular, relative to \DeltaCrit, the smallest value of $\Delta$ at which the eigenvalues of \UQAOA~wrap around at $f=0$ or $f=1$ and results in an isolated degeneracy, as discussed in the previous section. 

%\tad{Check these paragraphs summarize our \emph{explanations} rather than mainly 1) restating the behaviors we intend to explain without saying what the explanations are, and 2) repeating the one-sentence summary of explanations given above in the introduction to Section 5. I.e., this subsection should provide the reader with some understanding of what the explanations are without needing to read the detail subsections.}

\textbf{I. Discrete adiabatic region: } From the discrete adiabatic theorem, for very large $p$ (far right of diagram), the output state has high support on whatever cost eigenstate the mixer ground eigenstate is connected to. Below \DeltaCrit~this is the cost ground eigenstate, and above \DeltaCrit~ this is some cost excited eigenstate.

\textbf{II. Small angle approximation: } In the bottom part of the diagram, the performance is described using a small-angle expansion of \UQAOA~ operator. For these small angles, the QAOA schedule approximates a continuous anneal, with increasing $\Delta$ and $p$ associated with a longer anneal time and hence better performance. 
%in Section~\ref{s:framework_application} using a small-angle approximation framework. 

\textbf{III. Tracking eigenvector swap across eigenvalue gaps:} Above \DeltaCrit, changing eigenstate connections due to wrap-around produce avoided crossings. The gap size of these avoided crossings shrinks exponentially in how indirectly the mixer (or cost) couples the eigenstates whose eigenvalues have wrapped around. The changing eigenvectors continuously exchange places over a narrow interval (or ``swap") at the gap, with the swap taking place over smaller intervals for smaller gaps. Consequently, in order for performance to deteriorate with increasing $p$, the discretization must be fine enough for QAOA to sample the interval over which this swap happens. For finer discretization, the interaction at the gap may be modeled with a discrete version of the Landau-Zener formula. Finally, the value of $p$ at which performance deteriorates is determined by the $p$ at which reduced diabatic transitions to the ground state at the gap outweigh improvements in tracking the changing eigenvectors up to the gap.

\textbf{IV. Initial State dominated region: } Along the left edge of the performance diagram, $p$ is so small that the squared overlap of the QAOA output state only slightly improves on the initial overlap of the mixer ground state with the cost ground eigenspace. Thus the choice of initial mixer ground state is the dominant determiner of performance. This follows from previous observations that QAOA exhibits limited performance unless $p$ is at least of order $\log(n)$ for $n$ qubits~\cite{zhou2020, bravyi2019,hastings2019}.

%\tad{Since this is the only place we explain \region{Initial State}, can we clarify what this bound means? As stated here, it just seems to say that, for a given problem instance, there exists some value of $p$ below which QAOA performance is ``limited'' -- that's always the case except for trivial problems -- such as Hamming weight ramp -- where QAOA can solve the problem exactly with $p=1$. Presumably there is a stronger statement here with a bound related to scaling of problem size, such as ``$p$ must be at least of order $\log n$''. This will be useful to refer back to in discussing the scaling diagram applying only for $p\gg 1$, in particular for placing $\log n$ at the left edge of the horizontal axis.}

\subsection{Discrete Adiabatic Region}
\label{s:adiabatic_region}
The behavior near the right edge of the QAOA performance diagram arises from the discrete adiabatic theorem and connections between \UQAOA\ eigenstates. Below \DeltaCrit, in the generic case where there are no degeneracies with low-lying excited states in the middle, the QAOA output state has squared overlap with the cost ground eigenspace near one. This is because the mixer ground eigenstate is connected to the cost ground eigenstate, so by the discrete adiabatic theorem~\cite{dranov98,costa2021}, the state vector tracks the continuously changing eigenvectors from mixer ground state to cost ground state. 

As discussed in Section~\ref{s:QAOA operator behavior}, above \DeltaCrit\ the ground eigenstate of \Hmixer\ is connected to the maximal excited eigenstate of \Hcost, and so the discrete adiabatic theorem implies that the QAOA output state has squared overlap near one with the maximal cost excited state and consequently a high expected energy. As $\Delta$ continues to increase, the mixer ground eigenstate forms new connections to other excited eigenstates of \Hcost, including some low-lying ones, as eigenvalues at both $f=0$ and $f=1$ continue to go around the complex unit circle. Depending on the spacing of the cost and mixer energies, for all but very small systems (as in Fig.~\ref{f:qaoa_diagram_comparison}(A)) the connection between mixer ground and cost ground eigenstates might never be established again, or re-established only for very small intervals of $\Delta$. 

The situation is more complicated when the mixer or cost ground eigenvalue is degenerate. In this case, different sub-spaces of the mixer ground eigenspace can connect to different excited cost eigenspaces, and a degenerate cost eigenvalue can mean that wrap-around with a few eigenvalues (multiplicity equal to the ground space dimension) must happen at $f=1$ or $f=0$ to fully sever connection with the mixer ground eigenspace, as discussed in Section~\ref{s:degenerate_case}. While most of the quantitative analysis in this paper concerns the case when the unitary eigenvalues wrapping around are non-degenerate, the same framework can qualitatively describe performance diagrams for the degenerate case.

\subsection{Small Angle Region} %- behavior in bottom portion}
\label{s:framework_application}

We explain the lower-left region of the QAOA performance diagram through an application of small-angle approximations for QAOA. Exact expressions as power series in the algorithm parameters for QAOA probabilities and expectation values are derived in~\cite{hadfield2021analytical} using the Heisenberg representation of quantum mechanics. 
Keeping only the leading-order terms results in 
easily computable approximations valid when all parameters have relatively small magnitudes~\cite[Sec.~4]{hadfield2021analytical}, as is the case for the lower portion of the QAOA performance diagram. 

As described in \Appendix{framework}, the small-angle approximation relates QAOA performance to sums over combinations of QAOA parameters and expectations of cost operators.
  
The form of these sums indicates why the performance diagram behavior is general with respect to problems, mixers and schedules: because behavior as a function of $p$ and $\Delta$ enters only through the angle sums, so as long as coefficients -- from cost and mixer operators -- do not change sign, the framework predicts the same qualitative behavior at the bottom of the performance diagram. 

The small-angle expansion of~\cite{hadfield2021analytical}  
converges to the exact value in the regime $\Delta \ll 1/p$ (generalizing the case $\Delta \ll 1$ in~\cite{hadfield2021analytical}, which applies for fixed $p$).
The theory describes the qualitative behavior of QAOA for somewhat larger parameters~\cite[Fig. 4]{hadfield2021analytical},
including the qualitative behavior of the lower portion of the QAOA performance diagram even for $\Delta$ comparable to $1/p$. See \Section{Ising_example} for an example.

On a more conceptual level, for small enough $\Delta$, we can interpret QAOA as the implementation of a Trotterized, continuous adiabatic evolution between two interpolated Hamiltonians. As such, performance improves as $p$ increases, i.e., corresponding to a slower anneal, while increasing $\Delta$ at fixed $p$ corresponds roughly to increased evolution time.

\subsection{Eigenvector Tracking and Performance Near the Ridge}
\label{s:performance_ridge}

This subsection explains that the \region{Ridge} region for $\Delta > \DeltaCrit$ arises from the QAOA state vector stepping over the gap in eigenvalues of the \UQAOA\ operator when $p$ is not too large. This region ends when $p$ is sufficiently large that QAOA tracks the swapping eigenvector at the gap, leading to the adiabatic limit.

\subsubsection{Tracking the swapping eigenvector}
\label{s:holonomy_application}

As discussed in Section~\ref{s:dat}, the step size in parameter space (here given by $1/p$) required to track the changing eigenvectors of \UQAOA\ depends on the eigenvalue gaps encountered along the path through parameter space, with smaller gaps requiring smaller step sizes (here, larger $p$). When $p$ is too small, the eigenvector swap taking place at the avoided crossing from wrap-around above $\DeltaCrit$ is missed entirely, causing the state vector to track the ``wrong'' eigenvector to the ground state. Once $p$ is large enough, the discretization is fine enough to track an eigenvector up to and through the swap, causing the overlap of the state vector with the maximal excited state to increase.

While we leave a precise estimate of this size of $p$ for future work, we show in \Appendix{ptrack_derivation} the width of the interval over which the eigenvector exchange (or ``swap") occurs shrinks exponentially with how indirectly the two swapping cost eigenstates are coupled by the mixer.

Overall, larger steps through parameter space (i.e., smaller $p$) can allow QAOA to step over the small interval over which the eigenvectors continuously swap. For a reader familiar with the continuous adiabatic regime and annealing, this situation should be surprising; picking QAOA angles large enough that the  adiabatic limit performs poorly (by going to the wrong eigenvector) can nevertheless improve QAOA performance by allowing QAOA to deviate from the adiabatic limit with larger parameter steps.

\subsubsection{Landau-Zener Formula explains declining ridge slope on the right}
\label{s:landau_zener}

\begin{figure*}
    \centering
    \includegraphics[width=\textwidth]{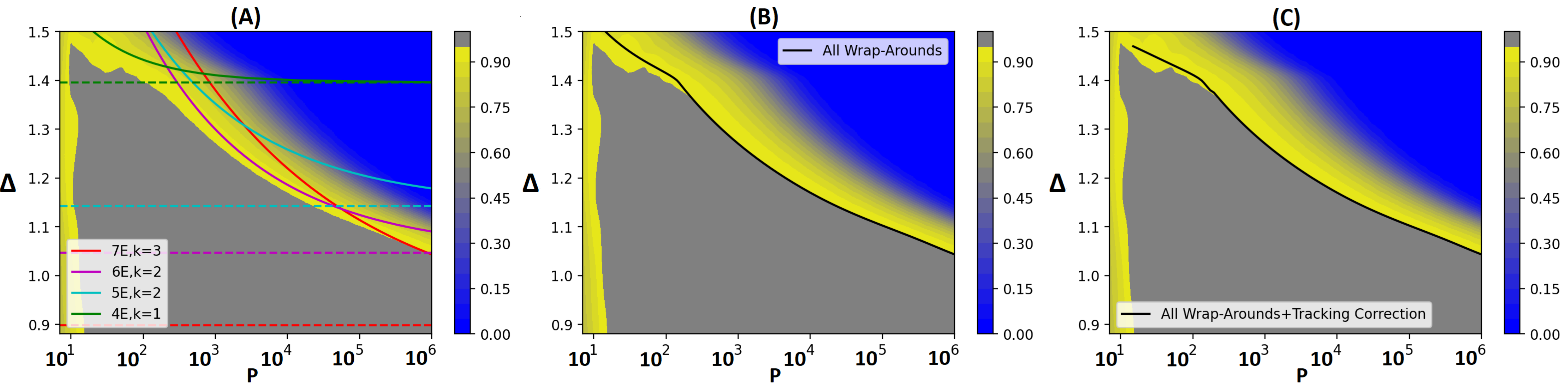}
    \caption{QAOA performance diagrams showing squared overlap with the cost ground state for the Hamiltonian pair from Fig.~\ref{f:eight_level_avoided_crossing_example}. The gray region covers the set of parameters for which the squared overlap of the output QAOA state with the cost ground state is at least 0.95. (A) The horizontal lines denote the values of $\Delta$ at which wrap-arounds occur (i.e. $\Delta^*$) between the cost ground state and the seventh (7E), sixth (6E), fifth (5E), and fourth (4E) excited states, in order of increasing $\Delta^*$, as well as their coupling distance to the cost ground state ($k$). The curves plot the values of $p$ for which Eq.~\ref{e:P_p} predicts a 95$\%$ rate of diabatic transition for each wrap-around. At each $\Delta$, the curve with the smallest $p$ determines the overall shape of the contour. (B) The curve shows the predicted value of $p$ where an overall 95$\%$ transition to the cost ground state is predicted from diabatic transitions across \emph{all} the gaps. (C) Same as (B) but numerically correcting for imperfect QAOA tracking of the eigenvectors up to the swaps at the gaps. Calculations to generate the curves are in Appendix~\ref{s:dlz example derivation}. }
    \label{f:dlz performance benchmark}
\end{figure*}

Performance deterioration with $p$ (the slope of the ridge after the decline starts) can be approximated with a discrete version of the Landau-Zener formula, due to interactions at the gap.

For $p$ sufficiently large, the discretization is fine enough to track the changing eigenvectors up to the avoided crossing, and the dynamics in the neighbourhood of the avoided crossing can be well-estimated by an off-diagonal perturbation in the ``effective" two-level Hamiltonian \cite{cohen_2020}. In \Appendix{discrete_landau_zener_derivation}, we use this approximation to derive the probability of a diabatic transition to the cost ground state  
\begin{equation}\label{e:P_p}
    P \approx \exp(- \frac{B_0}{1+\delta/\Delta^*} \delta^{2k} p)~,
\end{equation}
where $\delta=\Delta-\Delta^*$ with $\Delta^*$ being the value at which wrap-around occurs at $f=1$, and $B_0$ is a positive constant  given by Eq.~\ref{e:b0 definition}, independent of $\delta$ but dependent on $\cEff$, which, along with the coupling distance $k$, is defined in Section~\ref{s:nature_of_degeneracies} and explicitly provided in Eq.~\ref{e:effective_coupling}. While the general Discrete Landau-Zener approach is useful for avoided crossings emerging anywhere in the parameter space, the above formula is specifically derived for wrap-around at $f=1$ (or a lightly modified version at $f=0$, with \Hcost~and \Hmixer~dependencies exchanged).

%\tad{Section 4.4 notes isolated degeneracies for $0<f<1$ occur for larger Delta than first wrap around at $f=1$, hence suggesting they aren't relevant for the Ridge boundary. If, in practice, that boundary comes from intermediate-energy wrap around rather than the first (from highest energy state), could degeneracies for $0<f<1$ contribute? Or can we generalize the argument from Section 4.4 to apply to intermediate-energy wrap arounds, which also first occur at $f=1$ (or $f=0$ depending on spectrum of cost or mixer) rather than intermediate $f$?}
Section~\ref{s:nature_of_degeneracies} notes that wrap-around at $f=0,1$ is the source of the first isolated degeneracies as $\Delta$ increases from zero. Isolated degeneracies can occur for intermediate values $f$ for larger $\Delta$. However, as these do not change the overall eigenstate connections, they are only relevant for narrow ranges of $\Delta$. Thus the dominant source of performance deterioration arises from wrap-around at $f=0,1$.

For $\Delta$ above the smallest value of $\Delta^*$  (i.e., $\DeltaCrit$), QAOA leads to the ground state of $\Hcost$ only if the state vector undergoes a diabatic transition at the avoided crossing due to wrap-around. As $\Delta$ increases and other wrap-arounds take place, QAOA leads to the ground state only if \textit{multiple} diabatic transitions occur across the gaps. If one gap has a diabatic transition rate far lower than the others, it is the dominant source of performance deterioration. Therefore, for constant step size, the right boundary of the \region{RIDGE} region is determined by the largest gap. 
%\tad{Is the relevant quantity how rapidly the gap grows with $\Delta$ or its size at a given $\Delta$? Whether QAOA steps over the gap on a constant-$\Delta$ path presumably depends on the size of the gap, not how it changes with $\Delta$?? E.g., could we have a slower-growing gap (e.g., due to $k=1$) that is nevertheless larger than a faster growing gap (e.g., with $k=2$) over some range of $\Delta$ due to different constants of proportionality in the growth $\delta^k$}

Section~\ref{s:nature_of_degeneracies} explains that as $\Delta$ increases above the value $\Delta^*$ at which the wrap-around takes place, each respective gap grows as $\delta^{k}$. Eq.~\ref{e:P_p} suggests that the rate of diabatic transition decreases at each gap, pushing the right boundary of the performance ridge to the left and performance deteriorates faster with increasing $p$. 
This means that distantly-coupled states (higher $k$) not only require exponentially larger $p$ for QAOA to notice the swap, but the effect from this swap is very small and grows very slowly with increasing $p$. For instance in \Fig{qaoa_diagram_comparison}(A) where the wrapping states are directly coupled, performance deteriorates rapidly just above $\DeltaCrit$. By contrast, in \Fig{qaoa_diagram_comparison}(B), the third-order coupled states have a slow performance decay even after discretization is fine enough to track an eigenvector up to the first swap.

\Fig{dlz performance benchmark} shows the predictions of Eq.~\ref{e:P_p} for the performance diagram of \Fig{qaoa_diagram_comparison}(B).
Eq.~\ref{e:P_p} gives upper bounds for the values of $p$ at which a specified success probability $P$ can be obtained (excepting edge cases where the ground state eigenstates become accidentally reconnected). This is because these curves indicate the value of $p$ where a fraction $P$ of the QAOA state \textit{at the respective avoided crossing} goes on to the ground state. Any deviations from this happen because there were \textit{even more} deviations prior to the avoided crossing. For example. this could be caused by small gaps with low-energy excited states, other avoided crossings due to wrap-around, $p$ too small to closely track the changing eigenvector up to the avoided crossing, etc. This is why all curves in Fig.~\ref{f:dlz performance benchmark}(A) overestimate the $p$ at which the $P_0=0.95$ success contour occurs, due to poor tracking at small $p$ and due to other wrap-around avoided crossings at large $p$. Fig.~\ref{f:dlz performance benchmark}(B) corrects for the latter cause by taking all wrap-around gaps into account, but still overshoots the $P_0=0.95$ contour at small $p$ due to $p$ being too small to track the changing eigenvectors.  We numerically correct for this in Fig.~\ref{f:dlz performance benchmark}(C), giving a more accurate contour prediction.  From this, we can see that if $\Delta$ is large enough, the $p$ required to achieve any high diabatic transition to the ground state is so small that the state vector is unable to closely follow any eigenvector of the QAOA operator and consequently the performance degrades.

We can approximate the value of $p$ required to achieve a squared overlap with the ground state close to unity $ P = 1 - \varepsilon$
by inverting Eq.~\ref{e:P_p} and doing a Taylor expansion to obtain 
\begin{equation}
    p_{\varepsilon} \approx \frac{\varepsilon}{B_0} \frac{(1+\delta/\Delta^*)}{\delta^{2k}} =  - \frac{ \varepsilon ~ p_{0}}{\log{P_{0}}}~.
\label{e:lz_p_predictor}
\end{equation}
where $P_{0}$ is the probability of obtaining the ground state at a specific value $p_{0}$. %For example, $ P_{0}\approx 0.6585$ at $p_{0}=5000$ steps for $\Delta = 1.3$ in the example from \Fig{eight_level_avoided_crossing_example} and \Fig{dlz performance benchmark}. Eq.~\ref{e:lz_p_predictor} estimates that the number of steps needed for $ 95 \%$ probability of finding the ground state is $p_{\varepsilon}\approx 598$, which is quite close to correct answer of $p = 603$. 

\subsection{Summary}
\label{s:section5_summary}
In the adiabatic performance region, for $p$ exponentially large in system size, the performance is determined by changing eigenstate connections as $\Delta$ varies. In the small-angle region, 
an expansion approximation explains the increasing performance with fixed $p$ and increasing $\Delta$. Above $\DeltaCrit$, performance on the ridge is explained by whether $p$ is large enough to detect swaps with high-lying excited states, with the size of $p$ required exponential in how indirectly coupled the states are. 

Our discussion explains why QAOA performance diagrams have the same qualitative behavior for the metric of expected cost of the output state, e.g., as seen in \Section{Ising_example}. For sufficiently large $p$ and above $\DeltaCrit$, the output state gains support on high-lying excited states and thus a high expected energy. While these are intermixed with connections to low-lying excited states at narrow ranges of $\Delta$, there still are accumulated avoided crossings with eigenvalues that lead to those of high-lying cost excited states. Thus the same qualitative features of QAOA diagrams occur for several performance metrics.

\section{Examples}
\label{s:examples}

\subsection{Expected Cost Performance Diagram for an Ising Problem}
%\subsection{Expected value of QAOA output state and an Ising Example}
\label{s:Ising_example}

As described in \Section{qaoa_overview}, overlap with the ground state and expected cost are two metrics of QAOA performance. Previous examples of QAOA performance diagrams focused on ground state overlap~\cite{kremenetski21diagram}. In this section, we show expected cost gives similar behaviors by using a spin Ising problem as an illustrative example. The Ising spin problem is to find the ground state of $n$ spins $s_i$, $i=1,\ldots,n$ each of which is $\pm1$. The fully-connected version illustrated here has Hamiltonian
\begin{equation}
H= \sum_i h_i s_i + \sum_{i<j} J_{i,j} s_i s_j~,
\label{e:Ising Hamiltonian}
\end{equation}
with the $h_i$ and $J_{i,j}$ selected independently and uniformly at random in the range $-1$ to $1$ for a random instance. 

\begin{figure}[htb]
    \includegraphics[width=0.45\textwidth]{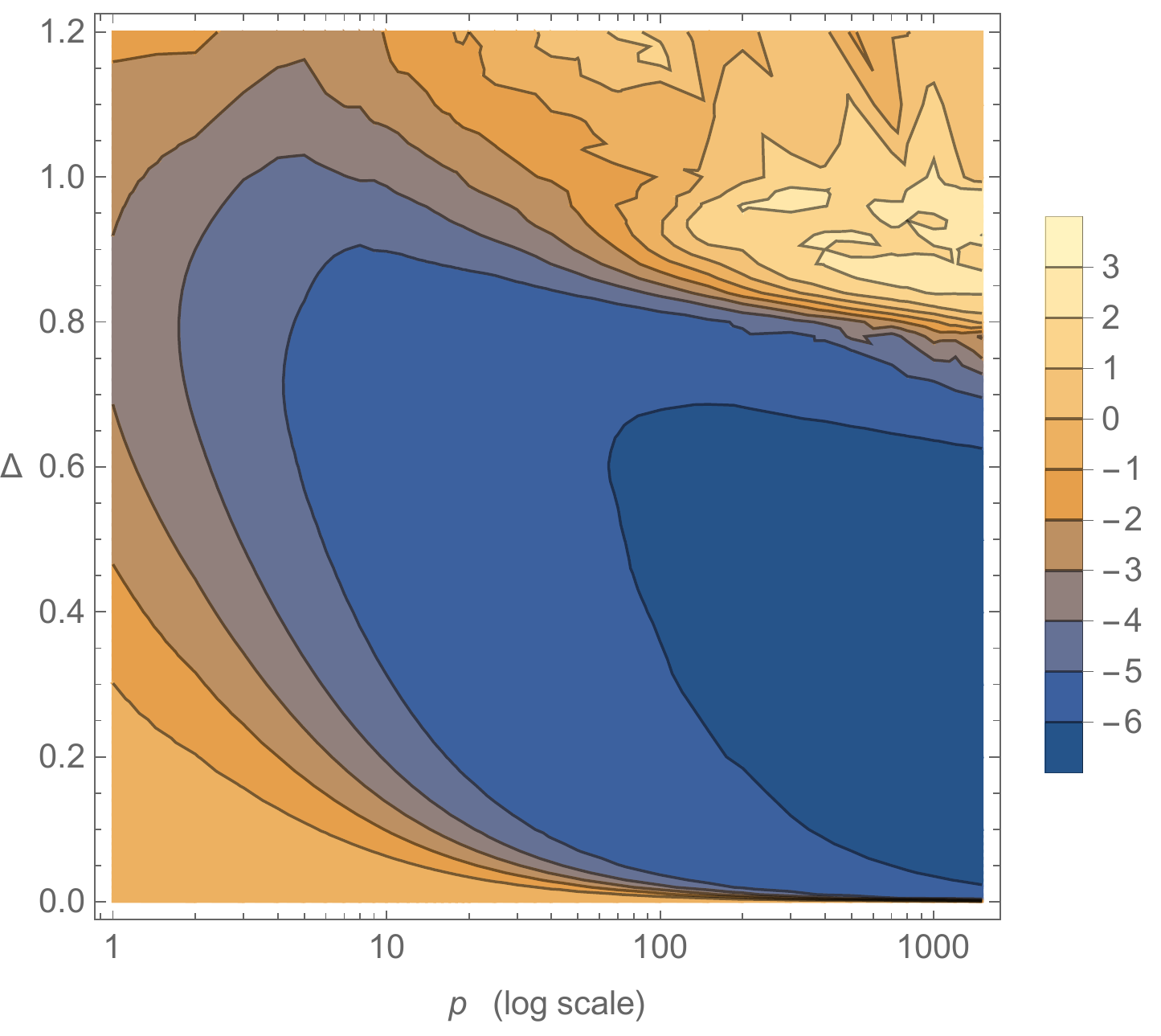}
    \caption{Expected cost after completing QAOA as a function of QAOA parameters $\Delta$ and $p$, for a 6-variable fully-connected Ising spin system instance taken from~\cite[Fig.~2]{kremenetski21diagram}. The ground state has expected cost $-6.13$. The initial uniform superposition has expected cost $\left<C\right>=0$. %\sh{[SH: where can I/the reader find the instance cost Ham?]}
    }
    \label{f:Ising_phase_diagram} 
\end{figure}

\Fig{Ising_phase_diagram} is the QAOA performance diagram of the Ising instance shown in~\cite[Fig.~2]{kremenetski21diagram}, but showing the expected cost instead of the squared overlap with the ground state. The diagram extends to $p \approx 1000$ to show the drop in performance above a critical value of $\Delta$, as seen in the upper-right portion of the figure. 

\begin{figure}[htb]
    \includegraphics[width=0.45\textwidth]{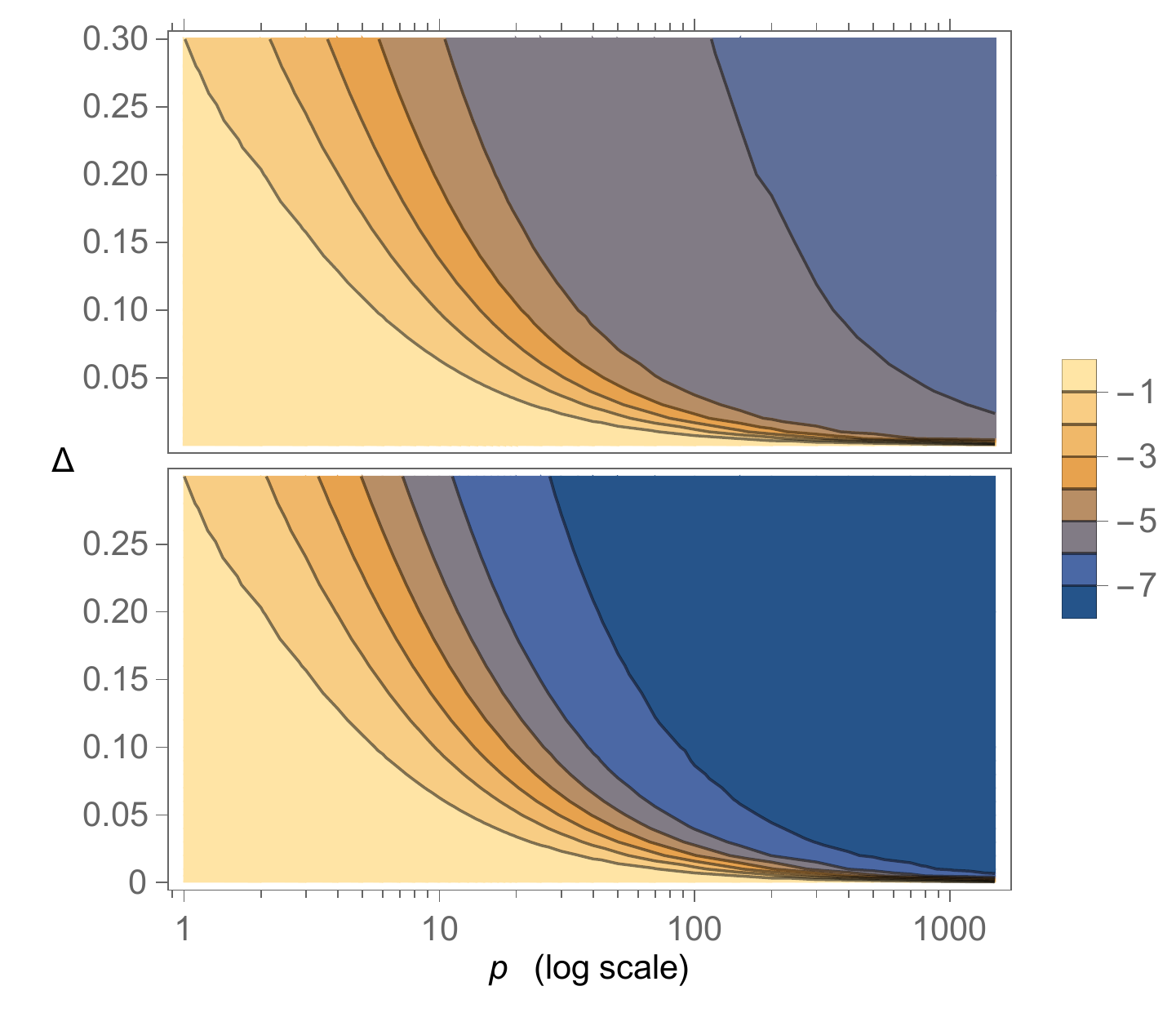}
    \caption{Actual (top) and predicted (bottom) expected cost for the Ising spin system shown in \Fig{Ising_phase_diagram}. }
    \label{f:Ising_phase_diagram and theory}
\end{figure}

\Fig{Ising_phase_diagram and theory} compares the expected cost of QAOA with the Pade rational function approximation based on the two-term expansion for the expected cost~\cite[Thm 4.2.1]{hadfield2021analytical}.

The range of $\Delta$ in this figure corresponds to the lower portion of \Fig{Ising_phase_diagram}.
Hence the the small-angle expansion of~\cite{hadfield2021analytical} 
%\sh{[SH: please make references here more precise i.e. refs to eqns in appendix? Or maybe give main eqn here and say details in appendix?]} 
matches the qualitative behavior of the lower portion of the performance diagram, namely high performance (i.e., small expected cost) occurs at smaller $p$ as $\Delta$ increases.
This two-term expansion somewhat underestimates the expected cost. 
%\sh{but.... [SH: what is takeaway here?]}.
Appendix~\ref{s:framework} provides additional details of these results.
%We provide additional details regarding the application of our results to Ising problems in Appendix~\ref{s:framework}.

\subsection{QAOA Performance Diagram for H\textsubscript{2} }
\label{s:h2_example}

%\sh{[SH: why is this section here? Suggest move elsewhere, maybe move Sec 5.4 and 5.5 to own section]}

\begin{figure}[htb]
    \centering
    \includegraphics[width=0.45\textwidth]{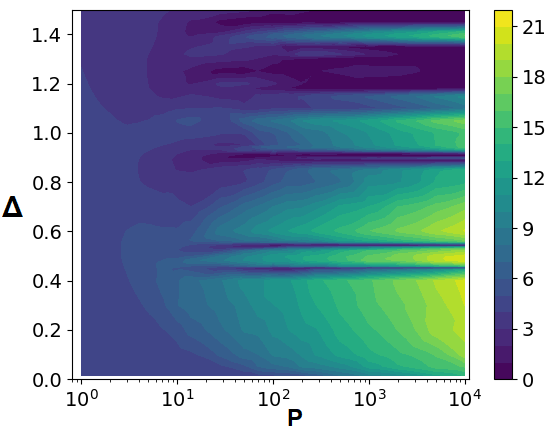}
    \caption{$-\ln|1-(\text{final squared overlap})|$ as a function of $\Delta$ and $p$.}
    \label{f:h2_plot}
\end{figure}

This section applies our analysis of QAOA to identifying the ground state of a hydrogen molecule. In this case, \Hcost\ is the full Hamiltonian prepared in the basis of Slater determinants using the full cc-pVTZ basis generated using Psi4~\cite{psi4}, while the mixer is simply the diagonal part of the cost Hamiltonian, as in~\cite{kremenetski21diagram}. Thus both the mixer and the initial state differ significantly from the common case of the $X$ mixer and uniform initial state.
Fig.~\ref{f:h2_plot} shows the resulting QAOA performance diagram.

In this case, $\DeltaCrit = 2\pi/13.9077 \approx 0.4518$ (where 13.9077 is the spectral range of this problem) due to wrap around of the eigenvalues of the cost ground and highest excited states.
Below \DeltaCrit\ is the behavior predicted in Section~\ref{s:framework_application}, with performance at fixed $p$ improving as $\Delta$ increases. The performance improvement is quite drastic in some cases; for instance, achieving $99.9\%$ squared overlap at $\Delta=0.05$ requires $p=70$, while at $\Delta=0.4$ it is only $p=10$.
%, approaching NISQ limits.  -- we don't discuss what range of p is `NISO limit', better to leave that for Discussion of possible benefit of the Ridge region

The wrapping cost ground and maximal excited states are directly coupled by the mixer, so the redirection of the ground state to the excited state is noticeable at small $p$. Since the expected energies of the cost wrapping states are similar for \Hmixer, Eq.~\ref{e:P_p}~does not directly apply. However, we expect an exponential dependence on $p$ and verify this by using Eq.~\ref{e:lz_p_predictor} to predict the value of $p$ at which the performance hits $99\%$ squared overlap, given an overlap of $0.1581$ at $p=5000$. This estimates a $99\%$ overlap at $p\approx 27$, which is close to the true value of $p=25$.

Slightly increasing $\Delta$ restores the connection between the two Hamiltonians' ground eigenstates - this is because the spectrum of the two Hamiltonians is so similar that the ground/maximal excited states of \Hmixer\ wrap around at $\Delta = 2\pi/13.8561 \approx 0.4534$ before any more cost eigenvalues wrap past each other, ``undoing'' the altered connection between the ground eigenstates of the two Hamiltonians. The cost Hamiltonian directly couples the highest/lowest eigenstates of the mixer, so this transition is clearly visible at low $p$, just as the first. Thus our topological perspective offers a more intuitive explanation for why performance deteriorates on such a narrow band of $\Delta$ than the usual angle size divergence analysis.

Above this value of $\Delta$, due to the weak connections between the cost eigenstates produced by the mixer (and vice versa), eigenvalues of indirectly-connected states wrapping around do not result in gaps that are noticeable at this scale. Eventually, above $\Delta \approx 1$, enough significant gaps from avoided crossings due to wrap-around have emerged that performance completely deteriorates.

\section{General Schedules and Problem Size Scaling}
\label{s:generality}

This section shows how the explanations of \Section{performance_diagrams} apply to additional cases of performance diagrams.
%\sh{[SH: at a glance I worry some context is lacking... where do we actually define / motivate performance diagrams? Only in the Intro then straight to explaining them in Sec 5? Is Intro motivation/explanation sufficient? ]}

\subsection{General schedules with gradually changing angles}
\label{s:continuous schedules}

The previous two sections illustrated how the small-angle approximation, eigenstate connections for $p$ in the adiabatic limit and small eigenvalue gaps explain performance diagrams for QAOA using the linear ramp of \Eq{gamma and beta}. This section generalizes these explanations to other schedules with gradually changing angles. This is of general interest as there is strong numerical evidence that optimized QAOA protocols approach an asymptotic continuous limit with large $p$~\cite{zhou2020,pagano2020, brady2021}.%of schedule.
%\sh{[SH: this para should be more crystal clear - which of "our results" apply here, and in which way?]}

One set of results applies to schedules in the discrete adiabatic limit. Schedules with gradually changing  $\beta_j, \gamma_j$ as defined in Section~\ref{s:qaoa_overview} correspond to discrete samplings of continuous paths in parameter space.  Section~\ref{s:topological_perspective} describes how the mixer and cost eigenstates connected by continuous paths through $f,\Delta$ parameter space are determined entirely by the initial and final parameter values along that path (assuming the path does not pass through degeneracies).  Applying the discrete adiabatic theorem, Section~\ref{s:adiabatic_region} showed which cost eigenstate the QAOA state approaches for $p \gg 1$, based only on $(\beta(0),\gamma(0))$ and $(\beta(1),\gamma(1))$. Specifically, the initial eigenstate determines the cost eigenstate that QAOA produces.
% For parameter paths where $\gamma(s) + \beta(s)\neq 0 $ for all $s$, it can be easier to determine these connections through a change into the $\Delta,f$-parameter space:
% \begin{equation}
%     \Delta(s) = \beta(s)+\gamma(s),\ \ f(s) = \frac{\gamma(s)}{\beta(s)+\gamma(s)}
% \end{equation}
% \tad{[In the next sentence, does `generality' specifically mean the change from $\gamma,\beta$ to $\Delta,f$ just mentioned? Otherwise, what justification to we have for the above claim that ``it can be easier to determine connenctions'' by making this change of variables?]}
For small angles, QAOA can be regarded as a first-order Trotter approximation to annealing~\cite{kocia2022}. One application of this generality of our approach is to analyze convergence of this approximation. In particular, fixed time-step, first-order Trotterizations of a continuous anneal correspond to fixed-$\Delta$ paths in our parameter space. Therefore, for a fixed choice of gapped, bounded Hamiltonians and continuous schedule sampled from to produce gradually changing angles, there exists a time step corresponding to \DeltaCrit~above which the product sequence of unitaries is guaranteed to fail to approximate continuous annealing. 

%\tad{By ``polynomial in system size'', do we mean to say our results also apply when $p$ is not large enough to reach the adiabatic limit? If so, I think a clearer contrast with the previous paragraph is to say that rather than ``polynomial in system size''??}
Our results also apply to schedules with gradually changing angles where the difference between consecutive angles is not small enough to produce behavior close to that of the discrete adiabatic limit. Section~\ref{s:performance_ridge} explains whether QAOA along a continuous path through parameter space follows a particular eigenvector swap, based on the gaps encountered along the path. In particular, to find the effects of gaps from wrap-arounds on nonlinear paths, it is enough to alter the calculus relative to $\Delta,f$ in Appendices~\ref{s:ptrack_derivation} and~\ref{s:discrete_landau_zener_derivation} to $\beta(s),\gamma(s)$, allowing an application of our analysis to nonlinear paths (since the location and emergence of gaps can be easily translated from the $\Delta,f$ parameterization in this paper). 
%\tad{[Is this note a separate point, or a reason why our analysis applies to nonlinear paths?]} 
The discrete Landau-Zener approach is helpful in analyzing diabatic transitions at avoided crossings near isolated degeneracies in parameter space, not just the ones at the edges. For avoided crossings in this setting, the formula is easier to use than the general formula of the Discrete Adiabatic Theorem in~\cite[Theorem 3]{costa2021}. %\tad{What does ``the formula'' refer to? A result in an Appendix modified as we indicate for nonlinear paths? I.e., can we use ``the resulting modified formula'' or ``the modified version of Eq.~XX'' to be more specific?}

Unlike the discrete adiabatic theorem and Landau-Zener formula, the small-angle framework applied in Section~\ref{s:framework_application} does not depend on the continuity of the parameter schedule and in some cases can be applied to small $p$. Thus the small-angle analysis, which applies to the lower portion of performance diagrams, generalizes beyond schedules with gradually changing angles. 
%Consequently, this analysis can be applied to construct performance diagrams with a valid lower portion for more general choices of schedule. 

\subsection{Sparse Local Constraint Satisfaction Problems: Performance Diagram and Problem Size}
\label{s:scaling}
%\tad{careful with ordering of limits $p \rightarrow \infty$, $n \rightarrow \infty$, $\Delta \rightarrow 0$ and the size of $p \Delta$: small as required by QAOA framework small-angle expansion, large as required for adiabatic limit.}

%\sh{[SH: Notation below is not rigorous which prevents me from assessing carefully]}

%\sh{[SH: I suggest trimming the section significantly, highlighting the main / most solid ideas, making more clear what is speculation or topic of future research]}

The performance diagrams and discussion of their properties summarized in \Section{section5_summary} focus on QAOA performance as a function of $p$ and $\Delta$ for a fixed problem. This section discusses how the behaviors scale with problem size for the $X$-mixer and typical cases of sparse local-constraint satisfaction problems, i.e., problems where each constraint deals with at most some constant number of variables and each variable appears in at most a constant number of constraints.

\begin{figure}[htb]
    \includegraphics[width=0.45\textwidth]{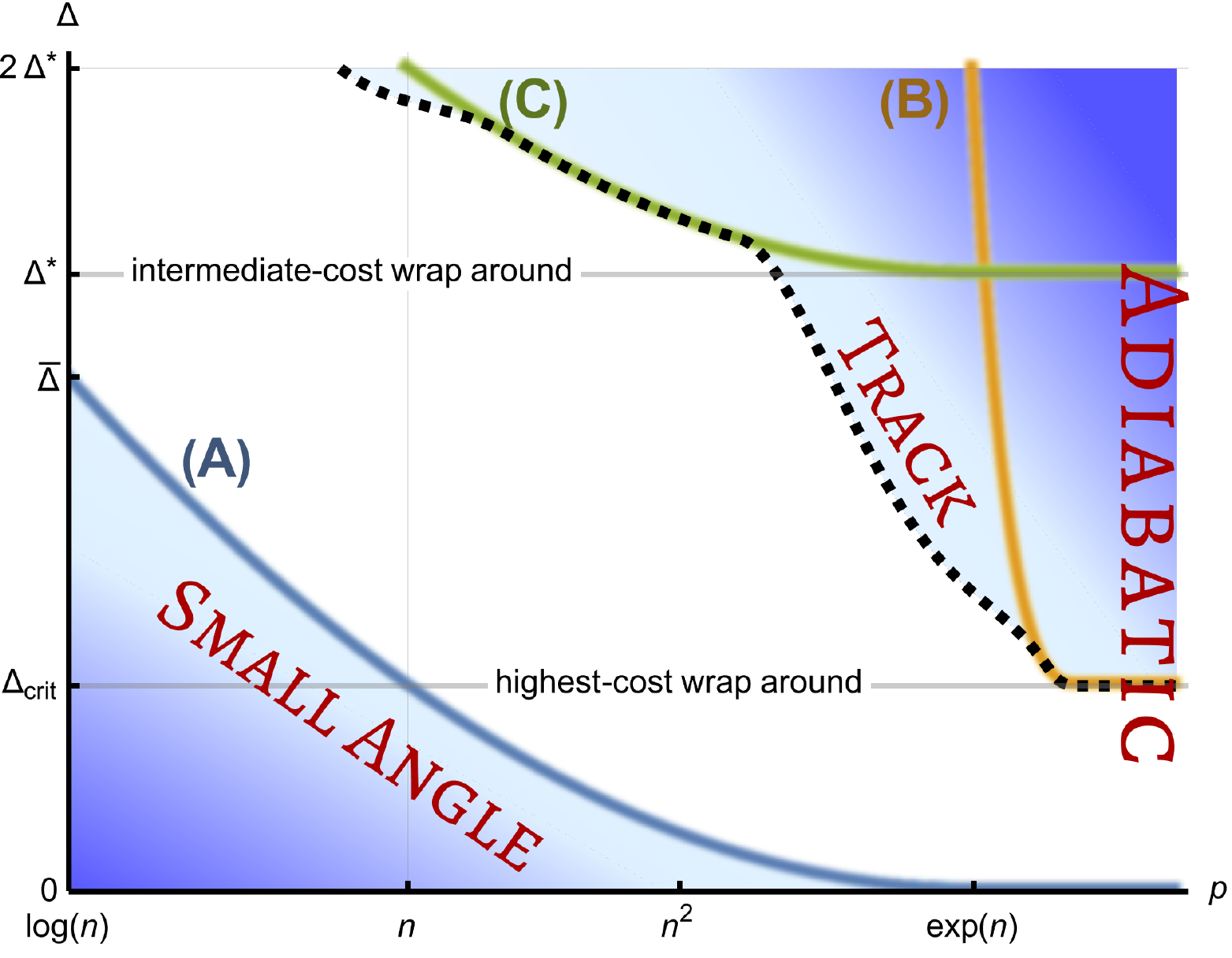}
    \caption{Schematic scaling of the QAOA performance diagram with problem size $n$. Curve (A) indicates the fuzzy, left boundary of the performance ridge from small angle analysis. Curves (B) and (C) mark the values of $p$ at which a fixed diabatic transition to the ground state is predicted solely due wrap-arounds at $\DeltaCrit$\ and $\Delta^*$ respectively. Here, $\DeltaCrit \propto 1/n$, $\bar{\Delta} \propto 1/\log(n)$ and $\Delta^* \propto \log(\log(n))/\log(n)$. These curves, along with those from all the wrap-arounds not shown in this range of $\Delta$, define an envelope contour (dashed) that forms the right, fuzzy boundary of the \region{Ridge} region. Curve (C) indicates where this boundary exists for $p$ polynomial (here, quadratic) in $n$. 
    }
    \label{f:performance diagram scaling}
\end{figure}

%\tad{For consistency with usual notation for CSPs, use `cost' instead of `energy' to describe values of Hamiltonians.}

Consider applying the $X$ mixer to a problem with size $n$ whose range of costs is $\Theta(n)$, e.g., as is typical of random hard problems such as $k$-SAT where the number of clauses, and hence maximum possible cost, is proportional to $n$~\cite{kirkpatrick94}. 
The largest eigenvalue of the $X$ mixer is also of this order in $n$.
Thus the first wrap around occurs at $\Delta \propto 1/n$.
Without loss  of generality, for this discussion we suppose that the cost Hamiltonian has a larger range of eigenvalues than the mixer, so this first wrap around is due to that of the highest cost.

%\sh{[Sh: I have trouble following where all the ln()s come from...]}
%\tad{Would it help to define a notation for those expressions so we explain it once by reference to appendix E and don't  have to repeat them multiple times (That also avoids the challenge for readers of repeats: they have to check each time whether they are the same or different combination of log's than seen previously)?}

%\tad{Vlad: check consistency between scaling diagram and text on use of $\Theta(..)$ vs $O(..)$. And between generic $\log$ and $\ln$ -- for asymptotic size they are the same. Can we simplify any of the expressions,  e.g., $O(\ln(n)/\ln(\ln(n)))^{O(n)}$ with notation or otherwise get a single $O$ (or $\Theta$) expression rather than nested?}

For such problems, \Fig{performance diagram scaling} illustrates a possible scaling for the \region{Ridge} region. %, extending somewhat above the first wrap around. 
%[SH: I have trouble following where the $p\sim n$ came from. See a mention below but hard to tell how strongly we are asserting it here..]
For large $p$, \Appendix{framework_scaling} shows that $\Delta \propto 1/p$ has constant QAOA performance, corresponding to the downward-sloping behavior of the bottom part of the \region{Ridge} region in the performance diagrams. 
These scaling arguments suggest the left-side of the ridge crosses $\DeltaCrit \propto 1/n$ with $p \propto n$, which arises from minimizing the expected cost from the first two terms of \cite[Theorem 4.2.1]{hadfield2021analytical} applied to random Ising problems as a function of $n$. $\bar{\Delta}$ indicates where this estimated lower bound intersects with the vertical axis at $p\propto \log(n)$. 

Above $\DeltaCrit$, for sufficiently large $p$ one reaches the adiabatic regime which has poor performance due to going to the wrong eigenstate, and so is outside the \region{Ridge} region.

For local constraint satisfaction problems, $b$ bit flips correspond to an $O(b)$ change in cost. In particular, the highest cost state, of order $n$, has $\Theta(n)$ bit flips compared to the ground state. Thus the coupling distance, defined in Section~\ref{s:nature_of_degeneracies}, between the highest cost and ground states is $k=\Theta(n)$. Consequently, when the highest cost states wrap around, from \Section{performance_ridge}, the $p$ at which significant deterioration occurs scales as $(1/\delta)^{\Theta(n)}$, and thus will not be noticeable save for exponentially large $p$ even at fixed values of $\Delta$ significantly smaller than 1.

%Wrap-arounds whose $p$ for substantial deterioration (e.g., less than $0.5$ squared overlap on ground state) at fixed $\delta = \Delta - \Delta^*$ scales  polynomially with $n$ correspond to states that are $k=O(\ln(n)/\ln(\ln(n)))$ bit flips away from the cost ground state \sh{[SH: why? I'm a bit lost here]}. For local constraint problems considered here, such states typically have cost $O(\ln(n)/\ln(\ln(n))$ above the ground state, so their wrap-arounds take place at $\Delta^* = 1/O(\ln(n)/\ln(\ln(n)))$. The same is true for wrap-arounds at $f=0$ with eigenvalues from the $X$-mixer, whose eigenstates also satisfy the property of $O(1)$ change in cost if the cost Hamiltonian couples them through $k=O(1)$.

%\tad{The previous paragraph uses $1/log(n)$, the next uses our revised $1/(log(n)/log(log(n))$. Delete the previous paragraph unless anything to merge with the next paragraph??}

In Appendix~\ref{s:derivation of performance scaling}, using the discrete Landau-Zener approximation from Section~\ref{s:landau_zener} and the lowest-order approximation of the gaps from Section~\ref{s:nature_of_degeneracies}, we estimate that when $p$ is some fixed polynomial function of $n$, the gaps that produce some fixed squared overlap $<1$ with the cost ground state first occur for $\Delta$ just above $\Delta^*\propto 1/w(n)$, where 
\begin{equation}
    w(n) := \frac{\log(n)}{\log(\log(n))}~.
\label{e:bound_to_ridge}
\end{equation}
In particular, for large $n$, we estimate that the wrap-around of states differing in $\propto w(n)$ bits from the ground state produces the gaps that predominantly contribute to an upper boundary to the right edge of the \region{Ridge} region, compared to all other wrap-arounds at $f=0,1$, for $p$ polynomial in $n$. Since even one such wrap-around is enough to produce this effect at $f=0$ or $f=1$ (see Appendix~\ref{s:derivation of performance scaling}), it does not matter that many excited states differing by $\propto w(n)$ bit flips might have a slower-scaling energy difference. 

According to this analysis, the left and right boundaries of the \region{Ridge} region do not intersect when $n$ is large.
%This $\Delta^*$ curve does not intersect with the $\Delta \propto 1/n$ in the case of large $n$. 
The wrap-around with low-lying excited states with $O(1)$ difference in cost from the ground state does not contribute to the right boundary, as they occur at higher values of $\Delta$. As we estimate in Appendix~\ref{s:derivation of performance scaling}, the wrap-around between cost or mixer states with $\sim w(n)$ coupling distance dominates in size over gaps occurring at smaller $\Delta$ with greater coupling distance. %\tad{What are ``previous gaps''? Presumably those occurring at smaller $\Delta$??}

This analysis suggests the general scaling of the lower-left and upper-right fuzzy boundaries of the \region{Ridge} region.
In particular, it suggests that for the class of problems used for this example, scaling the performance diagram axes by $n$ horizontally and $\propto w(n)$ vertically will result in similar diagrams for different $n$,
similar to overlapping behaviors seen with finite-size scaling in the context of phase transitions in combinatorial search such as satisfiability~\cite{kirkpatrick94}. Importantly, the estimation from our analysis approximates an \emph{upper boundary} to the \region{Ridge} region from just one effect (wrap-around at $f=0,1$) - other effects might need to be taken into account to predict the scaling of its actual location.

\section{Discussion}
\label{s:discussion}

%\tad{This paragraph is a ``related work''. Is there some aspect of future work to mention in this context? Otherwise it seems out of place in a Future Work section. One fix is to rename this section `Discussion', which covers both related work and future work.}
The deterioration of QAOA performance at large angles was observed in~\cite{wurtz_and_love_counteradiabatic} and attributed to a large Trotter error in comparison to the (continuous) adiabatic limit. We identify a more complex phenomenon: the poor performance is not necessarily due to adiabaticity breaking down, but rather due to the change in connectivity between initial and final eigenstates. Our application of the discrete adiabatic theorem to schedules with gradually changing angles shows QAOA convergence~\cite{binkowski2023} and supplements analysis of digitized continuous drives~\cite{kocia2022,hatomura2023}.

Our discussion indicates the \emph{location}, i.e., ranges of $p$ and $\Delta$, of the \region{Ridge} region. An open question is the \emph{absolute performance} of QAOA on the ridge. Numerical observations of the performance diagrams indicate performance on the ridge just above \DeltaCrit\ is comparable to that for small $\Delta$ and much larger $p$, but it remains to be seen how this behavior scales with larger problem sizes.

\begin{figure}
    \centering
    \includegraphics[width=0.45\textwidth]{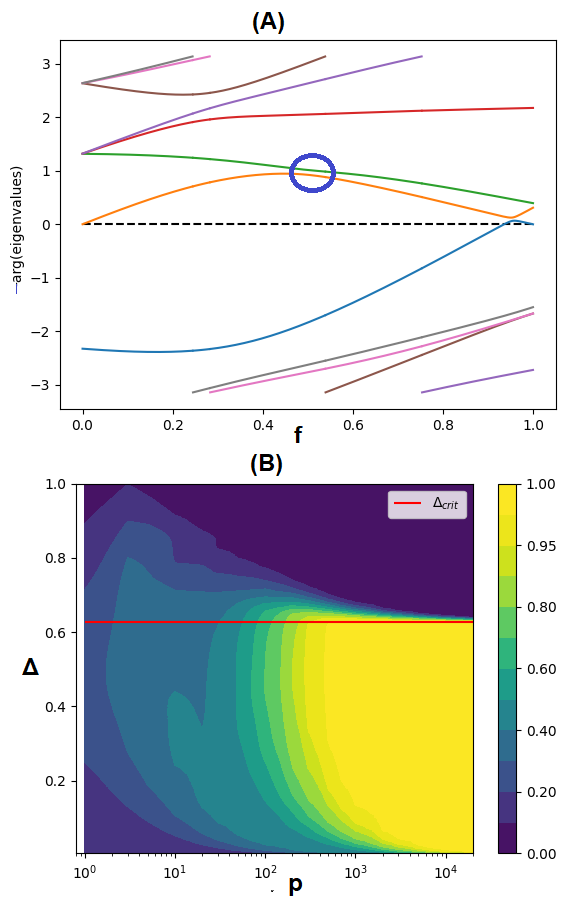}
    \caption{(A) Negative arguments of \UQAOA\ eigenvalues, with the $X$-mixer and cost energies (in order of bitstring value) 10, 5.4, 7.18,  0.6, 0, 7, 7, 3.35 at $\Delta=0.66>\DeltaCrit=2\pi/10$ which has a small gap with a low-lying energy state circled in blue. (B) Performance diagram for choice of mixer and cost in (A) displayed as squared overlap with cost ground state.}
    \label{f:small_gap_and_performance}
\end{figure}

In particular, there may be small gaps among low-energy states that occur at values of $f$ other than the location of the gap in the highest-energy wrap around just above \DeltaCrit. Then, the absolute performance in the \region{Ridge} region might be low, as the state vector fails to go to the ground state eigenvector at small $p$ due to these other small gaps. \Fig{small_gap_and_performance} illustrates this behavior with a three-qubit example. Here, $\Delta$ is slightly above \DeltaCrit\ and so there is a small eigenvalue gap near $f=0.95$ from wrap-around which is comparable in size to the eigenvalue gap between the ground and low-lying energies. Consequently, absolute performance on the QAOA ridge is limited by the gap from wrap-around for problems whose overall minimum gap is at $f=1$ (prior numerical investigation has found this to commonly be the case for chemistry problems~\cite{adiabatic_kremenetski2021} as seen in Section~\ref{s:h2_example}) and for problems where the gap with low-lying eigenstates is not as small as the gap formed from wrap-around (generically, this is true for $\Delta$ sufficiently close to \DeltaCrit). Therefore, another important follow-up is determining the typical performance on the ridge for different classes of problems. Further, the qualitative argument used to identify the location of the left \region{Ridge} region in the performance diagram is based on the an extrapolation of \cite[Theorem 4.2.1]{hadfield2021analytical} to angle sizes where higher-order terms are not necessarily small. A direction for future work here is to numerically test the applicability of this analysis to various classes of problems.

Another promising direction for future work is utilizing our approach to aid the design of schedules for low-depth circuits, which is of interest because noise in the implementation of QAOA on hardware can significantly degrade performance as the depth of the circuit grows~\cite{marshall2020,xue2019}. One could use the topological perspective to construct nonlinear schedules that exploit the structure of multiple wrap-arounds, i.e., large steps to avoid tracking swapping eigenvectors through gaps and small steps to follow eigenvectors through gaps, while using the discrete Landau-Zener approach (potentially applying discretized versions of current multi-state models \cite{sinitsyn2013,sinitsyn2014}) to estimate parameter regions that have low or high diabatic transitions to the ground state at a particular depth. This could potentially allow for shortcuts to adiabatic behavior~\cite{campbell2017, gueryodelin2019, hegade2021}. Even simpler design %variations 
improvements are possible - as argued and observed numerically in our work, it is often possible to reduce the circuit depth $p$, while increasing the angle size $\Delta$, with only marginal sacrifice in QAOA performance.  

Another observation from our work is that judiciously picking the start and endpoints in parameter space provides an alternative means of state preparation. Starting with one cost eigenstate above an isolated degeneracy at $f=1$, then taking an open half-tour around the degeneracy leads to a different eigenstate. In the cases where the highest cost excited state is easy to identify and prepare (as is the case for many Ising instances~\cite{ising_encodings}), this leads to a version of QAOA where the mixer is not required to have an easy-to-prepare ground state and can be optimized to reflect structure of the cost problem. Exploiting such structure might lead to advancements~\cite{bennett1997}, but whether such a protocol can improve on traditional approaches requires further investigation. 

%Finally, as noted in \Section{continuous schedules}, the small-angle framework of~\cite{hadfield2021analytical} does not require schedules with gradually changing angles whereas the wrap-around analysis does. Thus performance diagrams may be qualitatively different above \DeltaCrit\ if the angles vary substantially across $j$. \sout{, e.g., substantially different angles for even and odd QAOA steps.}

%\section{Conclusion}
\paragraph*{Conclusion.}
%In this work, 
%In conclusion, 
In this work we have presented a new lens of analysis for QAOA circuits with gradually varying unitaries. Using the discrete adiabatic theorem, we recover the usual expected behavior at sufficiently small angles corresponding to Trotterized annealing, as well as capture the fundamentally different 
behavior leading to novel phenomena due to the wrap-around of unitary eigenvalues for larger angles. The latter results in changing connections between cost and mixer Hamiltonian eigenstates and diabatic transitions at small gaps due to these wrap-arounds. We have used this and other techniques such as perturbation theory to explain a surprising qualitative generality in QAOA performance diagrams from optimization to chemistry problems. %and various schedules. 
Our findings represent a hitherto unexplored perspective on QAOA which carries implications for parameter schedule design, limitations of the annealing perspective on QAOA, and performance characterization. %While fully characterizing the relationship between QAOA performance, circuit depth, and problem size appears challenging, our work shows encouraging steps in this important direction.

\section*{Acknowledgments}
  %\sout{N.M.T., V.K.} 
  We are grateful for support from NASA Ames Research Center.    We acknowledge funding from the NASA ARMD Transformational Tools and Technology (TTT) Project. 
    %\sout{V.K. is thankful for support from NASA Academic Mission Services, Contract No. NNA16BD14C.} 
    %
Part of this work is funded by U.S. Department of Energy, Office of Science, National Quantum Information Science Research Centers, Co-Design Center for Quantum Advantage under Contract No. DE-SC0012704. 
  VK, TH, and SH 
  were supported by the NASA Academic Mission Services, Contract No. NNA16BD14C. AA is supported by Yoichiro Nambu Graduate Fellowship courtesy of Department of Physics, University of Chicago. 
  Calculations were performed as part of the XSEDE computational Project No. TG-MCA93S030 on  Bridges-2 at the Pittsburgh supercomputer center. We also would like to thank Lucas Brady, Carlos Mejuto Zaera, Ruslan Shaydulin, Thomas Watts, Haoran Lu, Alexander Avdoshkin, and Andrew Yates for providing feedback on earlier drafts of this paper.

\bibliography{refs}

% \newcommand{\changelocaltocdepth}[1]{%
%   \addtocontents{toc}{\protect\setcounter{tocdepth}{#1}}%
%   \setcounter{tocdepth}{#1}  }
% \changelocaltocdepth{1} %this command to stop TOC from showing subsections for Appendices
% %
% % however not working for some reason!?
%No longer relevant, TOC cut.

% %%%%%%%%%%
% % Supplementary material
% %%%%%%%%%% Merge with supplemental materials %%%%%%%%%%

\pagebreak
\widetext
\begin{center}
\textbf{\large Appendix}  
% for journal submission we may convert to separate Supplement document, with its own set of cross references and care to distinguish cites to main text vs the supplement

%\tad{Citation numbers in this section start with S; is that so this supplement will be a separate paper with its own references, separate from those of the main paper? Or is this an appendix, to appear after the main text and share references with the main text?}\\
%\sh{SH: Why was Appendix changed to SM? Normally only do this if for specific journals..}
%\anuj{i fixed the references issue and in the future this can be easily changed by modifying the renew commands in the tex file. Vlad and I switched it to supplementary material with single column formatting since at some point Norm mentioned that we might want to submit to PRX/PRX Quantum. } \sh{SH: fair enough, maybe we should try to decide on an initial choice today and then make consistent throughout...}

\end{center}

\onecolumngrid

\appendix

% temporarily change subsection numbering
% copied from change made at start of the doc for numbering in main text; that fix is reset by \appendix so repeat it here 
% only fixing subsections to retain the change of sections to letters
% from https://tex.stackexchange.com/questions/87131/numbering-sections
\renewcommand{\thesubsection}{\thesection.\arabic{subsection}}
\renewcommand{\thesubsubsection}{\thesubsection.\arabic{subsubsection}}

%SH: adding this so that appendix subsection not displayed in TOC
%**Not working, fix or cut TOC
%\setcounter{tocdepth}{-1}
%\newcommand{\changelocaltocdepth}[1]{%
%  \addtocontents{toc}{\protect\setcounter{tocdepth}{#1}}%
%  \setcounter{tocdepth}{#1}%
%}
%\changelocaltocdepth{-10}

\section{Application of QAOA small-angle analysis}\label{s:framework}

The results for QAOA in the small-parameter regime of \cite[Theorem 4.2.1]{hadfield2021analytical} describe behavior at the bottom of the performance diagram. This theorem expresses the expected cost of the final QAOA state as a series in polynomials of the QAOA angles, with coefficients determined by %the cost operators 
initial state expectation values of operators derived from the cost function via commutators~\cite{hadfield2021analytical}. %the theorem. %\sh{The zeroth order term is the initial %state cost expectation value.}
For %the cost expectation of 
level-$p$ QAOA, to leading-order
\begin{equation}
    \langle \Hcost \rangle_p =  \langle \Hcost \rangle_0 \,-\, ( 
    %\sum_{i,j>i} 
    \sum_{i,j\geq i} \gamma_i \beta_j  )  \, \langle \, [\Hcost, [\Hmixer,\Hcost]] \, \rangle_0  + \dots~,
\end{equation}
where the terms not shown to the right are proportional to increasingly higher order products of the $\gamma_i,\beta_j$, with the next nonzero contribution proportional to a quartic polynomial of the QAOA angles.

\subsection{QAOA schedules} \label{s:framework_schedules}

For the ramp schedules of \Eq{gamma and beta}, the %angle sums 
parameter polynomials of \cite[Theorem 4.2.1]{hadfield2021analytical} are functions of $p$ and $\Delta$.  
The angle sum appearing in the 
leading order term is 
\begin{equation}
    \sum_{i,j\geq i} \gamma_i \beta_j = \frac{p(6+5p+p^2)}{24(1+p)} \Delta^2~.
    \label{e:framework angle sum for ramp}
\end{equation}

The small-angle analysis applies to arbitrary parameter schedules. In particular, \cite[Remark 4.2.1]{hadfield2021analytical} indicates the angle sum terms scale as powers of $\Delta p$ in the same way as the ramp when $\Delta$ is interpreted as a characteristic magnitude of the angles. 
Thus the dependence on $p$ and $\Delta$ in the theorem does not depend on specific details of the QAOA schedule, thereby indicating why the observed QAOA performance diagram behavior is generic in the regimes of validity of \cite[Theorem 4.2.1]{hadfield2021analytical}, in particular when $p \Delta \ll 1$ (for fixed problem size).

\subsection{Ising problem} \label{s:framework_ising}

Finding the ground state of a Hamiltonian is a minimization problem, so we use $-X$ as the mixer, so the initial uniform superposition is the ground state of the mixer. This contrasts with the use of the $+X$ mixer in \cite{hadfield2021analytical}. This change of sign of mixer is equivalent to changing the sign of $\beta$ instead. In applying this theorem to finding the ground state of the Ising problem, we make this sign change in the angle sums.

For the Ising Hamiltonian of \Eq{Ising Hamiltonian}, the %first term 
leading-order correction to $\langle \Hcost\rangle_p-\langle \Hcost\rangle_0$ %\tad{\Hcost\ instead of $H_C$?}
of~\cite[Theorem 4.2.1]{hadfield2021analytical} is 
\begin{equation}\label{e:small-angle first term}
    -\frac{p(6+5p+p^2)}{6(1+p)} \Delta^2 
    \left( \sum_i h_i^2 + 2 \sum_{i<j} J_{i,j}^2 \right)~.
\end{equation}
The random Ising problems considered here have $\left<h_i^2\right>=\left<J_{i,j}^2\right>=1/3$, so averaging over this problem class gives 
%The average of the first term over this class of problems is
\begin{equation}
    -\frac{\Delta ^2 n^2 p \left(p^2+5 p+6\right)}{18 (p+1)}~,
\end{equation}
which equals $-n^2 (\Delta p)^2/18$ for large $p$,.
A similar evaluation of the angle sums and costs involved in the second term of \cite[Theorem 4.2.1]{hadfield2021analytical} for large $p$ gives $n^3 (\Delta p)^4/504$.

These expressions allow generalizing the range of validity of this series expansion beyond $\Delta \ll 1$ for fixed $p$ and a given problem (hence fixed $n$) discussed in \cite{hadfield2021analytical}.
Specifically, the second term %\tad{and presumably all higher terms??} 
is small compared to the first when $\Delta \ll 1/(p \sqrt n)$.
%, indicated schematically as the shaded region in \Fig{performance diagram scaling}.
%\tad{Revise this text if we remove the shaded region from \Fig{performance diagram scaling}.}

The first term is negative, indicating $\left<C\right>$ decreases, i.e., QAOA performance improves, when moving from the lower left toward the upper right of the diagram, until the second, and higher, terms become important for $\Delta \sim 1/(p \sqrt n)$ and lead to increasing $\left<C\right>$. This suggests the \region{Ridge} region in the performance diagram occurs for $\Delta \sim 1/(p \sqrt n)$ or larger.
On the other hand, $\Delta p \rightarrow \infty$ gives the adiabatic limit, which gives low performance above \DeltaCrit.

%\tad{This scaling argument is for average behavior of the class of random Ising problems discussed here, i.e., $h_i$ and $J_{i,j}$ uniformly distributed over a range of values independent of $n$. Can we relate this to per-instance behavior, i.e., the extent to which Ising instances can perform much better or worse than the average for this class?}

A similar analysis applies to sparsely connected Ising problems that correspond to local CSPs discussed in \Section{scaling}. One such class of problems is when each pair of spins is independently connected with probability $q=m/(n-1)$ with $m$ a fixed constant, independent of $n$. Thus, on average, a spin is connected to $(n-1)q=m$ other spins. The couplings $J_{i,j}$ for connected spins are randomly selected in the same way as for the fully connected case described in \Section{Ising_example}.

For the small-angle analysis, the sparse Ising problem only changes the expectation values of the coupling appearing in the cost operators, e.g., in \Eq{small-angle first term}. In particular, $\left<J_{i,j}^2\right>=q/3$ and, for large $p$, the leading-order term is $-n(m+1) (\Delta p)^2/18$.
Similarly, the second term of \cite[Theorem 4.2.1]{hadfield2021analytical} is $n(5m^2+28m+8) (\Delta p)^4/2520$. 
Both these terms are linear in $n$, and the second term is small compared to the first when $\Delta \ll 1/p$. This indicates a lower bound for the \region{ridge} region of these sparsely connected problems occurs with $\Delta \sim 1/p$.

%\tad{This new discussion of sparsely connected Ising is an analysis of a type of local-constraint satisfaction problem using the $X$ mixer (since the X mixer is the case treated in the small-angle theory). So this topic could instead be a subsection of Appendix E; but seems more natural as part of Appendix A on small-angle analysis. Could we change the title of Appendix E and/or add a topic sentence at the start of that appendix so it's clear that Appendix E is restricted to aspects of local CSPs distinct from small-angle analysis?}

\subsection{Scaling from small-angle analysis} \label{s:framework_scaling}

Any problem whose cost operator expectations and maximum energy depending on $n$ in the same way as the Ising example will have the same scaling for the location of the \region{Ridge} region in terms of $n$ as the Ising example.
% lemma A.0.2 in original version of framework paper changed to B.0.2 in Sept 2022 revision
This includes any QUBO whose cost coefficients are independent of $n$~\cite[Lemma B.0.2]{hadfield2021analytical}.

Moreover, the angle sums in the series expansion of~\cite[Theorem 4.2.1]{hadfield2021analytical} are proportional to powers of $\Delta p$ for large $p$ (e.g., the angle sum given in \Appendix{framework_schedules}). 
For higher order terms to be small requires $\Delta p \ll 1$, rather than just $\Delta \ll 1$ used in \cite{hadfield2021analytical} in the context of fixed $p$. Thus, for large $p$, $\Delta \propto 1/p$ has constant QAOA performance, corresponding to the downward-sloping behavior of the \region{Ridge} region for small-to-intermediate $p$ in the performance diagrams. 

The schematic scaling shown in \Fig{performance diagram scaling} is given by combining these observations for sparsely connected problems, and extrapolating the small-angle series from $\Delta$ so small that QAOA operators are close to the identity so the QAOA output state is close to the initial state. This emphasizes the qualitative behavior for moderately large $p$.

For such problems, as discussed in \Section{scaling}, \DeltaCrit\ is of order $1/n$. This suggests the \region{Ridge} region occurs when $1/n$ is at least of order $1/p$, i.e., $p$ is at least of order $n$.

%\tad{Possibly relate this scaling to observations that $p \sim \log n$ is required for QAOA to sample the full problem, e.g., to distinguish a tree from a graph with long cycles.}
A consequence of this discussion of the size of the terms in \cite[Theorem 4.2.1]{hadfield2021analytical} is that $p=O(\log n)$ is not large enough for QAOA to reach the \region{Ridge} region since we have $\Delta$ at most $O(1/n)$ to avoid wrap-around so $\Delta p$ is then at most of order $\log(n)/n$, which is small compared to 1. 
% this detail is for fully-connected problems: giving the first and second terms of the expansion of order $-(\log n)^2$ and $(\log n)^4/n$, respectively. 
Thus the first term dominates, indicating performance increases with increasing $p$ beyond $\log n$ so these parameters are below the \region{Ridge} region.

For problems whose costs scale with $n$ differently than the Ising example, a similar analysis can use the corresponding scaling of expectations of cost operators appearing in \cite[Theorem 4.2.1]{hadfield2021analytical}, and the $n$ dependence of the maximum energy to determine scaling of \DeltaCrit. Using these scaling behaviors will give corresponding estimates for the location of the \region{Ridge} region.

Specifically, the small-angle analysis can indicate the scaling of cost Hamiltonians that have different connectivity among the problem variables than the fully and sparsely-connected cases discussed here. 
For instance, a local Hamiltonian, where each variable only interacts with $O(1)$ other variables instead of fully connected interactions, will have substantially fewer terms in the cost expectations leading to slower growth with $n$. For problems with pairwise interactions among variables, \cite[Lemma B.0.2]{hadfield2021analytical} gives the cost expectations in terms of sums over the interaction strengths. Similarly, \cite[Lemma 2.5.3]{hadfield2021analytical} includes the generalization to interactions among triples of variables, which is relevant for the scaling of performance diagrams for 3-SAT which have qualitatively similar features to those seen with pairwise interactions~\cite{kremenetski21diagram}. The number of nonzero terms in these sums determine their scaling with respect to problem size $n$.

\subsection{Molecular ground state problems}

Our QAOA construction for the electronic structure problem~\cite{kremenetski21diagram} is also considered in \cite[Sec. 5.3]{hadfield2021analytical}. 
As indicated there,
it is relatively straightforward to extend the result and proof of \cite[Theorem 4.2.1]{hadfield2021analytical} to the quantum chemistry setting, which results in a similar term at leading order to the Ising case. 
%
%\sh{SH: I will try to fill something in here - should be same first term as in (S3), different second part...}

\section{Results for isolated degeneracies}
\label{s:terminal_degeneracies}

Here we provide supporting results for the claims regarding isolated degeneracies of Section
~\ref{s:QAOA operator behavior}.

\subsection{Formula for gap from wrap-around of two non-degenerate eigenstates}
\label{s:gap_section}
We derive a  perturbative formula for the size of the gap resulting from wrap-around of two non-degenerate cost eigenstates at $f=1$ based on $\Delta$ and the couplings of cost eigenstates by the mixer. It may be trivially modified to account for wrap-around at $f=0$. 

\subsubsection{Formula for gap size}
We first introduce some helpful definitions. 
\begin{definition}
Two cost eigenstates $\ket{x},\ket{y}$ are said to be \emph{``coupled"} by the mixer if there exists an ordered  tuple of cost eigenstates $S = \{\ket{s_0=x},\ket{s_1},...,\ket{s_{n}}\}$ such that 
\begin{equation} P(S) = \left(\prod_{j=0}^{n-1}\bra{s_j}\Hmixer\ket{s_{j+1}} \right)\bra{s_{n}}\Hmixer\ket{y}\neq 0~.
\label{eq:mixer_path_contributions}
\end{equation}

We call such this set a \emph{``coupling path"}, and the cardinality of this set the \emph{``length"} of the path. 
\end{definition}
For example, with the $X$-mixer, states $\ket{000}$ and $\ket{111}$ have a coupling path $S=\{\ket{000}, \ket{001}, \ket{011} \}$ of length three. This is not the only coupling path between the two states, others exist of the same length (e.g., $\{\ket{000},\ket{100},\ket{101}\}$) or different length (e.g., $\{\ket{000},\ket{001},\ket{101},\ket{100},\ket{110} \}$).

\begin{definition}
Two cost eigenstates $\ket{x},\ket{y}$ are said to be \emph{``k-coupled"} by the mixer if the shortest coupling path between them is of length $k$. 
\end{definition}

For the $X$-mixer example, the states $\ket{000}$ and $\ket{111}$ are 3-coupled, because the X-mixer directly couples two bitstrings with a Hamming distance of one, so all the shortest coupling paths between these two bitstrings have a length of 3. In general, for the case of the $X$-mixer and computational ($Z$) basis, states whose bitstrings have a Hamming distance of $k$ are $k$-coupled. We now recursively define a multivariate polynomial:
\begin{equation}
    Q^n(b_1,...b_n) = \frac{1}{(n+1)!}-\sum_{m=1}^n \frac{1+ib_m}{2m!}Q^{n-m}(b_{m+1},...,b_n),\ \ \ Q^0 = 1,\ \ \ \ b_i\in \mathbb{R}~.
    \label{eq: q_polynomial_def}
\end{equation}
The first few orders of this polynomial are shown in Table~\ref{table:eq_tab}. \\
\\
\\
\\

%\tad{What is $\Delta^*$ in Eq.~\ref{eq:energy_simplification}? Presumably the wrap around $\Delta$ for the gap under consideration; but that isn't stated until later (with the gap formula). Perhaps clearer to move def of $a_j$ and $\cEff$ to just after the gap formula, e.g., using ``where $c\Eff = $ and $a_j=$'' so in context, $\Delta^*$ refers to the value mentioned with the gap formula.}

%\sh{Why not get rid of $D_x$ terms? simplifies everything, zero anyways for most reasonable mixers, would make section more readable}

%\tad{What is ``cost energies'' in the gap formula? Can we use either ``costs'' or ``energies'' -- in either case, we mean eigenvalues of \Hcost.}

\textbf{Gap Formula:}
\textit{Let $\Delta = \Delta^*$ such that $e^{-i\Delta^*E_x} = e^{-i\Delta^*E_y}$ for exactly two distinct, non-degenerate energies of \Hcost\ $E_x,E_y$ and suppose the corresponding \Hcost\ eigenstates are $k$-coupled by the mixer. Then, for a path at $\Delta = \Delta^*+\delta$, $k\geq 2$ and using Eq.~\ref{e: Dj_Ej_Gamma_definition}, the gap magnitude $|g|$ from the avoided crossing due to wrap-around is
\begin{equation}
    |g| = 2|\cEff|\left|\frac{\delta}{\Gamma-1}\right|^k + O(\delta^{k+1})~,
\label{e:main_gap_equation}
\end{equation}
if $D_x - D_y\neq E_x-E_y$\\
}

We introduce notation:
\begin{equation}
    D_{s_j} = \bra{s_j} \Hmixer\ket{s_j},\ \ \ E_{s_j} = \bra{s_j}\Hcost\ket{s_j},\ \ \ \ 
    \Gamma = \frac{D_x-D_y}{E_x-E_y}~,
\label{e: Dj_Ej_Gamma_definition}
\end{equation} 
where $\ket{s_j}$ are cost eigenstates. In practice, $\Gamma=0$ for many common mixers (zero diagonals), which simplifies our formulas. Next, we define $a_j$ for the $j$-th term in some fixed coupling path $S$ 
\begin{equation}
a^S_j = \cot\left(\frac{\Delta^*}{2}(E_{s_j} - E_x)\right)~.
\label{eq:energy_simplification}
\end{equation}

Finally, we define an ``effective'' coupling between states $\ket{x},\ket{y}$ as:
\begin{equation}
    \cEff = \sum_{S} P(S)Q^{k-1}(a_1^S,...,a_{k-1}^S)~, 
    \label{e:effective_coupling}
\end{equation}
where the sum is over all $k-$coupling paths between the states.\\

If $\cEff\neq 0$, the degeneracy at $f=1,\Delta=\Delta^*$ is guaranteed to be isolated. As this is generically the case (though see examples of exceptions in Appendix~\ref{s:gap size for most wrap-arounds}), it follows that the degeneracy is generically isolated and the resulting gap from wrap-around goes as $\theta(\delta^k)$ as $\delta\rightarrow 0$. That is, the gap size shrinks exponentially with coupling distance. Further, the size of the resulting gap depends to lowest order only on the energies of the intermediate coupling states, so if $c_{x,y}^{\text{eff}}=0$, the resulting gap goes as $O(\delta^{k+1})$ as $\delta\rightarrow 0$, and it is not a guarantee that the degeneracy at $(1,\Delta^*)$ is continuous (not isolated). Finally, while $\cEff$ is difficult to compute in general, it depends only on $\Hmixer$ and $\Hcost$, allowing a characterization of how quickly the gap grows as $\Delta$ increases above $\Delta^*$. Also, given the simple recursive formula in Eq.\ref{eq: q_polynomial_def} which we use to determine $\cEff$, this formula is simpler than conventional perturbation theory \cite{sakurai,kato_linear_op_perturbation} when $k$ becomes large. 
An area of future work is to
compute ``average'' values of $\cEff$ and its scaling with $k$ for different families of mixer/cost Hamiltonians, as is done in Appendix~\ref{s:local_constraint_satisfaction_scaling}. In deriving the formula, we also get a first-order estimate for the gap location $f=1+\epsilon$ in a fixed-$\Delta$ path, satisfying $\delta/(\Delta^*\epsilon)=-1+(D_x-D_y)/(E_x-E_y)+O(\delta)$.

\begin{table}
\begin{center}
\begin{tabular}{ |c|c| } 
 \hline
 \textbf{k=1} & $1$ \\ 
 \hline
 \textbf{k=2} & $-\frac{i}{2}b_1$ \\ 
 \hline
 \textbf{k=3} & $-(\frac{1}{4}b_1b_2+\frac{1}{12}) $\\ 
 \hline
 \textbf{k=4} & $i(\frac{1}{8}b_1b_2b_3+\frac{1}{24}(b_1+b_3) )$\\
\hline
 \textbf{k=5} & $\frac{1}{16}b_1b_2b_3b_4+\frac{1}{48}(b_1b_2+b_1b_4+b_3b_4)+\frac{1}{120}$\\
 \hline
\end{tabular}
\end{center}
\caption{Simplified  versions of $Q^{k-1}(b_1,...,b_{k-1})$ from Eq.~\ref{eq: q_polynomial_def} \textbf{for some fixed coupling path} for $k=1,...,5$.}
\label{table:eq_tab}
\end{table}

%RESOLVED:\tad{Caption of the table \ref{table:eq_tab} says ``$a_j$ as defined by Eq.~\ref{eq:energy_simplification}''. Why does this matter? Aren't these polynomials defined for arbitrary choices of variables so the table is correct for any values of the $a_j$? To emphasize this, would it be clearer to use the same variables as used in the definition of $Q$, i.e., $b_j$? }

\subsubsection{Deriving Gap Formula}
\label{s:gap_derivation}

%RESOLVED\tad{ Eq.~\ref{e:main_gap_equation} \emph{is} the gap formula so  I think delete ``from'' from the phrase: ``derive the gap formula \emph{from} Eq.~\ref{e:main_gap_equation}'', otherwise it seems like we're deriving the gap formula from itself?}

To derive the gap formula, Eq.~\ref{e:main_gap_equation}, consider wrap-around that takes place at $(f=1,\Delta=\Delta^*)$ such that $e^{-i\Delta^*E_x}=e^{-i\Delta^*E_y}$ for non-degenerate cost eigenstates $\ket{x},\ket{y}$. Suppose that at $\Delta = \Delta^*+\delta $ the gap $g$ from avoided crossing due to wrap-around occurs at $f=1+\epsilon$, then without loss of generality the $N\times N$ QAOA unitary (as defined in Eq.~\ref{e:unitary}) may be expressed as:
$$\UQAOA(1+\epsilon, \Delta^*+\delta) = \lambda_1\ket{v_1}\bra{v_1} + (\lambda_1+g)\ket{v_2}\bra{v_2} + \sum_{j= 3}^{N} \lambda_j \ket{v_j}\bra{v_j}~,$$
where $\lambda_j$ and $\ket{v_j}$ are the eigenvalues and eigenvectors respectively of the QAOA unitary, and $\ket{v_1},\ket{v_2}$ are the eigenvectors involved in the eigenvector swap. These eigenvectors and  the gap $g$ depend on $\epsilon, \delta$. We define distance $r=\sqrt{\epsilon^2+\delta^2}$ in parameter space from the point of degeneracy at $(f=1,\Delta=\Delta^*)$. The difference between $\UQAOA(1+\epsilon,\Delta^*+\delta) $ and
\begin{equation} \mathcal{U} := \UQAOA(1,\Delta^*)~, \end{equation}
goes as $O(r)$, $r\rightarrow 0$. From perturbation theory \cite{kato_linear_op_perturbation}, the difference between $\ket{v_j}$ and the nearest eigenvector of $\mathcal{U}$ goes as $O(r)$, as does the difference between the eigenvalues of $\UQAOA(1+\epsilon,\Delta^*+\delta)$ and the nearest eigenvalues of $\mathcal{U}$. As discussed in Section~\ref{s:eigenvector swap}, $\ket{v_1}$ and $\ket{v_2}$ lie predominantly within the subspace spanned by $\ket{x}$ and $\ket{y}$, though $\ket{x}$ and $\ket{y}$ can have $O(r)$ magnitude of overlap with other $\ket{v_j}$. Since $g=O(r)$, we cannot simply compute the submatrix of \UQAOA~on $\ket{x}$ and $\ket{y}$, as the order of the gap size could be of the same order as overlap with the other eigenvectors of $\UQAOA$. To resolve this, we introduce the ``residual matrix" of a unitary $U$, $R^{\vec{l}}(U)$:
\begin{equation}
    R^{\vec{l}}(U) = \prod_{j=1}^{|\vec{l}|} (U-l_jI)~.
\end{equation}
By choosing $\vec{l} = \{\lambda_1, \lambda_3,...,\lambda_N\}$, we have:
\begin{equation}
    \RQAOA = R^{\vec{l}}(\UQAOA(1+\epsilon,\Delta^*+\delta)) =  g \prod_{j=3}^N(\lambda_1+g-\lambda_j)\ket{v_2}\bra{v_2} = g \prod_{j=3}^N(\lambda_1-\lambda_j)\ket{v_2}\bra{v_2} + O(g^2)~.
\end{equation}
The remaining term $O(g^2)$ is negligible provided that $|g| \ll |\lambda_1-\lambda_j|,\ \forall j\neq 1,2 $. As each $\lambda_j$ is perturbatively within $O(r)$ of some eigenvalue $e^{-i\Delta^*E_{cost}}$ of $e^{-i\Delta^*\Hcost}$, it is from here that the requirement of $\ket{x},\ket{y}$ being non-degenerate with other eigenstates of $\Hcost$ originates, and that we expect our gap formula from Eq.~\ref{e:main_gap_equation} to break down when the predicted eigenvalue is of the same order as separation of $e^{-i\Delta^*E_x}=e^{-i\Delta^*E_y}$ of other eigenvalues of $e^{-i\Delta^*\Hcost}$.  

Now, $\ket{v_2} = c_x\ket{x}+c_y\ket{y}+\gamma\ket{v^{\perp}}$ where $\ket{v^\perp}$ is orthogonal to $\ket{x},\ket{y}$ and $|\gamma|=O(r)$. Consequently, we have: 
\begin{equation}
\sqrt{|\bra{x}\RQAOA\ket{x}|^2 + |\bra{y}\RQAOA\ket{y}|^2 + |\bra{x}\RQAOA\ket{y}|^2 + |\bra{y}\RQAOA\ket{x}|^2} = \left| \left(\prod_{j=3}^N(\lambda_1-\lambda_j) + O(r)\right) g\right| = |Kg|~,
\end{equation}

\begin{equation}
    \frac{\sqrt{|\bra{x}\RQAOA\ket{x}|^2 + |\bra{y}\RQAOA\ket{y}|^2 + |\bra{x}\RQAOA\ket{y}|^2 + |\bra{y}\RQAOA\ket{x}|^2}}{\left| \prod_{j=3}^N(\lambda_1-\lambda_j)\right|} = |g|(1+O(r))~,
\label{e:gap_estimate_residual_matrix}
\end{equation}
and 
\begin{equation}
    \frac{|\bra{x}\RQAOA\ket{x}|^2}{|K|^2|c_x|^4} = \ 
    \frac{|\bra{y}\RQAOA\ket{y}|^2}{|K|^2|c_y|^4} =\ 
    \frac{|\bra{y}\RQAOA\ket{x}|^2}{|K|^2|c_x|^2|c_y|^2} = \frac{|\bra{x}\RQAOA\ket{y}|^2}{|K|^2|c_x|^2|c_y|^2} = |g|^2~.
\label{e:constraints_on_R}
\end{equation}

Perturbation theory implies that the leading order change comes from the diagonal terms. Therefore, for $k\geq 2$ for fixed $r$, the choice of gap is minimized when we choose $\epsilon,\delta$ such that the first-order-in-$r$ components of $\bra{x}\RQAOA\ket{x}$ and $\bra{y} \RQAOA\ket{y}$ vanish. Computing these values will yield a first-order approximation of the gap location. First, we enforce that both terms cancel out their first order in $r$ contributions:
$$\bra{x}\RQAOA\ket{x} = (\bra{x}i\left(\Delta^*\epsilon \Hmixer - (\Delta^*\epsilon + \delta)\Hcost  \right)\mathcal{U}\ket{x} + e^{-i\Delta^*E_x}-\lambda_1)\prod_{j\neq x,y}(e^{-i\Delta^*E_x}-e^{-i\Delta^*E_j}) + O(r^2)~,$$
with an analogous equation holding for $\ket{y}$. Thus, by solving the following equation
$$\Delta^*\epsilon D_x - (\Delta^*\epsilon+\delta)E_x = O(r^2) = \Delta^*\epsilon D_y - (\Delta^*\epsilon+\delta)E_y ~;$$
we obtain a relation between $\delta$ and $\epsilon$ 
\begin{equation}
    \frac{\delta}{\Delta^*\epsilon} = \frac{D_x-D_y}{E_x-E_y} - 1 + O(r) = \Gamma-1+O(r)~.
\label{e:wrap_around_gap_location}
\end{equation}
Here, $E_j,D_j,\Gamma$ are given by Eq.~\ref{e: Dj_Ej_Gamma_definition}. This provides us with a first-order estimate of the gap location from the avoided crossing. As an aside, if we regard the ``unperturbed'' energies in the computational basis near $f=1$ as being given by $\Delta (1-f)D_{x,y} + \Delta f E_{x,y}$ a slightly more accurate approximation would replace $\Delta^*$ with $\Delta$ in Eq.~\ref{e:wrap_around_gap_location}. However, we will proceed with this first-order estimate as we are interested in the lowest-order estimate for the gap size. This approximation is only valid in the regime where $|\Gamma-1|$ is large compared to $r$.

The terms appearing in  Eq.~\ref{e:constraints_on_R}~are not independent since $|c_x|^2+|c_y|^2 = 1 + O(r)$. Thus for fixed $r$, we try to choose $\epsilon,\delta$ and $|c_x|^2$ such that any one of them is minimized (to lowest order in $r$, $K$ is fixed with $\lambda_j=e^{-i\Delta^*E_j}$). We will now focus on $\bra{x}\RQAOA \ket{y}$ term

$$\bra{x}\RQAOA\ket{y} = \bra{x}\prod_{j=1,3}^N\left(e^{i\Delta \epsilon\Hmixer}e^{-i(\Delta\epsilon+\delta)\Hcost}\mathcal{U}-\lambda_j \right)\ket{y} ~,$$

$$\bra{x}\RQAOA\ket{y} = \bra{x}\prod_{j=1,3}^N\left((I+(\sum_{n=1}^\infty \frac{(i\Delta \epsilon \Hmixer)^n}{n!}))(I+(\sum_{m=1}^\infty \frac{(-i(\Delta \epsilon + \delta) \Hcost)^m}{m!}))\mathcal{U}-\lambda_j \right)\ket{y} ~.$$

Since $\ket{x},\ket{y}$ are $k$-coupled to each other, we must choose at least $k$ total powers of $\Hmixer$ from the above product for the contribution to the product to be non-zero. Since we're working in the eigenbasis of $\Hcost$, the \Hcost~ terms are diagonal and we only have to consider the terms coming from the Taylor expansion of $e^{i\Delta \epsilon\Hmixer}$. Therefore, the lowest power in $\epsilon,\delta$ contributions to this term come from only considering total powers of $k$ coming entirely from powers of $\Hmixer$, 
$$\bra{x}\RQAOA\ket{y} = \bra{x}\prod_{j=1,3}^N\left((\sum_{n=1}^k \frac{(i\Delta \epsilon \Hmixer)^n}{n!}))\mathcal{U}+(\mathcal{U}-\lambda_j) \right)\ket{y} + O(r^{k+1})~.$$

To identify the relevant terms, we fix a particular coupling path $S = \{\ket{x},\ket{s_1},...,\ket{s_{k-1}}\}$. Every time we apply a term from the product to the ket on the right, we choose whether  to apply $(\mathcal{U}-\lambda_j)$ or a power of $\Hmixer$ to hop further along the coupling path to $\ket{x}$. Crucially, since each term in the product commutes with every other term, we can order the $\lambda_j$ however we want when computing the contributions of each set of hops along each coupling path (e.g. hops of $\ket{x}\rightarrow \ket{s_1}\rightarrow \ket{s_{k-1}}\rightarrow \ket{y} $ from the coupling path $S$). A hop over $m$ states would require choosing a power $\frac{(i\Delta\epsilon)^m}{m!}\Hmixer^m$ of the mixer. 

%RESOLVED:\tad{Possible typo: Check use of $M$ vs $m$ in this paragraph: is ``$M$ hops'' supposed to be ``$m$ hops''? And for $M$ in the range of $j$? Later in the paragraph $M$ is said to be an ordered subset of $S$, rather than a number??}
Recall that each $\lambda_j$ is within $O(r)$ of an eigenvalue of $\mathcal{U}$ and that all states on our coupling paths are eigenstates of $\mathcal{U}$. Exploiting the freedom in $\lambda_j$ ordering, to evaluate the contribution of $m$ hops along indices $I$ of a particular coupling path $S$, we order the $\lambda_j$'s to apply them in the order that the hops visit the states with the closest eigenvalues. This forces us to choose to hop further along the path for the $m$ multiples that we encounter (otherwise we get an extra $O(r)$ contribution from $(\mathcal{U}-\lambda_j)$, which we ignore as we are only considering the lowest order terms in $r$), giving us a multiple $e^{-i\Delta^*E_s}$ along the way for the state $s$ visited. After that, our path has already arrived at $\ket{x}$ and thereafter we pick only $(\mathcal{U}-\lambda_j)$, giving us, to lowest order in $r$, multiples of form $(e^{-i\Delta^*E_x}-e^{-i\Delta^*E_q})$ where $E_q$ are the energies of the states not visited in the hops over the coupling path. In other words, for a fixed coupling path $S$, (and letting $B_{ij}=\bra{i}\Hmixer\ket{j}$), the contribution is: 
$$B_{x,s_1}...B_{s_{k-1},y} (i\Delta\epsilon)^k\sum_{M\subseteq S} \left(\prod_{q\notin M}\left(e^{-i\Delta^*E_x}-e^{-i\Delta^*E_q} \right) \prod_{s\in M } \frac{e^{-i\Delta^* E_s}}{m!}\right) + O(r^{k+1})~,$$ 
% keep the following as part of the same paragraph
where the summation is over all ordered subsets $M$ of $S$ and $m$ is the hopping distance in $S$ between consecutive elements of $M$. Dividing out $\prod_{j\neq x,y}e^{-i\Delta^*E_x}-e^{-i\Delta^*E_j}$ from Eq.\ref{e:gap_estimate_residual_matrix} gives:
$$B_{x,s_1}...B_{s_{k-1},y} (i\Delta\epsilon)^ke^{-i\Delta^*E_x}\sum_{M\subseteq S} \left(\prod_{s\in M,s\neq x } \frac{e^{-i\Delta^* E_s}}{(e^{-i\Delta^*E_x}-e^{-i\Delta^*E_s})m!}\right) + O(r^{k+1})~.$$ 
We simplify this expression by noting that $\frac{\exp{(-i\Delta^*E_{s_j})}}{\exp(-i\Delta^*E_x) - \exp(-i\Delta^*E_{s_j})} = -\frac{1+i\cot(\frac{\Delta^*}{2}(E_{s_j}-E_x))}{2}  = -\frac{1+ia^S_j}{2}$ and by organizing the summation recursively as sums of all hops starting with $s_1$ plus all those starting with $s_2$, etc., giving:
$$B_{x,s_1}...B_{s_{k-1},y}\left(\frac{1}{k!}-\sum_{m=1}^{k-1} \frac{1+ia^S_{m}}{2m!}Q^{k-m-1}(a^S_{m+1},..a^S_{k-1})\right)~,$$
with function $Q$ as defined in Eq.~\ref{eq: q_polynomial_def}. Therefore we have 
$$\frac{|\bra{x}\RQAOA\ket{y}|}{|\prod_{j\neq 1,2}(\lambda_1 -\lambda_j)|} = (\Delta^*\epsilon)^k\left|\sum_S P(S)Q^{k-1}(a_1^S,...,a_{k-1}^S)\right|+O(r^{k+1}) = (\Delta^*\epsilon)^k |\cEff| + O(r^{k+1})~,$$
with $P(S)$ as defined in Eq.~\ref{eq:mixer_path_contributions} and $\cEff$ as in Eq.\ref{e:effective_coupling}. 

The resulting expression, to lowest order, only depends on $\epsilon$. Therefore, for a given $\epsilon$, we have a degree of freedom to optimize $|c_x|$ through varying $\delta$. Note that from Eq.~\ref{e:constraints_on_R}, the gap is minimized when $|c_x|^2=|c_y|^2 = 1/2+ O(r)$, i.e. when the swap of eigenvectors is half-way through. It's important to note that fixing $\epsilon=\epsilon_{fix}$ and varying $\delta$ which (we know to be of $O(\epsilon)$) to minimize the gap size at $\delta= \delta_{opt}$ doesn't necessarily yield the same gap as then fixing $\delta_{opt}$ and varying $\epsilon$, unless that minimum already happens to be very close to $\epsilon_{fix}$. In this latter case, the numerators will be approximately the same and therefore the denominators will be both approximately $1/4$ (this is as large as they can get). From perturbation theory, we expect the exchange of eigenvectors to take place over $O(\epsilon^2)$ interval at fixed $\delta$ for $k\geq 2$, and since our $\epsilon_{fix},\delta_{opt}$ is in the middle of the exchange, the optimal $\epsilon$ for $\delta_{opt}$ is approximately $\epsilon_{fix}$. \Appendix{ptrack_derivation}, %Section C.1, 
shows this interval goes as $O(\epsilon^k)$. Thus, for $k\geq 2$, we can approximate the denominator as $1/4$, and the minimum gap location of the avoided crossing coincides with the middle of the eigenvector swap in the perturbative limit, i.e. the exchange of eigenvectors occurs approximately symmetrically about the avoided crossing from  wrap-around. For $k=1$ a different approach is required, as shown in \Appendix{special case}. 

For the $k\geq 2$ case, we have $|c_x|^2=|c_y|^2 = 1/2+ O(r)$ and therefore $|\bra{x}\RQAOA\ket{x}|^2 = |\bra{y}\RQAOA\ket{y}|^2 = |\bra{x}\RQAOA\ket{y}|^2(1+O(r))  $, changing the left-hand side of Eq.~\ref{e:gap_estimate_residual_matrix} to be:
$$\frac{\sqrt{4|\bra{x}\RQAOA\ket{y}|^2}}{\left|\prod_{j=3}^N(\lambda_1-\lambda_j)\right|} = |g|(1+O(r))~,$$
\begin{equation}
    \implies \frac{2(\Delta^*\epsilon)^k|\cEff|}{|g|} = 1+O(r)~.
\label{e:penultimate_gap_estimate_residual_matrix}
\end{equation}
Plugging in the relation from Eq.~\ref{e:wrap_around_gap_location} gives Eq.~\ref{e:main_gap_equation}.

\subsubsection{Special Case: k=1}
\label{s:special case}
When the wrapping states $\ket{x}$ and $\ket{y}$ are directly coupled, we can compute the gap from the submatrix of $\RQAOA$ on the $\ket{x},\ket{y}$ subspace. To lowest order in $r$, the terms are:
$$\bra{x}\RQAOA\ket{x} = (\bra{x}i\left(\Delta^*\epsilon \Hmixer - (\Delta^*\epsilon + \delta)\Hcost  \right)\mathcal{U}\ket{x} + e^{-i\Delta^*E_x}-\lambda_1)\prod_{j\neq x,y}(e^{-i\Delta^*E_x}-e^{-i\Delta^*E_j}) + O(r^2)~,$$
$$\bra{y}\RQAOA\ket{y} = (\bra{y}i\left(\Delta^*\epsilon \Hmixer - (\Delta^*\epsilon + \delta)\Hcost  \right)\mathcal{U}\ket{y} + e^{-i\Delta^*E_y}-\lambda_1)\prod_{j\neq x,y}(e^{-i\Delta^*E_y}-e^{-i\Delta^*E_j}) + O(r^2)~,$$
$$\bra{x}\RQAOA\ket{y} = (\bra{x}i\left(\Delta^*\epsilon \Hmixer  \right)\mathcal{U}\ket{x})\prod_{j\neq x,y}(e^{-i\Delta^*E_x}-e^{-i\Delta^*E_j}) + O(r^2) = \bra{x}\RQAOA\ket{y}~.$$
We define $c:=\bra{x}\Hmixer\ket{y}$ for the ease of notation. The eigenvalues of this submatrix of $\RQAOA$ are approximately $0$ and $\prod_{j\neq x,y}(e^{-i\Delta^*E_x}-e^{-i\Delta^*E_j})g $ so by calculating the minimum gap  between the eigenvalues of this residual matrix (fixing $\delta$ and changing $\epsilon$ to minimize the gap), we approximate the overall gap from the avoided crossing due to wrap-around. Dividing out the common product term and the imaginary unit, the gap between eigenvalues of a two-by-two matrix with diagonals $a,b$ and equal off-diagonals $c$ is given by $\sqrt{(a-b)^2+4c^2}$, so that

\begin{equation}
g(\epsilon,\delta) = \sqrt{\left(\Delta^*\epsilon(\Gamma-1)-\delta \right)^2(E_x-E_y)^2 + (2\Delta^*\epsilon c)^2}~.
\label{e:quadratic gap}
\end{equation}
The value of $\epsilon$ which minimizes the expression in Eq.~\ref{e:quadratic gap} is given by 
\begin{equation}
\epsilon_{min} = \frac{\delta}{\Delta|\Gamma-1|(1+\left(\frac{2c}{|\Gamma-1|(E_x-E_y)}\right)^2)}~.
\label{e:minimizing epsilon}
\end{equation}
Plugging Eq.~\ref{e:minimizing epsilon} into Eq.~\ref{e:quadratic gap} gives us the gap equation for the $k=1$ case:
\begin{equation}
g(\delta) \approx \frac{2c\delta}{\sqrt{(\Gamma-1)^2+(\frac{2c}{(E_x-E_y)})^2}}~.
\label{e:k1 gap}
\end{equation}

\subsection{Feasible regions for intermediate degeneracies}
\label{s:intermediate_degeneracies}
In Section~\ref{s:nature_of_degeneracies} we mention that no isolated degeneracies due to wrap-around occur at $0<f<1$ below the $\Delta=\Delta^*_{\text{wrap}}$ at which the first wrap-around occurs at $f=0,1$ (from the previous section, generically we have $\Delta^*_{\text{wrap}}=\DeltaCrit$). This appendix proves this statement. The proof also shows that, below $\Delta^*_{\text{wrap}}$, the eigenvalue curves connecting maximal excited eigenvalues are always closest to eigenvalue curves connecting ground eigenvalues at the edges $f=0,1$, implying that below $\Delta^*_{\text{wrap}}$, no swaps take place either, i.e. no narrow avoided crossings between the two curves.

%RESOLVED\tad{What is the relation between $\Delta_{\text{wrap}}^*$ in this theorem and \DeltaCrit?  Presumably \DeltaCrit\ can be larger if the first wrap around happens to be a continuous, rather than isolated degeneracy?? In that case, does the theorem actually prove what we claim in the previous paragraph, which makes a claim about $\Delta < \DeltaCrit$. If \DeltaCrit\ is larger than $\Delta_{\text{wrap}}^*$ mentioned in the theorem, does the theorem apply? Do we need to exclude the case of continuous degeneracy at first wrap around, thereby focus on the case where $\Delta_{\text{wrap}}^* = \DeltaCrit$}

\begin{theorem}
Let $\Delta_{\text{wrap}}^*$ be the smallest value of $\Delta>0$ at which a pair of eigenvalues of $\UQAOA(f,\Delta)$~ wraps around at either $f=0$ or $f=1$. Then, if $(f^*,\Delta^*)$ is a point of isolated degeneracy due to wrap-around between the two eigenvalues, with $0\leq f^*\leq 1$, we must have $\Delta^*\geq \Delta_{\text{wrap}}^* $
\label{theorem1}
\end{theorem}
%[\sh{SH: assumptions should be clear. Haven't checked proof carefully but can likely be shortenned. I can see including this Thm but less clear on utility of prev 2..}]
\textbf{Proof:} This is trivially true for $f^*=0,1$ and we will prove it holds for $0<f^*<1$ as well. To do so, we first recall the definition of the QAOA Hamiltonian, \HQAOA, given by Eq.~\ref{e:hqaoa}, choosing \HQAOA~to be equal to \Hmixer~at $f=0$, equal to \Hcost~at $f=1$, and to have continuous eigenvalues in between. Prior to the wrap-around of eigenvalues (i.e. when we can choose a branch cut such that the eigenvalue curves do not pass through the cut boundaries), this can always be done. 

Due to ordering and geometry, the first eigenvalue curves of \UQAOA~that interact with each other due to wrap-around are the eigenvalue curves connecting maximal excited eigenvalues and the eigenvalue curves connecting ground eigenvalues. Therefore, if $\Delta^*<\Delta_{\text{wrap}}^*$ and $0<f^*<1$, then \HQAOA~has either a lower ground state energy than \Hmixer~and \Hcost, or a higher maximal energy than \Hmixer~and \Hcost. We shall show that this is impossible.

First, we shift the spectra of \Hcost~and \Hmixer~for convenience so that, for each Hamiltonian, respectively, the highest and lowest energies are equal in magnitude and opposite in sign. Note that this is equivalent to multiplying \UQAOA~by a global phase, which cannot affect the formation of degeneracies. In this setting, \HQAOA\ will have a lower minimal energy or higher maximal energy if and only if it has an energy with a greater magnitude than that of any energy of \Hmixer~or \Hcost. 

Let $\lambda_{Q}(f, \Delta)$ be the largest eigenvalue of $\HQAOA$ at $(f,\Delta)$, and let $\ket{\psi_{Q}(f,\Delta)}$ be the corresponding eigenvector (choice of eigenvector is continuous in the absence of degeneracies). Finally, we define $\lambda_m$ as the largest magnitude of any eigenvalue of the mixer and $\lambda_c$ as the largest magnitude of any eigenvalue of the cost, and recall that $\Hinst=f\Hcost+(1-f)\Hmixer$.

Suppose \Hmixer~and \Hcost~share either a minimal or a maximal eigenspace.
%\sh{[SH:dont you need minimal eigenspace here too?]}.
Then the maximum magnitude energy of both \Hinst~and $H_Q$ is $(1-f)\lambda_m+f\lambda_c$ (since those eigenspaces are simultaneously diagonalizable) and therefore cannot be larger in magnitude than $\max(\lambda_m,\lambda_c)$. We will now consider the case where \Hmixer~ and \Hcost~do not share ground/highest excited eigenspaces. In this instance, $\max_{\psi}|\langle \Hinst \rangle_\psi|<(1-f)\lambda_m + f\lambda_c $ when $0<f<1$.

To begin with, we take the partial derivative with respect to $\Delta$ of both sides Eq.\ref{e:hqaoa}:
$$-i\left(H_Q+\Delta\frac{\partial H_Q}{\partial \Delta} \right)e^{-i \Delta H_Q} = -i\left(e^{-i\Delta(1-f)\Hmixer}(1-f)\Hmixer e^{-i\Delta f\Hcost} + e^{-i\Delta(1-f)\Hmixer}(f)\Hcost e^{-i\Delta f\Hcost} \right)~.$$
Dividing out the imaginary constant, and substituting in \Hinst~gives us:
$$\left(H_Q+\Delta\frac{\partial H_Q}{\partial \Delta} \right)e^{-i \Delta H_Q} = e^{-i\Delta(1-f)\Hmixer}\Hinst e^{-i\Delta f\Hcost}~,$$

$$\left(\lambda_Q +\Delta \bra{\psi_Q} \frac{\partial H_Q}{\partial \Delta} \ket{\psi_Q}\right)e^{-i\Delta\lambda_Q}  = \bra{\psi_Q}e^{-i\Delta(1-f)\Hmixer}\Hinst e^{-i\Delta f\Hcost}\ket{\psi_Q}~.$$
We simplify this expression using the Hellmann-Feynmann Theorem~\cite{feynman1939} and defining $\ket{\chi} = e^{+i\Delta(1-f)\Hmixer}\ket{\psi_Q}$ and $\ket{\phi} = e^{-i\Delta f\Hcost}\ket{\psi_Q}$ :
$$\left(\lambda_Q+\Delta\frac{\partial}{\partial \Delta}\lambda_Q\right)e^{-i\Delta\lambda_Q} =  \bra{\chi}\Hinst \ket{\phi}~.$$
Note that the matrix elements of \UQAOA~depend analytically on $\Delta, f$, and therefore so do the coefficients of \UQAOA's characteristic polynomial. From this, we know the eigenvalues of \UQAOA~vary continuously with $\Delta$. From the Implicit Function Theorem, the eigenvalue is also a continuously differentiable function of the characteristic polynomial coefficients (and thus of $\Delta, f$) if it is non-degenerate. Since we consider the eigenvalue near an isolated degeneracy, with $\Hmixer$\ and $\Hcost$\ not sharing a maximal eigenspace, for any fixed $f$ and choice of $\Delta$ there must exist some open interval $(\Delta_{\text{diff}},\Delta)$ along which $\lambda_Q$ is non-degenerate and hence differentiable. We suppose that we operate on such an interval moving forward. We now take the absolute value of both sides and note that the right hand side is at most the largest eigenvalue of $\Hinst$:
%From this, the same must be true of our $\lambda_Q$. Since we are considering the case where \Hmixer~and \Hcost~do not share a maximal eigenspace, we proceed \textbf{\textit{under the assumption}} that $\lambda_Q(f,\Delta)$ at fixed $0<f<1$ is degenerate for at most finitely many values of $\Delta$ on any finite interval over $\Delta$ \sh{[sH: ??]}\vkadd{Since we have a two-dimensional parameter space, and assume our Hamiltonians to not share a ground/maximal eigenspace, this is a very weak assumption}.  Thus, while $\lambda_Q$ is always a continuous function of $f,\Delta$, its derivative with respect to $\Delta$ is undefined for at most finitely many values of $\Delta$. 

\begin{equation}
\left|\left(\lambda_Q+\Delta\frac{\partial}{\partial \Delta}\lambda_Q\right)\right| = |\bra{\chi}\Hinst\ket{\phi} | \leq \max_\psi(|\langle \Hinst \rangle|) < (1-f)\lambda_m+f\lambda_c ~.
\label{e:main_bound}
\end{equation}

We now use this bound for a proof by contradiction. Suppose there exist $f_0,\Delta_0$ such that $|\lambda_Q(f_0,\Delta_0)| = (1-f_0)\lambda_m + f_0\lambda_c>0$ at $0<f_0<1$ for an eigenvalue $\lambda_Q(f,\Delta)$ for either the ground or highest excited eigenvalues of $H_Q$. Without loss of generality, suppose $\lambda_Q(f_0,\Delta_0) >0$. Note that for any fixed $f_0$, there must exist a smallest $\Delta_0$ for which this equality is attained, since in the limit of $\Delta_0\rightarrow 0$, we have $\lambda_Q <(1-f_0)\lambda_m+f_0\lambda_c$ due to \HQAOA\ approximating \Hinst. From now on, we fix $f_0$ and consider $\Delta_0$ as the smallest value of $\Delta$ where this equality is attained.

Next, we will show through a proof by contradiction that there must exist constants $l_1>0,\delta_1>0$ s.t. $|\partial \lambda_Q/\partial \Delta| \geq l_1,\ \forall \Delta\in (\Delta_0-\delta_1, \Delta_0)$. Suppose that no positive lower bound exists for $|\partial \lambda_Q/\partial \Delta|$ on any such interval. Let $$L:=(1-f_0)\lambda_m+f_0\lambda_c  - \langle \Hinst\rangle_{max}>0.$$ Note that since $\lambda_Q$ is continuous,  $$\exists \delta_0>0 \text{ s.t}. \forall \Delta\in(\Delta_0-\delta_0, \Delta_0), 
|\lambda_Q(f_0,\Delta)-(1-f_0)\lambda_m-f_0\lambda_c| < L/3.$$ On this same interval, by hypothesis we can find a $\Delta$ such that $$|\partial \lambda_Q(f_0,\Delta)/\partial \Delta| < L/3\Delta_0<L/3\Delta.$$ From this, we get:
\begin{equation}
(1-f_0)\lambda_m + f_0\lambda_c \leq \left|\lambda_Q+\Delta \frac{\partial \lambda_Q}{\partial \Delta}\right| + \Delta\left| -\frac{\partial \lambda_Q}{\partial \Delta}\right| + \left|(1-f_0)\lambda_m+f_0\lambda_c-\lambda_Q \right| \leq \langle \Hinst\rangle_{max} + \frac{L}{3}+\frac{L}{3} < (1-f_0)\lambda_m+f_0\lambda_c.
\end{equation}
This gives us a contradiction. So $\exists l_1,\delta_1>0$ such that $|\partial \lambda_Q/\partial \Delta|>l_1, \forall \Delta\in (\Delta_0-\delta_1, \Delta_0)$. Note that this derivative must be positive, since $\Delta_0$ is the  smallest value of $\Delta$ where $\lambda_Q$ attains this large of a value. Consequently, we must have $|\lambda_Q+\Delta \partial \lambda_Q/\partial \Delta| = \lambda_Q +\Delta\partial\lambda_Q/\partial \Delta$. From here, we get the final step of our proof - that  
\begin{equation}
    \forall \Delta\in(\Delta_0-\delta_1, \Delta_0),
    0< (\Delta_0-\delta_1)l_1 < \Delta \frac{\partial \lambda_Q}{\partial \Delta} < (1-f_0)\lambda_m+f_0\lambda_c - \lambda_Q.
\end{equation}
Since $\lambda_Q$ is continuous, this gives us a contradiction, since it cannot equal $(1-f_0)\lambda_m+f_0\lambda_c$ at $f=f_0,\Delta=\Delta_0$.

Therefore, we cannot have a $\lambda_Q$ with magnitude greater than or equal to $(1-f)\lambda_m+f\lambda_c$ for any $0<f<1$ when \Hmixer~ and \Hcost~ do not share a maximal eigenspace before the wrap-around at the edges $f=0$ or $f=1$. Together with the case we've already considered where they do share one, this concludes the proof. 

\section{Deriving tracking estimates and describing the performance ridge}
\label{s:swap interval and dlz}
Here we characterize the value of $p$ at which overlap with the state prepared in the discrete adiabatic limit starts increasing with increasing $p$ using estimates on the eigenvector swap interval and discretized Landau-Zener to derive equations for performance deterioration as discussed in Section~\ref{s:performance_ridge}.

\subsection{Estimating width of interval over which eigenvector swap happens}
\label{s:ptrack_derivation}
Consider the eigenvectors of $\UQAOA$ swapping along a fixed-$\Delta$ path in parameter space. If close to $f=1$, we can approximate this as a swap between cost eigenstates $\ket{x}$ and $\ket{y}$, as the eigenvectors of $\UQAOA$ will be very close to those of $\Hcost$. We first derive an estimate of the length $s(|c_x|)$ of the interval in the schedule over which a swapping eigenvector changes from having $|c_x|^2$ squared overlap with $\ket{x}$ initial eigenvector to $1-|c_x|^2$ overlap with it. While a direct estimate of the critical discretization required to start gradually tracking the eigenvector swap is left for future work, the inverse of this interval is a rough proxy for its scaling. 

%\subsubsection{Estimating the swap interval}
Our initial approximation, as mentioned above, is that the swap happens precisely between states $\ket{x}$ and $\ket{y}$ (we expect this approximation to be most valid when $c_x$ is not too close to 0 or 1, i.e. the difference between $|c_x|^2$ and 0 or 1 
%RESOLVED\tad{distance between what? Is $\delta$ this distance or something else that makes sense to compare with $\delta$?}
must be above order $\delta^2$). We also assume that support on other cost eigenstates is negligible - this assumption is justified provided that the eigenvalue gap $g(|c_x|)$ between the two swapping eigenvectors remains much smaller than the gaps between them and other eigenvalues. We denote the two swapping eigenvectors $\ket{v_1(c_x)},\ket{v_2(c_x)}$ as follows:
\begin{equation}
\begin{split}
\ket{v_1(c_x)} = c_x\ket{x} + \sqrt{1-c_x^2}e^{i\phi(c_x)}\ket{y}~,\\ 
\ket{v_2(c_x)} = \sqrt{1-c_x^2}\ket{x}-c_ye^{-i\phi(c_x)}\ket{c_y}~,\end{split}
\label{eq:v_1_v_2_def}
\end{equation}
with $c_x,c_y \in \mathbb{R}^++\{0\}$ and $c_x^2+c_y^2=1$. The \UQAOA~operator restricted to the subspace of these two eigenvectors, $\widetilde{\UQAOA}(c_x)$ is given by
\begin{equation}
    \widetilde{\UQAOA}(c_x) = (\lambda_1(c_x)+g(c_x))\ket{v_1(c_x)}\bra{v_1(c_x)} + \lambda_1(c_x)\ket{v_2(c_x)}\bra{v_2(c_x)}~,
    \label{e:restricted_uqaoa_1}
\end{equation}
with $\lambda_1(c_x)$ being the eigenvalue associated with $\ket{v_2(c_x)}$. Now, we step symmetrically to the other side of the minimum gap from the avoided crossing, where the two eigenvectors will have exchanged places in relative overlap. Since the gap is formed from an avoided crossing between two locally linear spectral curves, the avoided crossing has a slant, so both eigenvalues gain an extra phase of $e^{i\xi}$. By symmetry, both are otherwise the same as on the other side of the gap. The \UQAOA~operator restricted to the subspace of the two eigenvectors is therefore:
\begin{equation}
    \widetilde{\UQAOA}(\sqrt{1-c_x^2}) = e^{i\xi}(\lambda_1(c_x)+g(c_x))\ket{v_1(\sqrt{1-c_x^2})}\bra{v_1(\sqrt{1-c_x^2})} + e^{i\xi}\lambda_1(c_x)\ket{v_2(\sqrt{1-c_x^2})}\bra{v_2(\sqrt{1-c_x^2})}~.
    \label{e:restricted_uqaoa_2}
\end{equation}

\begin{figure}
    \centering
    \includegraphics[width=\textwidth]{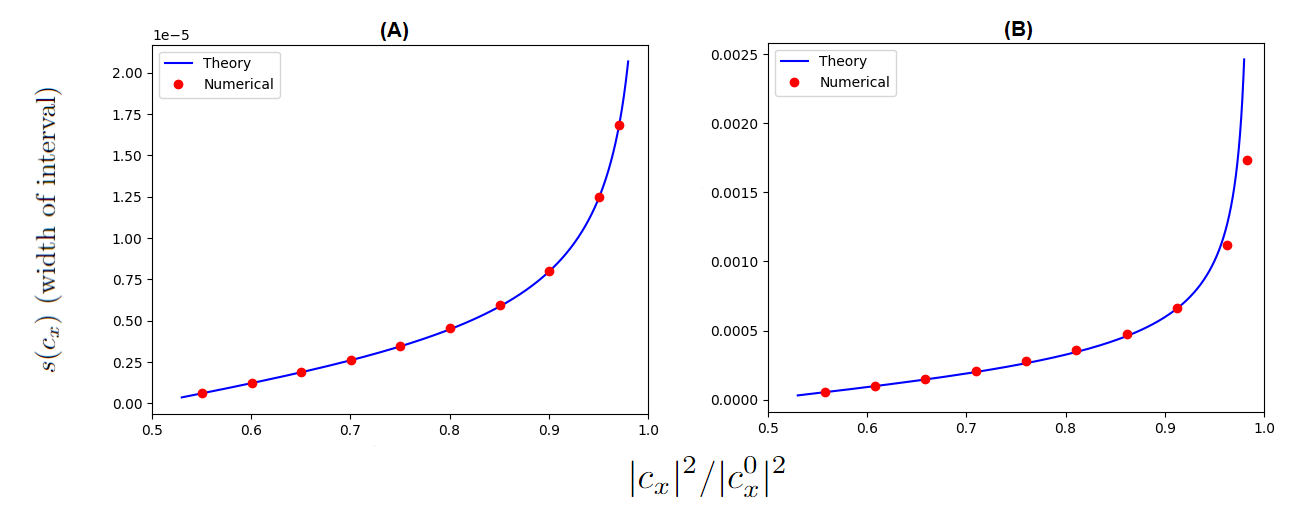}
    \caption{$|c_x|^2/|c_x^0|^2$ vs. the width of interval over which the swapping eigenvector goes from having $|c_x|^2/|c_x^0|^2$ overlap with one eigenvector to having $|c_x|^2/|c_x^0|^2$ overlap with the other eigenvector. Numerical estimates taken for $|c_x|^2 = 0.55,0.6,0.65,0.7,0.75,0.8,0.85,0.9,0.95,0.97$. Theory given by Eq.~\ref{e:swap_interval_equation_normalized}, with $\Delta = \DeltaCrit+0.05$. (A) The Hamiltonian pair from Fig.~\ref{f:eight_level_avoided_crossing_example}. For this example, $|\cEff| \approx 0.5^3\times 0.6315$ and $|c_x^0|^2\approx 0.9995$. (B) The Hamiltonian pair from Fig.~\ref{f:Ising_phase_diagram}. For this example, $|\cEff|\approx 0.2818$ and $|c_x^0|^2\approx 0.9867$.}
    \label{fig:s_theory_numerical_example}
\end{figure}
% The difference between the two operators (on this restricted sub-space) is given by:
% $$\left(\lambda_1(c_x)+g(c_x) \right)\left(c_x^2\ket{x}\bra{x}+c_x\sqrt{1-c_x^2}(e^{i\phi(c_x)}\ket{y}\bra{x}+e^{-i\phi(c_x)}\ket{x}\bra{y})+(1-c_x^2)\ket{y}\bra{y} \right)$$
% $$+\lambda_1(c_x)\left( (1-c_x^2)\ket{x}\bra{x} -c_x\sqrt{1-c_x^2}(e^{-i\phi(c_x)}\ket{y}\bra{x}+e^{i\phi(c_x)}\ket{x}\bra{y})+c_x^2\ket{y}\bra{y} \right) $$
% $$-e^{i\xi}\left(\lambda_1(c_x)+g(c_x) \right)\left((1-c_x^2)\ket{x}\bra{x}+c_x\sqrt{1-c_x^2}(e^{i\phi(\sqrt{1-c_x^2})}\ket{y}\bra{x}+e^{-i\phi(\sqrt{1-c_x^2})}\ket{x}\bra{y})+c_x^2\ket{y}\bra{y} \right) $$
% $$-e^{i\xi}\lambda_1(c_x)\left( c_x^2\ket{x}\bra{x} -c_x\sqrt{1-c_x^2}(e^{-i\phi(\sqrt{1-c_x^2})}\ket{y}\bra{x}+e^{i\phi(\sqrt{1-c_x^2})}\ket{x}\bra{y})+(1-c_x^2)\ket{y}\bra{y} \right) $$
We take the difference between Eq.~\ref{e:restricted_uqaoa_1} and Eq.~\ref{e:restricted_uqaoa_2} using Eq.~\ref{eq:v_1_v_2_def}. The resulting contributions for $\ket{x}\bra{x}$ and $\ket{y}\bra{y}$ are:
\begin{equation}
\left((1-e^{i\xi})\lambda_1(c_x)+(c_x^2-e^{i\xi}(1-c_x^2))g(c_x) \right)\ket{x}\bra{x} - \left((1-e^{i\xi})\lambda_1(c_x)+(c_x^2-e^{i\xi}(1-c_x^2))g(c_x) \right)\ket{y}\bra{y}~.
\label{e:ketx_brax_terms}
\end{equation}
Since $g(c_x)$ is the (very small) difference between two unitary eigenvalues, it is almost entirely orthogonal to them in their vector representations (i.e. the $\mathbb{R}^2$ representation of $\mathbb{C}$), and so is almost entirely orthogonal to the vector representation of $\lambda_1(c_x)$). 
%RESOLVED\tad{Eigenvalues are numbers: how are they orthogonal? Instead of orthogonal numbers, do you mean the corresponding eigenvectors are nearly orthogonal?}
Consequently, the absolute value of each coefficient is minimized when $\xi=0$, i.e., when the avoided crossing doesn't have a slant. Since the avoided crossing we consider takes place near $f=1$, by shifting all the energies of \Hcost\
%RESOLVED\tad{as noted above, what are ``cost energies''? Can we just use ``costs'' or ``energies''?}
by some constant, we can either shift the slant further or reverse it - meaning at some intermediate energy shift, we can eliminate the slant entirely. To do so, we compute (and choose an energy shift to minimize) the following expression:
\begin{equation}
|\bra{x}\UQAOA(f+s,\Delta)-\UQAOA(f,\Delta) \ket{x}|^2 + |\bra{y}\UQAOA(f+s, \Delta)-\UQAOA( f, \Delta)\ket{y}|^2 ~.
\label{e:minimizing_term}
\end{equation}
For $s\ll\Delta (||\Hmixer||+||\Hcost||)$, we can approximate the difference as
\begin{equation}
    \UQAOA(f+s,\Delta)-\UQAOA(f,\Delta) \approx -is\Delta(\Hcost - \Hmixer)~.
\label{e:uqaoa_diff}
\end{equation}
Using Eq.\ref{e:uqaoa_diff}, and Eq.\ref{e: Dj_Ej_Gamma_definition}, we can approximate Eq.\ref{e:minimizing_term} as
\begin{equation}
    s^2\Delta^2\left((E_x-D_x)^2+(E_y-D_y)^2\right)~.
\label{e:minimizing_term_simplified} 
\end{equation}
Recalling that we want to shift the energies of \Hcost 
%RESOLVED\tad{What are ``energies of the cost'' (used here and later in the paragraph)? Eigenvalues of \Hcost?}
by some constant $\Lambda$ to minimize this quantity, some simple calculus shows this to be obtained for $\Lambda = -\frac{E_x+E_y-D_x-D_y}{2}$. Under this shift, the avoided crossing is no longer slanted (while the gap $g(c_x)$ is unaffected as shifting the energies of the cost contributes only a global phase), so simplifying Eq.\ref{e:ketx_brax_terms} with $\xi=0$ and plugging in the shift to Eq.\ref{e:minimizing_term_simplified} yields
$$2|g(c_x)|^2|2c_x^2-1|^2 \approx 2s^2\Delta^2\left(\frac{(E_x-D_x)-(E_y-D_y)}{2} \right)^2~.$$
From our main gap equation in Eq.\ref{e:main_gap_equation} and Eq.\ref{e:constraints_on_R}, we can replace $|g(c_x)|$ with $|\cEff|(\delta/((D_x-D_y)/(E_x-E_y)-1))^k/(c_x\sqrt{1-c_x^2})$. Solving for $s$, we obtain
\begin{equation}
    s(c_x) = \frac{ 2|\cEff| }{\Delta |(E_x-D_x)-(E_y-D_y)|}\left(\frac{|2c_x^2-1|}{c_x\sqrt{1-c_x^2}}\right)\left|\frac{\delta}{\Gamma-1} \right|^k~.
    \label{e:swap_interval_equation_unnormalized}
\end{equation}

Further away from $f=1$, the swap won't take place between precisely $\ket{x}$ and $\ket{y}$, but rather between eigenvectors with strong support on these two eigenvectors. To estimate the eigenvector composition outside of this avoided crossing, non-degenerate perturbation theory gives
\begin{equation}
    |c_x^0|^2 \approx 1- \sum_j \left(\frac{\Delta (1-f^*) (\Hmixer)_{x,j}}{2\sin\left(\frac{\Delta}{2}f^*(E_j-E_x)\right)}\right)^2~,
    \label{eq:first_order_eigenvector_correction}
\end{equation}
where $f^*$ is the location of the gap. Since the eigenvector changes from first order perturbation to the operator from $\Hmixer$ near $f=1$ take place over much larger interval than most of the swap near the avoided crossing (for $k\geq 2$), we approximate $c_x^0$ at the gap location itself. Defining $|\widetilde{c_x}|^2 = |c_x|^2/|c_x^0|^2$, we get a ``normalized'' version of the swap interval formula:
\begin{equation}
    s(c_x) = \frac{ 2|\cEff| }{\Delta |(E_x-D_x)-(E_y-D_y)|}\left(\frac{|2\widetilde{c_x}^2-1|}{\widetilde{c_x}\sqrt{1-\widetilde{c_x}^2}}\right)\left|\frac{\delta}{\Gamma-1} \right|^k~,
    \label{e:swap_interval_equation_normalized}
\end{equation}

where $s(c_x)$ is the length of the interval at fixed $\Delta$ over which a swapping eigenvector goes from having squared overlap $|c_x|^2$ with $\ket{x}$ to having that same squared overlap $|c_x|^2$ with $\ket{y}$ instead. Fig.~\ref{fig:s_theory_numerical_example} shows the formula has good agreement with numerical simulation. As a perturbative formula, its accuracy is highest when it is near the value of $\Delta=\Delta^*$ at which wrap-around happens (Fig.~\ref{fig:s_theory_numerical_example}(B) has a larger relative error due to it having smaller $k$ compared to the eight-level example, since the approximation of a symmetric swap about the minimum gap of the avoided crossing is better suited the greater $k$ is than 1). Further, the unnormalized version tends toward underestimating the interval length (it predicts a finite interval between two high overlaps that might not be attained due to swap happening between eigenvectors not exactly $\ket{x},\ket{y}$), and the normalized version inevitably overestimates the interval length (it predicts an infinite interval length for a completed swap $|c_x|/|c_0|=1$). 

Therefore, the swap only takes place over some finite interval outside of which dynamics due to first-order non-degenerate perturbation theory dominate.

\subsection{Discrete Landau-Zener}
\label{s:discrete_landau_zener_derivation}
The eigenvector swap occurs in the neighbourhood of the gap from wrap-around. The other eigenvalues are well-separated from those associated with the swapping eigenvectors, as illustrated in  \Fig{eight_level_avoided_crossing_example}. 
%RESOLVED\tad{The figure doesn't have a bottom-half. Can we just say the figure itself, without specifying a part of it? Or say the inset in the figure?}
Thus the dynamics in the neighbourhood of the avoided crossing can be described by an off-diagonal perturbation in the effective two-level Hamiltonian \cite{cohen_2020}
\begin{equation}\label{e:eff_ham}
    H_{\textrm{eff}}= \begin{bmatrix} E_{1}(f) & a\\a^{\ast} & E_{2}(f) \end{bmatrix}~,
\end{equation}
where $E_{1}$ and $E_{2}$ are the unperturbed energies of the eigenvectors near the avoided crossing which vary with $f$, and $a$ is the size of the off-diagonal coupling. The energy gap at the avoided crossing is twice the value of $a$. The leading order change in energy difference can be approximated by a linear function \begin{equation}
    \mathcal{E} = E_{2} - E_{1} \equiv \Dot{\mathcal{E}} \delta j~,
\end{equation}
%\sh{[SH: huh? at minimum poor notation here]}
where $\delta j$ is number of steps %\tad{Is $j$ the number of steps (an integer)? $\delta j$ isn't an integer??} 
away from the avoided crossing and the total step count $j$ and $f$ are related by Eq.~\ref{e:f}. When the value of $p$ is large enough that the state vector follows a QAOA eigenvector, the Landau-Zener formula \cite{Wittig2005} gives the probability of a diabatic transition between the two eigenvectors
\begin{equation}\label{e:P_D}
    P = \exp(-\frac{2 \pi |a|^2}{|\Dot{\mathcal{E}}|})~.
\end{equation}
In our case, we can estimate the ``unperturbed'' energies as the eigenvalues of the cost and mixer diagonals in the cost (mixer) eigenbasis for wrap-around at $f=1$ ($f=0$). For wrap-around at $f=1$, the energies as a function of $f$ are simply 
\begin{equation}
    E_1(f) = D_x\Delta(1-f) + E_x\Delta f~,
\end{equation}
\begin{equation}
    E_2(f)=D_y\Delta(1-f) + E_y\Delta f~.
\end{equation} 

Assuming that the wrap-around between these two eigenvalues takes place for the first time at $\Delta=\Delta^*$ gives
\begin{equation}
    \frac{\partial \mathcal{E}}{\partial f} = \Delta((E_x-E_y)-(D_x-D_y)) = 2\pi(1+\delta/\Delta^*)(1-\Gamma)~,
\label{e:partial E partial f}
\end{equation}
where $E_j,D_j,\Gamma$ are given in Eq.~\ref{e: Dj_Ej_Gamma_definition} and $\delta=\Delta-\Delta^*$. This estimation of the ``unperturbed" energies is only valid for $\Gamma$ not close to 1. Otherwise higher-order corrections become more relevant. Further, $|a|$ is half the magnitude of the perturbed gap between the eigenvalues, and so from Eq.\ref{e:main_gap_equation} is given by  
\begin{equation}
|a| =  \left|\frac{\cEff}{(\Gamma -1)^k} \right|\left(\delta \right)^k = C_1\delta^k~,
\label{e:a_magnitude}
\end{equation}
where $C_1>0$ depends only on  ``normalized" Hamiltonians $\Delta^*\Hmixer$, $\Delta^*\Hcost$ and the magnitude of the product of off-diagonal coupling terms of \Hmixer.  

From Eq.~\ref{e:P_D}, the energy difference is a function of $f$ which in turn is inversely proportional to $p$ for a linear ramp and therefore the energy difference per QAOA step near the avoided crossing is given by 
\begin{equation}
   \Dot{\mathcal{E}} \simeq \frac{\delta \mathcal{E}}{\delta j} = \frac{\partial \mathcal{E}}{\partial f} \frac{\delta f}{\delta j} = \frac{\partial \mathcal{E}}{\partial f} \frac{1}{p}~,
\end{equation}
where the partial derivative is taken on either side of the avoided crossing. Using this in Eq.~\ref{e:P_D} shows that the probability has an exponential dependence on the number of steps as captured by Eq.~\ref{e:P_p}
\begin{equation}
    P = \exp(- 2 \pi p |a|^2 \abs{\frac{\partial f}{\partial \mathcal{E}}})~,
    \label{e:dlz_general_formula}
\end{equation}
\begin{equation}
    P = \exp(- \frac{B_0}{|1+(\delta/\Delta^*)|} \delta^{2k}   p )~,
    \label{e:dlz_edge_formula}
\end{equation}
where 
\begin{equation}
    B_0 = \frac{C_1^2}{|1-\Gamma|} = \frac{(\cEff)^2}{|1-\Gamma|^{2k+1}}~,
\label{e:b0 definition}
\end{equation}
with $\cEff$ as defined in Eq.~\ref{e:effective_coupling} and $\Gamma$ as defined in Eq.~\ref{e: Dj_Ej_Gamma_definition}.

For the special case of $k=1$, using Eq.~\ref{e:partial E partial f}~and Eq.~\ref{e:k1 gap}~in Eq.~\ref{e:dlz_general_formula}~with $c=\bra{x}\Hmixer\ket{y}$ gives:
\begin{equation}
P = \exp{\left(\frac{c^2}{(1-\Gamma)^2 +(2c/(E_x-E_y))^2}\frac{\delta^2}{|(1+(\delta/\Delta^*))(1-\Gamma)|} p \right)}~.
\label{e:k1 dlz}
\end{equation}

\section{Details for eight-level example}
\label{s:dlz example derivation}

\subsection{Computing discrete Landau-Zener formula for each wrap-around}

This appendix derives the curves in Fig~\ref{f:dlz performance benchmark}(A), which uses the Hamiltonian pair from Fig~\ref{f:eight_level_avoided_crossing_example}.
The ground state, $\ket{000}$, wraps around with four excited states at $f=1$ over the span of $\Delta$ considered in the plot. In particular, the states $\ket{111},\ket{110},\ket{101}$, and  $\ket{100}$. Each of these, in turn. wrap around with other states, but we are only interested in the gaps from wrap-around with the ground state as those are responsible for the deteriorating performance. First we need to compute $|\cEff|$ for each wrap-around. To do so, we can use the simplified expressions of Eq.~\ref{eq: q_polynomial_def} in Table~\ref{table:eq_tab} for $k=1,2,3$, as we use the X-mixer rescaled by 0.5 which directly couples states with Hamming distance 1. Further, as a consequence of this mixer choice, we have $\Gamma=0$ from Eq.~\ref{e: Dj_Ej_Gamma_definition}. The energies of the cost Hamiltonian are given by:
\begin{equation}
\{E_{000}=0,\ \ E_{001}=2.5,\ \ E_{010}=3,\ \ E_{011}=3.5,\ \ E_{100}=4.5,\ \ E_{101}=5.5,\ \ E_{110}=6,\ \ E_{111}=7 \}~.
\label{e:example cost energies}
\end{equation}
For the wrap-around between $\ket{000}$ and $\ket{111}$, we have $\Delta^* = 2\pi/7$. The coupling paths we must consider are all distinct ways to get from $\ket{000}$ to $\ket{111}$ in three bit-flips.  From Eq.~\ref{eq:mixer_path_contributions} we have $P(S)=0.5^3$ for all coupling 3-paths, and from Eq.~\ref{e:effective_coupling}, we have:
$$
    \cEffspecific^{000,111} = 0.5^3\left(\frac{1}{4}\cot\left(\frac{\Delta^*}{2}E_{001}\right)\cot\left(\frac{\Delta^*}{2}E_{011}\right)+\frac{1}{12} \right) + 0.5^3\left(\frac{1}{4}\cot\left(\frac{\Delta^*}{2}E_{001}\right)\cot\left(\frac{\Delta^*}{2}E_{101}\right)+\frac{1}{12} \right)
$$
$$
+0.5^3\left(\frac{1}{4}\cot\left(\frac{\Delta^*}{2}E_{010}\right)\cot\left(\frac{\Delta^*}{2}E_{011}\right)+\frac{1}{12} \right) +0.5^3\left(\frac{1}{4}\cot\left(\frac{\Delta^*}{2}E_{010}\right)\cot\left(\frac{\Delta^*}{2}E_{110}\right)+\frac{1}{12} \right)
$$
$$
+0.5^3\left(\frac{1}{4}\cot\left(\frac{\Delta^*}{2}E_{100}\right)\cot\left(\frac{\Delta^*}{2}E_{101}\right)+\frac{1}{12} \right) + 0.5^3\left(\frac{1}{4}\cot\left(\frac{\Delta^*}{2}E_{100}\right)\cot\left(\frac{\Delta^*}{2}E_{110}\right)+\frac{1}{12} \right)
$$
$$ \approx 0.5^3\times 0.6315\approx 0.0789~.$$
Applying Eq.~\ref{e:dlz_edge_formula}, and defining $\delta_{111} = \Delta-2\pi/7$, we have
$$P =\exp\left(-\ \frac{0.0789^2}{1+(7\delta/2\pi)} \delta_{111}^6 p \right)~.$$
Solving for $p$ that gives the value $P=0.95$ for this state gives the value $p=p_{111}$
\begin{equation}
p_{111} = -\frac{\ln(0.95)}{0.0789^2}\frac{(1+(7\delta_{111}/2\pi))}{\delta_{111}^6}~.
\label{e:p111}
\end{equation}
For the wrap around between $\ket{000}$ and $\ket{110},\ket{101}$, we have $\Delta^* = 2\pi/6$ and $\Delta^*=2\pi/5.5$ respectively. Considering all  possible paths between $\ket{000}$ and $\ket{110},\ket{101}$ in two bit-flips gives:
$$\cEffspecific^{000,110} = \frac{0.5^2}{2}\cot\left(\frac{\Delta^*}{2}E_{100} \right) + \frac{0.5^2}{2}\cot\left(\frac{\Delta^*}{2}E_{010} \right)  \approx 0.1250~,$$
$$\cEffspecific^{000,101} = \frac{0.5^2}{2}\cot\left(\frac{\Delta^*}{2}E_{100} \right) + \frac{0.5^2}{2}\cot\left(\frac{\Delta^*}{2}E_{001} \right)  \approx 0.1765~.$$
Solving for $p$ after plugging in $P=0.95$ and the relevant constants into Eq.~\ref{e:dlz_edge_formula} leads to 
\begin{equation}
p_{110} = -\frac{\ln(0.95)}{0.125^2}\frac{(1+(6\delta_{110}/2\pi))}{\delta_{110}^4}~,
\label{e:p110}
\end{equation}
\begin{equation}
p_{101} = -\frac{\ln(0.95)}{0.1765^2}\frac{(1+(5.5\delta_{101}/2\pi))}{\delta_{101}^4}~,
\label{e:p101}
\end{equation}
where $\delta_{110}=\Delta-2\pi/6$ and $\delta_{101} = \Delta-2\pi/5.5$. In the $k=1$ case we have $\Delta^*=2\pi/4.5$ and $\delta_{100}=\Delta-2\pi/4.5$. By applying Eq.~\ref{e:k1 dlz} and solving for $P=0.95$, we obtain
\begin{equation}
p_{100} = -\frac{\ln(0.95)(1+(1/4.5)^2)}{0.5^2} \frac{1+(4.5\delta_{100}/2\pi)}{\delta_{100}^2}~.
\label{e:p100}
\end{equation}
\begin{table}
\begin{center}
\begin{tabular}{ |c|c|c|c|c| }
\hline
Wrapping Excited State $\ket{y}$ & $\Delta^*$ & $\cEffspecific^{000,y}$ & $\Gamma$ & $B_0$\\
\hline
$\ket{111}$ & $2\pi/7$ & 0.0789 & 0 & $0.0789^2$\\
\hline
$\ket{110}$ & $2\pi/6$ & 0.125 & 0 & $0.125^2$\\
\hline
$\ket{101}$ & $2\pi/5.5$ & 0.1765 & 0 & $0.1765^2$\\
\hline
$\ket{100}$ & $2\pi/4.5$ & 0.5 & 0 & $0.5^2$\\
\hline
\end{tabular}
\caption{Parameters for the eight-level example}
\end{center}
\label{table:discrete_dlz_parameter_table}
\end{table}

Table~\ref{table:discrete_dlz_parameter_table} summaries the resulting parameters.

\subsection{Combined effect from all gaps}
In this section, we derive the curve displayed in Fig.~\ref{f:dlz performance benchmark}(B).

We want to estimate the value of $p$ at which the squared overlap of the QAOA output state with the ground state is 0.95. Under the assumption that $p$ is large enough for the QAOA state vector to track the eigenvectors up to the first few gaps, we get
\begin{equation}
0.95 = (0.95)^{p/p_{111}}(0.95)^{p/p_{110}}(0.95)^{p/p_{101}}(0.95)^{p/p_{100}} \quad,\quad p = (p_{111}^{-1}+p_{110}^{-1}+p_{101}^{-1}+p_{100}^{-1})^{-1}~,
\label{e:p from all gaps}
\end{equation}
with the $p_s$ terms given by Eqs.~\ref{e:p111}-\ref{e:p100}. For values of $\Delta$ below the $\Delta^*$ at which wrap-around with state $\ket{s}$ occurs, we take $p_s\rightarrow \infty$. As expected, the smallest $p_s$, if significantly smaller than the others, dominates the shape of the right edge of the performance ridge.

\subsection{Correcting for poor tracking}
Here, we derive the curve in Fig.~\ref{f:dlz performance benchmark}(C). We approximate the squared overlap due to non-wrap-around gaps as occurring for $1-R^{p}$ for some unknown constant $R$. This can be re-written as $1-0.05^{p/p_{0.95}}$ where we approximate $p_{0.95}$ as the $p$ at which there is a squared overlap of $0.95$ with the ground state at $\Delta = 0.895$, just below $\DeltaCrit\approx 0.8976$. This is numerically determined to be between 14 and 15 - we use 14.5 to generate the curve. The overall formula becomes
$$ 0.95 =  (0.95)^{p/p_{111}}(0.95)^{p/p_{110}}(0.95)^{p/p_{101}}(0.95)^{p/p_{100}}\left(1-0.05^{p/p_{0.95}} \right)~, $$
with $p$ given by 
\begin{equation}
p = \left(p_{111}^{-1}+p_{110}^{-1}+p_{101}^{-1}+p_{100}^{-1} + \frac{\ln(1-0.05^{p/p_{0.95}})}{p\ln(0.95)}\right)^{-1}~.
\label{e:p overall corrections}
\end{equation}
We numerically scanned values of $p$ near the ansatz from Eq.~\ref{e:p from all gaps} such that the two sides in Eq.~\ref{e:p overall corrections} are as equal as possible to produce Fig.~\ref{f:dlz performance benchmark}(C).

\section{Local-constraint satisfaction problems with the $X$-mixer: effects of wrap-around}
\label{s:local_constraint_satisfaction_scaling}

\subsection{Exponential scaling of gap size for most wrap-arounds in this setting}
\label{s:gap size for most wrap-arounds}

In this section, we examine a special case where the scaling of the gap size from wrap-around has an exponent potentially much larger than the coupling distance between the wrapping states. 

Consider a cost Hamiltonian $H=H_1 + H_2$ and the two sets of qubits, $Q_1$ and $Q_2$, that $H_1$ and $H_2$ respectively perform non-identity operations on. Suppose  $Q_1,Q_2\neq \varnothing, Q_1\cap Q_2 =\varnothing, $, with $H$ diagonal in the computational basis. Suppose wrap-around takes place between two cost eigenstates that differ by bit flips on a set of qubits $Q$ and that $Q_1\cap Q\neq \varnothing,\ Q_2\cap Q\neq \varnothing$. Suppose the wrapping cost eigenstates differ by energy $E$ on $H$ and by $E_1$ and $E_2$ on $H_1$ and $H_2$ respectively. Note that since $Q_1\cap Q_2 = \varnothing$, the qubits under the two Hamiltonians are separate systems, and so all changes in eigenstate connections between the mixer and $H$ must correspond to changes in eigenstate connections between the mixer and $H_1$ and $H_2$ restricted to $Q_1$ and $Q_2$ respectively. Consequently, the mixer eigenstate will only change its connection to the eigenstate whose eigenvalue is wrapping around at $\Delta=2\pi/E$ if it corresponds to a wrap-around for both $\Delta = 2n\pi/E_1$ and $\Delta = 2m\pi/E_2$ (otherwise the change will be to an eigenstate with only some or no bits flipped, i.e. to the same cost eigenstate or to a cost eigenstate with energy difference just $E_1$ or just $E_2$). 

Now consider the setting of two cost eigenstates $\ket{x}$ and $\ket{y}$ wrapping around with $\Hcost = \sum_{i,j} J_{ij}Z_iZ_j + \sum_i h_i Z_i$ where at most $cn$ of the coupling terms $J_{ij}$ are nonzero for some constant $c$. Suppose that these eigenstates differ by bit flips on qubits with indices $I$ (different combinations of flips on these qubits form the shortest coupling path between the states by the X-mixer). Therefore, for qubits $q_i$ with $i\notin I$, the single-qubit ket $\ket{q_i}\in \{\ket{0},\ket{1} \}$ will be the same for all states in all of these paths, and so for the specific purpose of calculating the contribution to the gap from the energies of the states along the shortest coupling path, we can simplify every Pauli operator in $\Hcost$ as follows:
$$\forall i \notin I,\ \ Z_i \rightarrow \bra{q_i} Z_i\ket{q_i}\mathbb{I}_i,\quad \forall i \in I,\ \ Z_i\rightarrow Z_i~.$$
If the $\Hcost$ couplings on the flipped qubits fail to form a connected set, then the resulting ``reduced'' Hamiltonian can be separated into two Hamiltonians acting non-trivially on disjoint, non-empty subsets of the qubits. Since we know that in this setting a continuous degeneracy almost always forms, the contributions of all the shortest $k-$coupling paths must cancel out to zero. Therefore, the contributions of coupling paths passing only through states whose bit-flipped qubits don't have \Hcost~couplings forming a connected set over the qubits must cancel out (i.e. $\cEff = 0$ from Eq.\ref{e:effective_coupling}~and Eq.\ref{e:main_gap_equation},~and any higher order terms involving only those states must cancel as well). \\

Let $\xi$ be the size of the smallest set of qubits that (1) has the qubits subject to bitflips along the shortest coupling paths of length $k$ as elements and (2) has the $\Hcost$~couplings form a connected graph over the qubits in the set. Since the bits not in the shortest coupling path have to be flipped twice (since they don't differ between the starting and final states), a total of $2(\xi-k) + k = 2\xi-k$ bit flips are required. Consequently, the gap from the wrap-around scales as $O(\delta^{2\xi-k})$.

For local constraint satisfaction problems on $n$ logical qubits, the probability of two qubits sharing a coupling goes as $\sim c/n$ for some constant $c$. From random graph theory~\cite{erdos1960}, we know that subgraphs of size significantly smaller than $n$ have a vanishingly small probability of being connected. From this, we can infer that almost all (though certainly not all, for an overall connected graph of couplings over all qubits) wrap-arounds involving two cost eigenstates differing by a sublinear-in-qubit-number of bit flips nevertheless have $\xi=\Theta(n)$ and therefore have gaps scaling exponentially in qubit number. 

This analysis is useful outside of local-constraint satisfaction problems. We can consider the case where the cost and mixer Hamiltonians are both decomposed into sums of Pauli strings. In this case, if $\ket{x}$ and $\ket{y}$ are the wrapping cost eigenstates, then we consider a set of qubits subject to constraints that (1) the set cannot be separated into two disjoint subsets where the every Pauli string of the cost and mixer acts non-trivially on at most one of the subsets and (2) the number of mixer Pauli strings required to flip bits to go from $\ket{x}$ to $\ket{y}$ is minimized to some number $\nu$. Then the gap from wrap-around scales as $O(\delta^\nu)$.

\subsection{Performance ridge: location scaling with problem size} 
\label{s:derivation of performance scaling}
In this section, we justify our claim that some fixed degree of performance deterioration due to wrap-around is attained at $p$ polynomial in $n$ (the number of qubits) from wrap-around with as much as a single excited state differing by $\Theta(w(n))$ bit flips and $\Theta(w(n))$ energy from the ground state, with $w(n):= \log(n)/\log(\log(n))$ from Eq.~\ref{e:bound_to_ridge}, which we call ``{quasi-logarithmically distant excited states}". This forms an upper boundary at $\Delta \propto 1/w(n)$ for where the right edge of the performance ridge can be for $p=n^h$ with a constant $h>0$, since performance deterioration occurs past this point due to wrap-arounds at $f=0,1$ (though we cannot say what happens to performance \emph{up} to this point). Since $X$-mixer eigenstates differing by $k$ flips (i.e. $\ket{+}\leftrightarrow\ket{-}$) are connected by the $\Theta(k)$ transitions by the $Z_iZ_j$ terms of $\Hcost$, this analysis applies equally to wrap-around by cost and mixer states. Further, it ought be noted that our analysis applies even to the case of $\Hcost$ with degenerate ground states, since the $X$-mixer has a non-degenerate ground state, and so even the wrap-around of a single mixer excited state with this quasi-logarithmic coupling distance is enough to change the connection to some distant cost excited state. 

We make our argument as follows. First, we show that the quasi-logarithmically distant excited states are the most distantly coupled set of states from the ground state for which, at $\delta= \Delta^*$ (i.e., $\Delta = 2\Delta^*$) a single gap is large enough to achieve some fixed, low diabatic transition rate to the ground state at $p = \Theta(n^h)$ for some constant $h>0$. Thereafter, we show that this result is unaffected by the a large number of very small gaps from more distant wrap-arounds, because they collectively do not result in some noticeable reduction in the diabatic transition to the ground state prior to the wrap-around of quasi-logarithmically distant excited states. That is, a ``death by a thousand cuts'' does not happen. 

\begin{figure}
    \centering
    \includegraphics[width=0.7\textwidth]{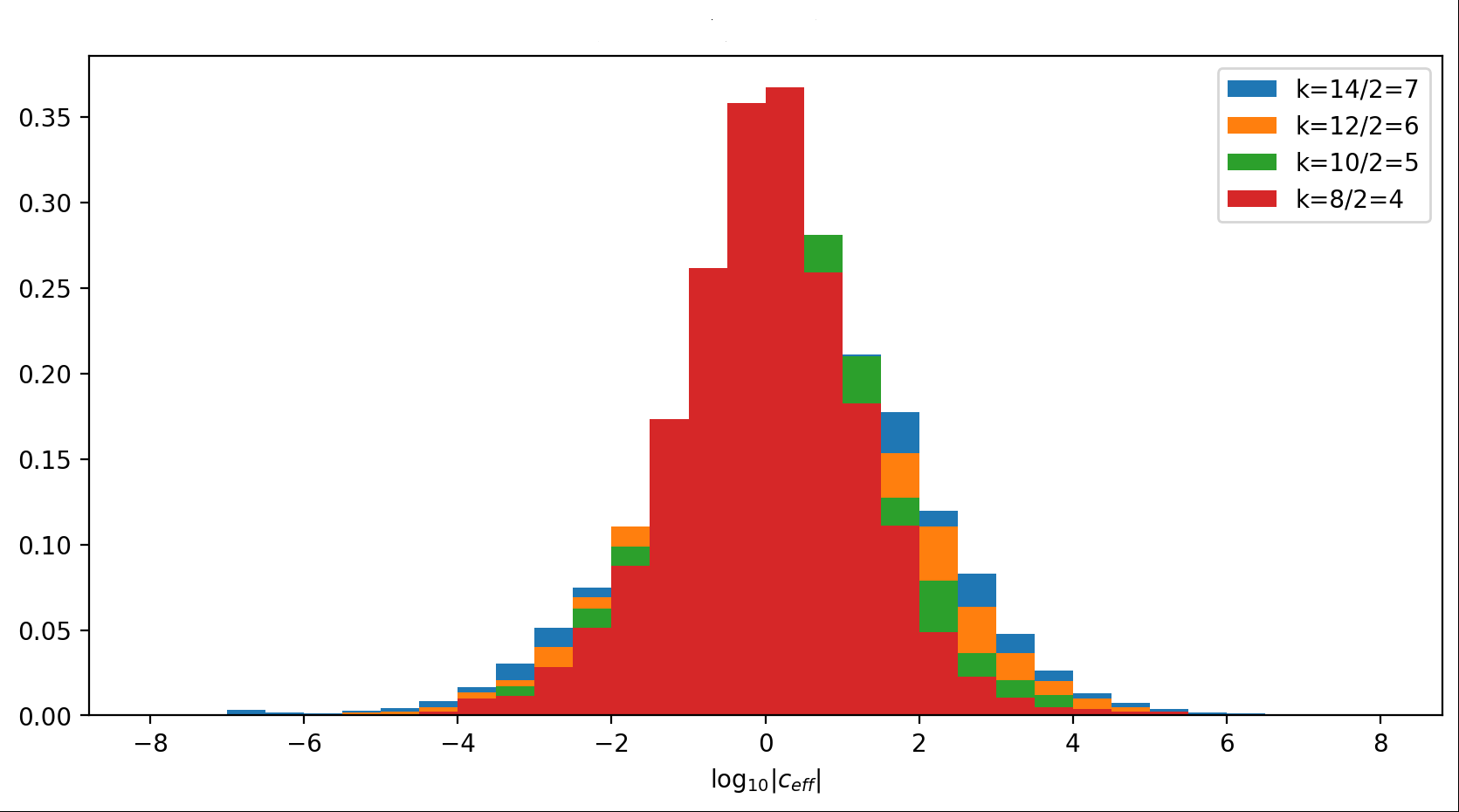}
    \caption{Distributions of $\log_{10}\left|\cEff\right|$ for non-zero values of \cEff. Each distribution is composed of all non-zero \cEff\ values from wrap-around between the ground state and states that it is $k$-coupled to in each of 100 randomly generated cost Hamiltonians (with coupling provided by the X-mixer). Here, $k=n/2$ for Hamiltonians on $n=8,10,12,14$ qubits. Each cost Hamiltonian was generated by introducing a $Z_iZ_j$ coupling between two qubits with probability $3/n$, with the weight of that coupling (and each single-qubit $Z_i$ operator) chosen uniformly from a distribution between -1 and 1. If a qubit was not coupled to any other qubit after the Hamiltonian was generated, a post-processing step couples that qubit to some other qubit that is chosen uniformly at random.  }
    \label{fig:ceff_distribution}
\end{figure}

The rigor of our argument has some limitations. Firstly, we use lowest-order estimates of the gap sizes to make these conclusions. For sufficiently large $\delta$, the lowest-order calculation \emph{over}estimates the gap size (the eigenvalues are constrained to be on the complex unit circle) and thus our argument on the irrelevance of distant excited states remains valid. Additionally, the values of $\delta$ we consider are much smaller than $1$, making the approximation reasonable. Another limitation is the estimation of a typical scaling of $B_0=(\cEff)^2$ from Eq.~\ref{e:dlz_edge_formula}. There is numerical evidence that this term scales exponentially with coupling distance (see Fig.~\ref{fig:ceff_distribution}), but this remains unproven. Finally, we assume that any higher-order transitions back to the ground state occur through several other gaps, and thus are negligible compared to the reduction in diabatic transitions to the ground state at the ``diabatic'' ground state eigenvalue curve's crossings with other excited states.

\subsubsection{Polynomially small gaps}
\label{s:polynomially small gaps}
For $\Delta = 2\Delta^*$, the exponent of Eq.~\ref{e:dlz_edge_formula} attains a non-growing diabatic transition to the ground state at $p=\chi n^h$ for $\chi,n>0$ if $B_0\delta ^{2k}\geq 2 n^{-h}$, where $\delta=\Delta^*=2\pi/(E_s k)$, $k$ is the coupling distance, and $E_s$ is some constant converting coupling distance to energy difference. Note that $2\pi/E_s$ and the exponential-in-$k$ $B_0$ can be absorbed by multiplying $\delta$ by an overall constant. Therefore, we establish the smallest possible scaling of $\delta$ with respect to $n$ by solving:
\begin{equation}
\left(\frac{1}{k}\right)^{k} \geq n^{-h/2} ~\implies~ -k\log(k) \geq -(h/2)\log(n)~\implies~k(n) \propto  \ln(n)/\ln(\ln(n))~. 
\label{e:polynomially scaling gaps inequality}
\end{equation}
The condition is satisfied with equality for $h=2$ if $k(n)=\exp(\mathcal{W}(\ln(n)))$ where $\mathcal{W}$ is the Lambert W function~\cite{dlmf}. Consequently, for large $n$, and arbitrary $h$, $k(n) \propto \ln(n)/\ln(\ln(n))$. 
For the $X$-mixer, and likely for the cost Hamiltonian as well, at least one excited state exists that has both a coupling distance and energy difference of $\propto w(n)$. Without loss of generality, we can limit our discussion to that of the \Hcost\  excited/ground states. 

\subsubsection{Irrelevance of distantly coupled excited states for the ridge boundary at polynomial \textit{p}}
\label{s:irrelevance of distantly coupled excited states}

We consider the diabatic transition to the ground  state occurring as a product of transitions from Eq.~\ref{e:dlz_edge_formula}. For the $X$-mixer, $\Gamma=0$, so $B_0=(\cEff)^2$. We can make some simplifying assumptions - since  \cEff\ has been observed to have exponential scaling in coupling distance, it will not affect the relevant scale of $\Delta$ at which which we see performance deterioration above some fixed constant, so can be neglected and set to 1. Additionally the denominator in the exponent of Eq.~\ref{e:dlz_edge_formula} can provide a factor of at most $n$ so will be ignored for determining polynomial scaling. We also ignore ``repeat" wrap-arounds, since there can only be $O(n)$ many for $\Delta<1$ for each gap, again, not affecting whether the scaling is polynomial. 

To show that a ``death by a thousand cuts" does not occur, we estimate the highest possible contributions to transitions away from the ground state for which the wraparound gaps could be responsible. The first overestimate has us use $\delta \rightarrow \Delta$, clearly an overestimate to the gap size from the usual $\delta=\Delta-\Delta^*$ for wrap-around. The other overestimate is that we treat the effect of all wrap-around gaps from excited states a coupling distance $k$ away from the ground state as having the same energy difference from the ground state as the excited state with the largest energy difference. Consequently, we treat these wrap-arounds one family (characterized by having the same coupling distance to the ground state) at a time - since there are only $n$ such families, this cannot change whether the scaling this gives us is polynomial.

We now deal with some constraints. Suppose we have $\Delta(n) = 1/f(n)$. The only excited states that have wrapped around by this point are those with energy difference $E$ from the ground state energy, such that $2\pi/E < 1/f(n)$. For both the $X-$ mixer and local constraint satisfaction problems, states differing by a coupling distance $k$ from the ground state $E(k) = O(k) = O(n/m)$ for flipping $1/m$ of all the bits. From $E(n/m)=O(n/m)$ and $ \frac{2\pi}{E}<1/f(n)$, we get the constraint that 
\begin{equation}
    f(n) = O(n/m) \rightarrow m\leq \mu_0\frac{n}{f(n)}~,
\label{e:constraint 1}
\end{equation}
for some fixed positive constant $\mu_0$ and sufficiently large $n$.  Now, note that there are ${n \choose n/m}$ such excited states. Assuming they all contribute to with scaling $\Delta^{2n/m}$, we get the following requirement
$$\log\left( {n \choose n/m}\times \left(\frac{1}{f(n)}\right)^{2n/m} \right) \geq \log(n^{-h})~.$$
Asymptotically, $\log\left({n \choose n/m }\right) = (n/m)(m\log(m/(m-1)) +\log(m)) +O(\log(n))$, and so keeping the highest-order terms relevant for scaling, in the limit of large $n$ we have 
\begin{equation}
\frac{n}{m}\left(\log(m)+m\log\left(\frac{m}{m-1}\right)-2\log(f(n)) \right) \geq -h\log(n)~.
\label{e:constraint 2}
\end{equation}
For constant $m$ and any growing $f(n)$, the terms inside the parentheses eventually become negative and the scaling inequality is violated for large enough $n$. So $m$ must grow with $n$ (and any excited states of coupling distance $\Theta(n)$ from the ground state are irrelevant), which means we can asymptotically neglect the $m\log(m/(m-1))$ term as going to 1. 

The last thing that we must take into account is the probability that a particular set of bitflips is connected on the graph formed by couplings from \Hcost. This is because if they are not connected, then the scaling from the gap must be worse than the coupling distance $k$, as detailed in Section~\ref{s:gap size for most wrap-arounds}. Any connected graph of $\xi$ qubits (in terms of $\Hcost$ couplings between its members) represents an excited state differing from the ground state by those $\xi$ flips whose wrap-around goes as $\Theta(\delta^\xi)$. Every such connected set has some \emph{disconnected}, smaller subgraphs on $k<\xi$ qubits corresponding to wrap-arounds from less distantly coupled excited states which \emph{do not} scale as $\sim \delta^k$ but rather as $\sim \delta^{2\xi-k}$. Note that these gaps (i) are much smaller than the gaps from wrap-arounds with $\xi$ coupling distance to the ground state, due to the extra $\delta^{\xi-k}$ factor and so are as negligible as higher order estimates of gaps and (ii) there are at most $2^{\xi}$ such states for each connected graph on $\xi$ qubits, and thus even if they weren't smaller, can be corrected for simply by rescaling $\delta$ by a constant of 2, not affecting scaling. Consequently, gaps from excited states differing by disconnected bit flips with the ground state may be neglected for the purposes of our analysis. 

We define $\mathcal{P}(n,m)$ as the probability that randomly selected set of $n/m$ qubits from $n$ qubits is connected. We find an upper bound to this quantity with random graph theory. If there are a total of $nc/2$ constraints introduced with constant $c\geq 2$ (which translate to qubit couplings), and the number of constraints per qubit bounded to be no more than some constant multiple of $c$, for large $n$ and large $m$ (we know from above that $m$ must grow with $n$), we approximate the probability of edges between our $n/m$ vertices as $p=c/n$ with a random graph of $G(v=n/m, p=c/n)$ - this is a graph on $v$ vertices where an edge is introduced between each pair of vertices with probability $p$. This serves as an upper bound as it counts amidst its possibilities connected graphs that are not possible with a bounded degree. Combining this with a simplification of Eq.~\ref{e:constraint 2} gives us a final constraint of:

\begin{equation}
\log\left(\mathcal{P}(n,m)\right) +\frac{n}{m} \left(\log(m) -  2\log(f(n))\right) \geq -h\log(n)~.
\label{e:polynomial_scaling_constraint}
\end{equation}

\subsubsection{Bounds on probability of couplings on flipped qubits forming a connected graph}

First, we construct an upper bound for the probability of a randomly generated graph $G(v,p)$ being connected. From Cayley's formula~\cite{bollobas2001}, we know that the number of possible distinct trees on a set of labeled $v$ vertices is $v^{v-2}$. Each tree represents a spanning tree for some connected set, so, letting $Q:=(v-1)(v-2)/2$, for every distinct tree, we can construct ${Q \choose l}$ connected graphs with $v-1+l$ edges, each of which has probability of $p^{v-1+l}(1-p)^{Q-v-l+1}$ of forming. Summing these probabilities together from $l=0$ to $l=Q$ gives us an upper bound for the probability of connectivity as each connected graph is counted this way, with most connected graphs being double-counted due to being constructable from several distinct spanning trees. Replacing $v^{v-2}$ with $v^{v-1}$ for simplicity gives us an upper bound expression
$$ (1-p)^Q(pv)^{v-1}\left(\sum_{l=0}^Q {Q \choose l} p^l(1-p)^{-l}\right) \leq (1-p)^Q(pv)^{v-1}\left(\sum_{l=0}^Q \frac{1}{l!}\left(\frac{Qp}{1-p}\right)^l\right)~.$$
In our setting $v=n/m$, $Q=O(n^2)$ and $p=c/n$ for $c\geq 2$, and thus we only need to consider Taylor expansions in $p$ up to second order for the bounds to apply in the large $n$ limit
$$(1-p)^Q(pv)^{v-1}\left(\sum_{l=0}^Q \frac{1}{l!}\left(\frac{Qp}{1-p}\right)^l\right) \leq \exp\left\{Q\left(-p-\frac{p^2}{2}\right) \right\}(pv)^{v-1}\exp\left\{ Qp(1+(3/2)p)\right\}= (pv)^{v-1} e^{Qp^2} \leq (pv)^{v-1} e^{(pv)^2}~.$$
Substituting $v=n/m$ and $p=c/n$, gives us an upper bound of 
\begin{equation}
\mathcal{P}(n, m) \leq \left(\frac{c}{m}\right)^{\frac{n}{m}-1}\exp\left\{\left(\frac{c}{m}\right)^2 \right\}~.
\label{e:probability upper bound}
\end{equation}
The upper bound can be further tightened using results on the minimum number of spanning trees of a graph~\cite{bogdanowicz2009}, but as we will show the above bound will be sufficient for our purposes.

\subsubsection{Performance ridge right boundary scaling from wrap-around gaps}

Re-introducing the upper bound from Eq.~\ref{e:probability upper bound} to account for the effects of Section~\ref{s:gap size for most wrap-arounds}\ into Eq.~\ref{e:polynomial_scaling_constraint}\ gives us:
$$\frac{n}{m}\left( \log(m) +\log(c/m) - \log((f(n))^2)\right)-\log(c/m)+(c/m)^2 \geq -h\log(n)~.$$
As the extra additive, growing terms are unimportant for our scaling, this gives us:
$$\frac{n}{m}\left(\log(c)-2\log(f(n)) \right)\geq \log\left(n^{-h}\right)~.$$
Since the term in the parentheses inevitably becomes negative (and we can simply neglect the constant inside as not affecting overall scaling), we want to choose as large of an $m$ as possible, namely from Eq.~\ref{e:constraint 1}, giving us
\begin{equation}
-f(n)\log(f(n))\geq -\widetilde{h}\log(n) ~,
\label{e:final scaling constraint}
\end{equation}
where $\widetilde{h}$ is a positive constant obtained by moving all the multiplicative constants from the left. This is the same setup as in Section~\ref{s:polynomially small gaps}'s Eq.~\ref{e:polynomially scaling gaps inequality}. Eq.~\ref{e:final scaling constraint}\ tells us that the fastest possible shrinking $\Delta(n)$ (inverse of fastest-growing $f(n)$) goes as $\propto w(n) = \log(\log(n))/\log(n)$ while Section~\ref{s:polynomially small gaps} explicitly provides examples of large gaps that emerge at this scale of $\Delta$.

For $p$ polynomial in $n$, this yields an approximation of the scale of $\Delta$ at by which  the upper portion of the performance ridge must occur due to the wrap-around effect scaling as $\propto \log(\log(n))/\log(n)$. Notably, we cannot give performance guarantees \emph{up to} this point, so performance could deteriorate well below this scale.  

\end{document}